\documentclass[submission, Phys]{SciPost}
\hypersetup{
    colorlinks,
    linkcolor={red!50!black},
    citecolor={blue!50!black},
    urlcolor={blue!80!black}
}

\usepackage[bitstream-charter]{mathdesign}
\usepackage{placeins}
\usepackage{bm}
\DeclareSymbolFont{usualmathcal}{OMS}{cmsy}{m}{n}
\DeclareSymbolFontAlphabet{\mathcal}{usualmathcal}

\begin{document}

\begin{center}{\Large \textbf{
Nonpertubative Many-Body Theory for the Two-Dimensional Hubbard Model at Low Temperature: From Weak to Strong Coupling Regimes\\
}}\end{center}

% TODO: write the author list here. Use first name (+ other initials) + surname format.
% Separate subsequent authors by a comma, omit comma and use "and" for the last author.
% Mark the corresponding author with a superscript star.
\begin{center}
Ruitao Xiao\textsuperscript{1$\star$},
Yingze Su\textsuperscript{1$\star$},
Junnian Xiong\textsuperscript{1},
Hui Li\textsuperscript{2},
Huaqing Huang\textsuperscript{1$\dagger$} and
Dingping Li\textsuperscript{1$\ddagger$}
\end{center}

\begin{center}
{\bf 1} School of Physics, Peking University, Beijing 100871, China
\\
{\bf 2} School of Physics, Zhejiang University, Hangzhou 310058, China
\\

$\star$ {\small \sf These authors contributed equally to this work.} \\
$\dagger$ {\small \sf huanghq07@pku.edu.cn}, \qquad
$\ddagger$ {\small \sf lidp@pku.edu.cn}
\end{center}

\begin{center}
\today
\end{center}

% For convenience during refereeing (optional),
% you can turn on line numbers by uncommenting the next line:
%\linenumbers

\section*{Abstract}
{\bf
% TODO: write your abstract here.
In theoretical studies of two-dimensional (2D) systems, the Mermin-Wagner theorem prevents continuous symmetry breaking at any finite temperature, thus forbidding a Landau phase transition at a critical temperature $T_c$.
The difficulty arises when many-body theoretical studies predict a Landau phase transition at finite temperatures, which contradicts the Mermin-Wagner theorem and is termed a pseudo phase transition.
To tackle this problem, we systematically develop a symmetrization scheme, defined as averaging physical quantities over all symmetry-breaking states, thus ensuring that it preserves the Mermin-Wagner theorem.
We apply the symmetrization scheme to the GW-covariance calculation for the 2D repulsive Hubbard model at half-filling in the intermediate-to-strong coupling regime and at low temperatures, obtaining the one-body Green's function and spin-spin correlation function, and benchmark them against Determinant Quantum Monte Carlo (DQMC) with good agreement.
The spin-spin correlation functions are approached within the covariance theory, a general method for calculating two-body correlation functions from a one-particle starting point, such as the GW formalism used here, which ensures the preservation of the fundamental fluctuation-dissipation
relation (FDR) and Ward-Takahashi identities (WTI).
With the FDR and WTI satisfied, we conjecture that the $\chi$-sum rule, a fundamental relation from the Pauli exclusion principle, can be used to probe the reliability of many-body methods, and demonstrate this by comparing the GW-covariance and mean-field-covariance approaches.
This work provides a novel framework to investigate the strong-coupling and doped regime of the 2D Hubbard model, which is believed to be applicable to real high-$T_c$ cuprate superconductors.
}

\vspace{10pt}
\noindent\rule{\textwidth}{1pt}
\tableofcontents\thispagestyle{fancy}
\noindent\rule{\textwidth}{1pt}
\vspace{10pt}

\section{Introduction}
\label{sec:intro}
The Hubbard model is considered the simplest model of interacting fermions on a lattice, but it is very difficult to solve or to obtain reasonable approximations across its entire parameter space (for example, temperature and doping).
There is sufficient evidence that a full understanding of it is crucial for exploring unconventional superconductivity (particularly in the context of high-$T_{\mathrm{c}}$ superconductivity), antiferromagnetism, the pseudogap phase, strange metal behavior, charge-density waves (CDWs), and many other phenomena.
Substantial advances have been made in recent years in investigating the Hubbard model across various regions of its phase diagram, due to the development of different analytic and computational methods \cite{arovas_hubbard_2022,qin_hubbard_2022,rohringer_diagrammatic_2018,schafer_tracking_2021,Hao_Coexistence_Science2024}.
The exact location of phase boundaries, for example, the superconducting transition temperature of cuprate superconductors, is still an extremely challenging task \cite{Sakai_Nonperturbative_2023,Scalapino_a_common_2012,qin_hubbard_2022,Gull_Continuous-time_2011,Maier_Dynamical_2018}.\par

A key requirement for any approach in many-body physics is the preservation of three fundamental relations (or identities).
First, the fluctuation-dissipation relation (FDR), an exact relation derived by Kubo, connects the response function (i.e., the susceptibility) to the spectral density of fluctuations or the two-body correlators in the frequency domain \cite{kubo_Statistical_I_1957,kubo_Statistical_II_1957,Callen_Irreversibility_1951,Mahan_Many-particle_2013}, with its time-domain representation given by the Kubo formula \cite{stefanucci_nonequilibrium_2013}.
Second, the Ward identities (WI) represent rigorous constraints between correlation functions, such as the one- and two-body Green’s functions or the vertex and inverse Green's functions, for systems with continuous symmetry. 
Specifically, the Ward-Takahashi identities (WTI) reflect current conservation and ensure the validity of the $f$-sum rules \cite{Benfatto_Ward_2005,stefanucci_nonequilibrium_2013}.
These $f$-sum rules are indispensable for theoretical consistency checks and experimental analysis, as they constrain frequency integrals of response functions by basic physical quantities such as electron mass and density. 
Third is the local moment sum rule, known as the $\chi$-sum rule. 
Unlike the conservation-derived $f$-sum rule, this rule embodies the Pauli exclusion principle by ensuring the square of the local spin occupation equals the occupation itself \cite{rohringer_diagrammatic_2018,Vilk_Non-Perspective_1997,Bergeron_Optical_2011,Tremblay_Improved_2023,Vilk_Antiferromagnetic_2024}.
The $\chi$-sum rule serves as an accuracy test for impurity solvers in methods such as dynamical mean-field theory (DMFT) and its diagrammatic extensions \cite{rohringer_diagrammatic_2018}, and has also been employed as a foundational constraint in formulating self-consistent many-body theories \cite{Vilk_Non-Perspective_1997,Bergeron_Optical_2011,Tremblay_Improved_2023,Vilk_Antiferromagnetic_2024}.
Consequently, a viable approximation should aim to maintain these three relations, especially as the physical justification of the FDR and WTI is inextricably linked to experimental analysis; any violation would undermine the theory's plausibility.

\subsection{Methods and difficulties at low temperatures}
Determinant Quantum Monte Carlo (DQMC) is a numerically exact, unbiased stochastic algorithm that simulates the many-body path integral by randomly sampling auxiliary field configurations.
It preserves the required relations and captures full spatial correlations, making it a powerful tool for studying strongly correlated fermionic systems.
However, the method is limited by its computational scaling and by the serious fermion sign problem, which appears in the most physically interesting regimes, specifically, in doped systems at low temperatures near the superconducting phase transition temperature \cite{Hirsch_Two-dimensional_1985,Chang_recent_2015,SmoQyDQMC.jl,sun_delay_2024,iglovikov_geometry_2015}.
Nevertheless, DQMC can provide exact results in regimes where the sign problem is mild.
In particular, the half-filled case is free of the sign problem, allowing access to non-perturbative exact results even at low temperatures and strong coupling \cite{Varney_Quantum_2009}.
These results can serve as benchmarks for developing approximation methods, which may then be extended with confidence to sign-problem regimes after being validated against DQMC data in sign-problem-free regions.\par

Diagrammatic Monte Carlo (DiagMC) is a numerical technique developed to circumvent the sign problem in DQMC.
It computes physical quantities through the stochastic sampling of Feynman diagrams \cite{Prokofev_Polaron_1998, Kris_Diagrammatic_2010}.
Calculations of DiagMC can be performed directly in the thermodynamic limit, thus avoiding the finite-size scaling analysis required in DQMC.
In addition, since its diagrammatic expansion is a perturbative series around the Hartree-Fock (HF) solutions, all three fundamental identities should be valid.
In the two-dimensional (2D) Hubbard model at $\beta=8$, DiagMC has revealed a convergence radius of $R=5.1$ at filling $n=0.875$, indicating a Kosterlitz-Thouless transition to superconductivity at $U=-5.1$ \cite{Rossi_Determinant_2017}, consistent with established phase diagrams presented in Ref.~\cite{Paiva_Critical_2004}.
However, it is precisely due to its perturbative nature that DiagMC is generally believed to capture the physics of the weak-coupling (Slater-like) branch, but fails to describe the strong-coupling (Mott-Heisenberg) branch \cite{Lenihan_Evaluating_2022, Garioud_symmetry-Broken_2024}.
In determining the antiferromagnetic (AF) critical temperature at half filling in the 3D Hubbard model, whether starting from the disordered or ordered phase, DiagMC has so far been limited to $U \leq 6$, where, given a bandwidth of $W = 12$, the ratio $U/W = 0.5$ places it outside the strongly coupled regime; in contrast, for $U>6$, there is increased difficulty in summing the perturbation series and a loss of accuracy in the critical region near the phase transition \cite{Lenihan_Evaluating_2022, Garioud_symmetry-Broken_2024}.
In typical cuprate superconductors, the Hubbard $U$ generally exceeds 8 (in units of hopping), and even at optimal doping the inverse transition temperature $\beta$ remains high, reaching values from above 30 up to 100.
Consequently, investigating the low-temperature regime of the 2D Hubbard model is particularly challenging, and to our knowledge, there have been no DiagMC calculations at such low temperatures for $U=8$.\par

Many-body approximation methods are non-perturbative techniques within the quantum many-body theory, such as the GW approximation \cite{migdal_interaction_1958, hedin_new_1965} and the two-particle self-consistent (TPSC) approach \cite{Vilk_Non-Perspective_1997,Bergeron_Optical_2011,Tremblay_Improved_2023,Vilk_Antiferromagnetic_2024}.
In our view, the many-body methods typically proceed by solving equations governing the Green's function (approximated through Feynman diagram truncation) and then employing perturbative corrections or other approaches to improve accuracy or estimate additional quantities.
Although these methods vary widely in their theoretical foundations and numerical reliability, they generally share two principal benefits: absence of the fermion sign problem and relatively low computational cost.
These strengths underpin their broad applicability across various parameter regimes, including those of low temperature and finite doping.
However, their approximate nature means that they cannot be expected to fulfill all three fundamental identities \cite{rohringer_diagrammatic_2018}.
A more critical limitation emerges in low-dimensional systems, where such approximations can predict spurious finite-temperature phase transitions.
For example, the AF phase transition in the 2D Hubbard model at half-filling.
These predictions are in direct violation of the famous Mermin-Wagner theorem \cite{mermin_absence_1966}, which renders the methods untrustworthy near pseudo-critical points and invalidates their results at lower temperatures.

The Mermin-Wagner theorem states that continuous symmetries cannot be spontaneously broken in 2D systems at non-zero temperature when interactions are short-ranged \cite{mermin_absence_1966}.
In the context of statistical field theory, a symmetry-breaking solution at low temperatures corresponds to a non-zero mean-field saddle point of the action. 
This saddle point gives rise to massless Goldstone modes in models with a continuous symmetry.
In 2D systems, these massless Goldstone modes induce infrared divergences in the perturbative expansion around the saddle point.
As a result, the broken phase is destabilized by these massless excitations and symmetry restoration occurs.
In the $\mathrm{O}(N)$ universality class, there is only one phase transition allowed in 2D: the Berezinskii-Kosterlitz-Thouless (BKT) transition with $\mathrm{O}(2)$ symmetry.
This transition involves the binding and unbinding of topological defects, such as vortices in the 2D system, as temperature increases from the low-temperature to the high-temperature regime.\par

This theoretical framework provides clear guidance for understanding the phase transition behavior of the 2D Hubbard model.
Although our understanding of the 2D Hubbard model remains incomplete, we can turn to the 2D $XY$ and Heisenberg models for insights into phase transition behavior.
Indeed, computational results for the 2D $XY$ and Heisenberg models validate the aforementioned theory.
In the 2D $XY$ model, which is of the $\mathrm{O}(2)$ universality class, the BKT transition has been confirmed by a tensor network approach utilizing higher-order singular value decomposition \cite{Yu_Tensor_2014}.
The system comprises a low-temperature branch with quasi-long-range order characterized by algebraic decay of correlations and a high-temperature branch with exponential decay of correlations.
In contrast, whether a phase transition exists in the $\mathrm{O}(3)$-symmetric 2D Heisenberg model remains an open question.
Nevertheless, recent studies using tensor network methods \cite{schmoll_classical_2021,Ueda_Tensor_2022,Burgelman_Contrasting_2023} and some Monte Carlo simulations \cite{Shenker_Monte-Carlo_1980,Tomita_Finite-size_2014} suggest that there is most probably no finite-temperature phase transition in the 2D Heisenberg model, supporting the scenario of asymptotic freedom proposed for the continuous model with $\mathrm{O}(3)$ symmetry.
Interestingly, since high-temperature expansions imply that the high-temperature phase is distinct from the low-temperature phase \cite{Stanley_Possibility_1966,Adler_High-temperature_1993}, the two regimes can also be viewed as thermodynamically distinct branches, even without a sharp finite-temperature phase transition.

\subsection{Lessons from statistical field theory}

To tackle the problem of infrared divergences and calculate physical quantities at low temperatures for a statistical field theory with continuous symmetry, Jevicki provided a crucial insight.
Jevicki noticed that the infrared divergences of Feynman diagrams for the ground-state energy of the 2D $\mathrm{O}(N)$ invariant $\sigma$-model, calculated perturbatively up to two loops, cancel exactly when summed over all diagrams \cite{Jevicki_Ground_PLB1977}.
The infrared divergences in 2D models with continuous symmetry stem from the presence of massless Goldstone excitations in the symmetry-broken phase associated with the classical vacuum around which the perturbation is performed.
After verifying the cancellation up to second-order perturbation, Elitzur later conjectured that, when summed over all Feynman diagrams, the infrared divergences in any $\mathrm{O}(N)$ invariant field correlations would also cancel to any order perturbation within the low-temperature expansion around the classical vacuum or saddle-point solution \cite{Elitzur_applicability_1983}.
Elitzur commented that these massless Goldstone excitations are generated by applying symmetry-generating operators to the classical broken vacuum, and since invariant quantities commute with the symmetry generators, they remain unaffected by their application to the vacuum.
It is thus expected that invariant quantities will decouple from Goldstone excitations and may consequently remain finite.
This conjecture was later proven up to arbitrary perturbative orders by David \cite{David_Cancellations_CMP1981}, demonstrating that the cancellation of infrared divergences in $\mathrm{O}(N)$ invariant functions occurs at non-exceptional momenta to any perturbative order.\par

The same idea was subsequently successfully applied to the vortex physics of type-II superconductors around the year 2000.
Almost sixty years ago, Eilenberger calculated the spectrum of harmonic excitations of the Abrikosov vortex lattice based on the Ginzburg-Landau theory of type-II superconductors in an external magnetic field \cite{Eilenberger_Thermodynamic_1967}.
Maki and Takayama later pointed out that the gapless mode is softer than the conventional Goldstone mode expected from the spontaneous breaking of both translational and $\mathrm{U}(1)$ symmetries, leading to infrared divergences in the perturbation expansion around the Abrikosov vortex lattice state \cite{Maki_Thermodynamic_1971}.
The cancellation of infrared divergences in the effective free energy up to two-loop order was first noted in Ref.~\cite{Rosenstein_First-principles_1999}, but the final result was obtained in Refs.~\cite{Li_Thermal_2001,Li_Thermal_1999} after including the Umklapp contribution in two-loop diagrams.
Therefore, the infrared divergences most probably cancel for any $\mathrm{U}(1)$ invariant function including the effective potential, and even for the Ginzburg-Landau theory of type-II superconductors in an external magnetic field.
Although $\mathrm{U}(1)$ invariance is restored due to the massless excitation (acoustic excitation mode), and although the infrared divergence cancels for $\mathrm{U}(1)$ invariant quantities such as the structure function as shown in Ref.~\cite{Li_Thermal_1999}, the breaking of translational symmetry (from the homogeneous solution to the vortex lattice solution) remains.
The phase transition corresponding to this translational symmetry breaking is typically first-order.
For vortex matter, it is referred to as the vortex melting transition.
By comparing the effective free energy of the vortex solid obtained in Refs.~\cite{Li_Thermal_2001,Li_Thermal_1999} with that of the vortex liquid, Refs.~\cite{Li_Melting_2002,Hikami_Magnetic-flux-lattice_1991} determined the melting line of the vortex lattice.
The effective free energy of liquid was obtained via Borel-Pad{\'e} resummation of the expansion series (up to nine-loop diagrams) given in Ref.~\cite{Hikami_Magnetic-flux-lattice_1991}.
Details of the theoretical calculations can be found in Refs.~\cite{Li_Supercooled_2004,Rosenstein_Ginzburg-Landau_2010}.
Subsequent experiments have verified the theoretical melting line predictions \cite{Kokubo_Dynamic_2007,Koshelev_Melting_2019}.\par

\subsection{Motivation for non-perturbative many-body methods}
In view of the literature review above, both many-body theory and statistical field theory for 2D systems share similar difficulties, and it is beneficial to borrow ideas from each other.
For the Hubbard model relevant to cuprate high-$T_{\mathrm{c}}$ superconductors, the most relevant coupling in the cuprates is the strong-coupling $U \geq 8$.
The phase transition temperatures (in unit of tunnelling amplitude $t$) are generally very small, typically $T \leq 1/30$, or equivalently, inverse temperatures $\beta \geq 30$.
This regime is virtually inaccessible to DQMC for the finite-doped 2D Hubbard model.
Even if we can address the problem using DiagMC calculations at very low temperatures, we speculate that $U$ cannot be as large as 8, because the perturbative nature of DiagMC (the Slater-like branch) makes it incapable of capturing the physics of the strong-coupling regime (the Mott-Heisenberg branch).
It is therefore necessary to seek a many-body theory capable of tackling the challenges posed by 2D systems with continuous symmetry, such as the 2D Hubbard model with spin rotational symmetry, at strong coupling and low temperature.\par

Although we cannot maintain all of three fundamental relations in a many-body method, we should aim to minimize the violation of these identities as much as possible.
Given the importance of the WTI (since the experimental consistency is verified by the $f$-sum rules), it is better to preserve the WTI due to its foundation in conservation laws.
Although the $\chi$-sum rule is not our primary requirement, it is desirable to keep its violation small for any reasonable approximation.
The simplest fermionic many-body method is the HF approximation, which also known as mean-field theory for fermionic many-body systems.
Relevant attempts have been made, yielding reasonable results in weakly correlated regions at low temperatures far from the critical point \cite{rosenstein_mean_2019}.
It was later recognized that for the HF approximation, the response function, which is related to both the two-body correlator and the vertex function, is given by the RPA formula, and satisfies the WTI \cite{Hui_Linear_PRB2023}.
However, the $\chi$-sum rule is badly violated in the strongly correlated Hubbard model.
Therefore, many-body theory should go beyond the HF approximation in order to minimize the violation of the $\chi$-sum rule for strongly correlated systems such as the Hubbard model.

This perspective also finds support from statistical field theory.
Unlike in fermionic systems, mean-field theory in bosonic statistical field theories such as the Ginzburg-Landau theory, corresponds to the perturbative expansion around the saddle point of the free energy.
For high-temperature superconductors, because the Ginzburg number is too large, the perturbative expansion around the mean-field solution is no longer reliable \cite{Rosenstein_Ginzburg-Landau_2010}.
In this context, the self-consistent HF theory is identified as the simplest non-perturbative approach and is equivalent to the Gaussian variational theory, rather than a mean-field theory.
The need for such a non-perturbative method is evidenced by phenomena like vortex physics \cite{Rosenstein_Ginzburg-Landau_2010}, with its success exemplified by the theoretical prediction of the spinodal line \cite{Rosenstein_Ginzburg-Landau_2010} and confirmed by experimental studies \cite{xiao_observation_2004, thakur_exploring_2005, Adesso_Transition_2006}.
However, in the symmetry-broken phase (where the expectation of order parameter field is nonzero), HF theory violates the Goldstone theorem, which predicts the existence of a massless mode in the spontaneous continuous symmetry breaking phase.
This massless mode can be recovered using the covariance theory for calculating correlation functions of the order parameter field, for example, the correlator in the vortex lattice state \cite{Li_Fluctuation_2012}.\par

Based on the issues identified, and the experience and lessons drawn, we implement the following strategies.
We systematically develop a symmetrization scheme, ensuring that our results preserve the Mermin-Wagner theorem.
We employ the GW approximation \cite{migdal_interaction_1958, hedin_new_1965}, a simple non-perturbative approach beyond the Hartree-Fock (mean-field) level for the one-body Green's function, along with the covariance theory \cite{Hui_Linear_PRB2023}, a systematic framework for obtaining two-body correlation functions that satisfy the FDR and WTI, to study the 2D Hubbard model.
We will compare the results with benchmark data from DQMC simulations of the half-filled 2D Hubbard model on a finite lattice at low temperature and strong coupling, and assess the violation level of the local moment sum rules (or the $\chi$-sum rule).\par

\subsection{Paper organization}
This paper is organized as follows:
In Sec.~\ref{chap:average}, we establish the symmetrization scheme.
In Sec.~\ref{covart} and Sec.~\ref{gwa}, we introduce the covariance theory and GW approximation.
In Sec.~\ref{model}, Sec.~\ref{applia} and Sec.~\ref{calcus}, we implement the symmetrization scheme and the GW-covariance approximation to the 2D Hubbard model on the A-B lattice; this approach predicts a pseudo paramagnetic-AF phase transition in the 2D Hubbard model, breaking the spin $\mathrm{SU}(2)$ symmetry.
In Sec.~\ref{chap:results}, we present the benchmark of our approach with the DQMC results.
In Sec.~\ref{chap:criterion}, we suggest a self-consistency criterion for many-body approximation methods.
Finally, in Sec.~\ref{chap:conclusion}, we provide a brief summary and discussion.\par

\section{\label{chap:average}General symmetrization scheme}

In the context of many-body physics, physical quantities can be expressed as statistical ensemble averages of certain physical functionals of the fields.
For example, such a functional can represent density-density fields, whose statistical ensemble average is referred to as the density-density correlation.
The group action/transformation is defined by specifying how the fundamental fields, such as the electron field, transform under each group element.
If the original action remains invariant under transformations of the fields by a group $\mathcal G$, the system is said to be $\mathcal G$-invariant or $\mathcal G$-symmetric.
At high temperatures, such a system is typically in what is known as a $\mathcal G$-invariant state, in which all physical quantities (or physical functionals of the fields) are invariant under $\mathcal G$.
However, at low temperatures, the state that minimizes the free energy may not be $\mathcal G$-invariant, in which case certain physical functionals of the fields are not invariant under $\mathcal G$.
When this happens, the system is said to be in a spontaneous symmetry-breaking phase.
For example, the Hubbard model is $\mathrm{SU}(2)$-invariant as its action remains unchanged under $\mathrm{SU}(2)$ transformations acting on the spin indices of the electron fields.
At high temperatures, the half-filled 3D Hubbard model with interaction is in a $\mathrm{SU}(2)$-invariant paramagnetic state, while at low temperatures it is in an AF-symmetry-breaking phase.
In the latter, certain physical functionals of the fields, such as the AF order parameter field, break the $\mathrm{SU}(2)$ symmetry.

In principle, symmetry breaking in low-dimensional systems may be summarized as follows.
For any finite lattice, there is neither spontaneous symmetry breaking nor phase transition, but only a crossover between the low- and high-temperature regimes.
Even in infinite lattice systems, continuous symmetry breaking does not occur in 2D.
For continuous systems in the $\mathrm{O}(N)$ universality class, there is similarly neither continuous symmetry breaking nor phase transition, but only a crossover between the low- and high-temperature regimes.
The only exception is the $\mathrm{O}(2)$ case, for which a finite-temperature BKT phase transition can occur in either continuous or infinite lattice systems.\par

In practice, however, many-body approximations may predict continuous symmetry breaking in finite 2D systems.
This contradiction prevents the application of many-body theory at temperatures below the instability point of the group-invariant state.
To tackle this difficulty, we introduce the symmetrization scheme.
As noted by Jevicki et al., even though continuous symmetry breaking does not occur in 2D at low temperatures, one should still begin from a symmetry-broken state, and the symmetry-invariant correlators and other symmetry-invariant physical quantities calculated within this symmetry-broken framework can nevertheless give reasonable quantitative estimates.
The idea of Jevicki et al. has been generalized in a way somewhat different from the original one to study the Hubbard model at half-filling and low temperature, and this generalized framework was confirmed to be valid by comparison with DQMC results \cite{rosenstein_mean_2019}.
Specifically, we can calculate not only symmetry-invariant physical quantities in the pseudo symmetry-breaking state but also those non-invariant quantities through the symmetrization.
In the following, we outline a general framework for performing symmetrization.
\par

We consider an arbitrary field functional, which may involve one- or two-body fields.
For simplicity in conveying our ideas, we consider a finite lattice with a discrete symmetry group $\mathcal G$.
Physical quantities can be approximately expressed as $F\equiv\langle\Psi|\mathcal{F}(\hat{\psi}_{\alpha},\hat{\psi}_{\beta}^{\dagger})|\Psi\rangle$, where $\hat{\psi}_{\alpha}$ and $\hat{\psi}_{\beta}^{\dagger}$ are electron field operators, $\Psi$ is a $\mathcal{G}$-symmetry-breaking state with the lowest free energy among all possible configurations, and $F$ represents the statistical ensemble average of the field-operator functional $\mathcal{F}$ over the state $\Psi$.
Due to symmetry, for any $g\in \mathcal{G}$, the state $g\Psi$ also has the lowest energy, just as $\Psi$ does.
However, $g\Psi$ and $\Psi$ are distinct states, corresponding to different order parameters.
Since the lattice is finite, there is no infinite energy barrier between these states.
Consequently, tunneling between them is possible.
Thus, all states in $\{ g\Psi \,|\, g\in \mathcal{G}\}$ may contribute to physical quantities, and the physical quantities should be expressed by the average value of the field functional over all possible symmetry-breaking states.
More precisely, we should consider 
\begin{equation}
    \bar{F}\equiv \frac{1}{|\mathcal{G}| }\sum_{g\in\mathcal{G}}\langle
        g\Psi | \mathcal{F}(\hat{\psi}_{\alpha},\hat{\psi}_{\beta}^{\dagger}) | g\Psi
    \rangle\label{eq:discrete average},
\end{equation}
where $|\mathcal{G}|$ denotes the number of group $\mathcal{G}$ elements. 
This is equivalent to $\bar{F}=\langle\Psi|\bar{\mathcal{F}}(\hat{\psi}_{\alpha},\hat{\psi}_{\beta}^{\dagger})|\Psi\rangle$ with
\begin{equation}
    \bar{\mathcal{F}}(\hat{\psi}_{\alpha},\hat{\psi}_{\beta}^{\dagger})\equiv\frac{1}{|\mathcal{G}|}\sum_{g\in\mathcal{G}}\mathcal{F}(g^\dagger\hat{\psi}_{\alpha}g,g\hat{\psi}_{\beta}^{\dagger}g^\dagger)\label{eq:discrete average of fields}, 
\end{equation}
where $g^\dagger\hat{\psi}_{\alpha}g$ and $g\hat{\psi}_{\beta}^{\dagger}g^\dagger$ represent the field transformations under the group $\mathcal{G}$.
The formula states that, for certain group, the physical quantity corresponds to the statistical ensemble average of the group-symmetric field functional over one group-symmetry-breaking state.
Since the $\bar{\mathcal{F}}(\hat{\psi}_{\alpha},\hat{\psi}_{\beta}^{\dagger})$ is $\mathcal{G}$-symmetric, its statistical ensemble average $\bar{F}$ over any spontaneous $\mathcal{G}$-symmetry-breaking state is the same.
In this case, the order parameter field will vanish after this symmetrization, while other quantities, such as the spin correlation function (as will be shown later), will become independent of the symmetry-breaking direction.\par

When dealing with continuous symmetries, the basic idea of taking average remains unchanged only the summation is replaced by an integral, i.e., the invariant Haar measure \cite{Aubert_Invariant_2003}.
Specifically, if $U$ represents the matrix representation of an element from a continuous symmetry group $\mathcal{G}$, the symmetrization procedure involves integration over the group manifold with the appropriate Haar measure
\begin{equation}
    \overline{\left<
        \hat{\psi}^\dagger_{\alpha_1}\dots\hat{\psi}^\dagger_{\alpha_n}
        \hat{\psi}_{\beta_1}\dots\hat{\psi}_{\beta_m}
    \right>}
    =\int \mathrm dU\,
    [U^*]^{\ \alpha'_1}_{\alpha_1}\dots[U^*]^{\ \alpha'_n}_{\alpha_n}
    U^{\ \beta'_1}_{\beta_1}\dots U^{\ \beta'_m}_{\beta_m}
    \left<
        \hat{\psi}^\dagger_{\alpha'_1}\dots\hat{\psi}^\dagger_{\alpha'_n}
        \hat{\psi}_{\beta'_1}\dots\hat{\psi}_{\beta'_m}
    \right>\label{haar measure}, 
\end{equation}
which is a generalization of Eq.~(\ref{eq:discrete average}) for continuous symmetric groups.\par

In summary, for any finite lattice with either discrete or continuous symmetry, as well as any 2D infinite lattice with continuous symmetry, symmetry-breaking solutions are not problematic, provided that the field functional is symmetrized first.
Physically speaking, we envisage that these systems exist in the form of different \textbf{domains} inside the sample, each characterized by their own ``order parameters''.
Inside a domain, fluctuations are considerable, yet the ``order parameter'' remains identifiable.
On the scale of the entire sample, these ``order parameters'' behave as short-range fluctuations.
Such ``order parameters'' can indeed be approximately observed experimentally \cite{hashimoto_energy_2014}.
However, long-range correlations between different domains are very weak, which corresponds to the absence of long-range order \cite{mermin_absence_1966}.
The symmetrization scheme is precisely the averaging over all domains from a macroscopic perspective.
Formally speaking, the symmetrization scheme is taking the average of the field functional $\mathcal{F}$ over all the group action $g$ acting on the electron fields.
This is equivalent to averaging the field functional $\mathcal{F}$ over all symmetry-breaking states.
The thermodynamic quantities of the system, particularly the free energy, remain invariant across different spontaneous symmetry-breaking states, and thus these states should be assigned equal weight to justify our symmetrization scheme.\par

In this work, we investigate the paramagnetic-to-AF phase transition (or crossover) using many-body approximations in the Hubbard model on the 2D square lattice, as an example to specifically illustrate the application of the symmetrization scheme.
This pseudo phase transition involves the breaking of both discrete translational symmetry and continuous $\mathrm{SU}(2)$ spin symmetry, and the continuous $\mathrm{SU}(2)$ spin symmetry needs to be restored via symmetrization.

\section{\label{generf}Formalism}

We will present the theories within the framework of functional integrals \cite{altland_condensed_2010,negele_quantum_2018,stoof_ultracold_2009}. 
We consider a generalized system with the Matsubara action:
\begin{equation}
    \begin{aligned}
        \mathcal{S}[\psi^*,\psi]=
        - & \sum_{\alpha_1\alpha_2}\int\mathrm d(12)\, \psi^{*}_{\alpha_1}(1)T_{\alpha_1\alpha_2}(1,2)\psi_{\alpha_2}(2) \\
        - & \frac{1}{2}\sum_{ab}\int\mathrm d(12) S^{a}(1)V^{ab}(1,2)S^{b}(2),
    \end{aligned} \label{action}
\end{equation}
where $\alpha=\uparrow,\downarrow$ means spin up and down, $\sigma^{0}$ is the identity matrix, $\sigma^{a}\,(a=x,y,z)$ are Pauli matrices, and the 
\begin{equation}
    S^{a}(1)=\sum_{\alpha\beta}\psi^{*}_{\alpha}(1)\sigma^{a}_{\alpha\beta}\psi_{\beta}(1)\label{charge and spin operator}
\end{equation}
represents charge operator for $a=0$ and spin operator for $a=x,y,z$, relatively. $\psi^*,\psi$ are Grassmannian fields. The numbers in parentheses denote different spacetime coordinates, as $1=(\tau_1, \bm{x}_1)$, $\int\mathrm d(1)\,\dot{=}\int_0^\beta\mathrm d\tau_1 \sum_{\bm{x}_1}$, where $\beta$ is the inverse temperature, $0\leq\tau_1<\beta$ is the Matsubara time, $\bm{x}_1$ is the space coordinate. $T_{\alpha_1\alpha_2}(1,2)$ is the quadratic term of the action,
\begin{equation}
    T_{\alpha_1\alpha_2}(1,2)
    =\delta_{\alpha_1\alpha_2}\delta(\tau_1,\tau_2)\delta_{\bm{x}_1,\bm{x}_2}(-\partial_{\tau_2})
    -\mathcal{K}_{0\alpha_1\alpha_2}(1,2)\label{kinetic matrix}
\end{equation}
where $\mathcal{K}_{0\alpha_1\alpha_2}(1,2)$ corresponds to the kinetic term of the grand Hamiltonian in Matsubara representation. 
$V^{ab}(1,2)$ is the interaction satisfying $V^{ab}(1,2)=V^{ba}(2,1)$. 
The grand partition function is
\begin{equation}
    \mathcal{Z}=\int\mathcal{D}[\psi^{*},\psi]\,\mathrm{e}^{-\mathcal{S}[\psi^*,\psi]}.
\end{equation}
Definition of the one-body Green's function is
\begin{equation}
    G_{\alpha_1\alpha_2}(1,2)=-\langle \psi_{\alpha_1}(1)\psi^*_{\alpha_2}(2) \rangle,
    \label{eq:define G}
\end{equation}
where $\langle \dots \rangle=\mathcal{Z}^{-1}\int\mathcal{D}[\psi^{*},\psi]
    \dots\mathrm{e}^{-\mathcal{S}}$ is the ensemble average. 
For convenience, denote the spin structure by matrix formation:
\begin{equation}
    \bm{G}\dot{=}\begin{pmatrix}
        G_{\uparrow\uparrow}   & G_{\uparrow\downarrow}   \\
        G_{\downarrow\uparrow} & G_{\downarrow\downarrow}
    \end{pmatrix}.
\end{equation}
The matrix product is $[\bm{X}\bm{Y}]_{\alpha_1\alpha_3}=\sum_{\alpha_2}X_{\alpha_1\alpha_2}Y_{\alpha_2\alpha_3}$, 
and the trace is $\mathrm{Tr}[\bm{G}]=G_{\uparrow\uparrow}+G_{\downarrow\downarrow}$. 
The inverse of the Green's function is defined by
\begin{equation}
    \sum_{\alpha_2}\int \mathrm d(2)\, G^{-1}_{\alpha_1\alpha_2}(1,2)G_{\alpha_2\alpha_3}(2,3)=\delta_{\alpha_1\alpha_3}\delta(1,3),
\end{equation}
thus the non-interacting Green's function is $G_{0\alpha_1\alpha_2}(1,2)=T^{-1}_{\alpha_1\alpha_2}(1,2)$. 
Definition of the connected two-body correlation function is
\begin{equation}
    \chi_{XY}(1,2)=\langle X(1)Y(2)\rangle_{\mathrm{C}}\label{corr_defin},
\end{equation}
where $X,Y$ are one-body operators, and $\langle\dots\rangle_{\mathrm{C}}$ represents the connected-diagram contribution in the ensemble average. \par

The following subsections are organized as follows:
Present the covariance theory as a general framework for treating two-body correlation functions.
Describe the GW approximation for calculating single-particle Green's functions, along with the GW-covariance approximation for obtaining two-body correlation functions.
Specialize the generalized action in Eq.~(\ref{action}) to the case of the 2D Hubbard model and introduce the A-B lattice to describe the AF state.
Apply the symmetrization scheme to this particular case.

\subsection{\label{covart}Covariance theory}
The covariance theory is a systematic framework for constructing two-body correlation functions from an approximate one-body Green's function \cite{Hui_Linear_PRB2023}. 
The key of the covariance theory is to derive the covariance equations (Eq.~(\ref{eq:covariance equation})) based on the chosen one-body approximation. 
Solving these equations for the vertex function $\Lambda$ directly yields the two-body correlation functions (using Eq.~(\ref{correlation})). 
The resulting two-body correlation functions inherently satisfy the FDR, as Kubo’s FDR \cite{kubo_Statistical_I_1957,kubo_Statistical_II_1957,Callen_Irreversibility_1951,Mahan_Many-particle_2013} in the context of response theory is formally equivalent to the definition of the correlation function given by functional differentiation in Eq.~(\ref{correlation}).
This framework imposes no specific constraints on the form of the self-energy in the defining equation, making it applicable to various one-body methods, provided the resulting covariance equations are solvable.
As a demonstrated application, the GW-covariance approximation has been shown to produce correlation functions satisfying the WTI \cite{Hui_Linear_PRB2023}. 
While this proof is specific to the GW case, it naturally leads to the broader conjecture that correlation functions yielded by the covariance theory will respect the WTI, provided the underlying one-body approach possesses the corresponding symmetry. \par

We start with the definition of connected correlation function Eq.~(\ref{corr_defin}). 
In general, an one-body operators $X$ have such quadratic structure as
\begin{equation}
    X(3)=\sum_{\alpha_1\alpha_2}\int\mathrm d(12)\psi_{\alpha_1}^*(1)K_{X\alpha_1\alpha_2}(1,2;3)\psi_{\alpha_2}(2),
\end{equation}
and we term $K_{X}$ the kernel of $X$.
The spin operator $S^a(3)$, for instance, corresponds to $\bm{K}_{S^a}(1,2;3)=\bm{\sigma}^a\delta(1,2)\delta(1,3)$. 
We consider an external source $\phi$ which is coupled to the operator $Y$, that is, modify the action $\mathcal{S}$ to $\mathcal{S} - \int\mathrm{d}(3)\,\phi(3) Y(3)$.
The correlation function is then given by
\begin{equation}
    \chi_{XY}(1,2)=\frac{\delta \langle X(1)\rangle}{\delta \phi(2)}=\int\mathrm d(34)\mathrm{Tr}\left[\bm{K}_{X}(3,4;1)\bm{\Lambda}_{\phi}(4,3;2)\right].\label{correlation}
\end{equation}
Here, we denote ${\delta \bm{G}(1,2)}/{\delta \phi(3)}$ by $\bm{\Lambda}_{\phi}(1,2;3)$, and denote ${\delta \bm{G}^{-1}(1,2)}/{\delta \phi(3)}$ by $\bm{\Gamma}_{\phi}(1,2;3)$. They can be related to each other by
\begin{equation}
    \bm{\Lambda}_\phi(1,2;3)=-\int\mathrm d(45)\bm{G}(1,4)\bm{\Gamma}_\phi(4,5;3)\bm{G}(5,2). \label{Lambda}
\end{equation}\par

The problem now becomes how to calculate the vertex $\Lambda_\phi$ (or $\Gamma_\phi$). Notice that the external source is of the same kind with the kinetic term in action, thus modifying the action as $\mathcal{S}\to \mathcal{S}-\int\mathrm d(3)\phi(3) Y(3)$ is equivalent to modifying the non-interacting Green's function as $G_0\to G_0[\phi]$, i.e.,
\begin{equation}
    \bm{G}^{-1}_{0}[\phi](1,2)= \bm{G}^{-1}_{0}(1,2)+\int d(3)\phi(3)\bm{K}_{Y}(1,2;3).
\end{equation}
Therefore, in the presence of the external source $\phi$, 
as long as the modification $G_0\to G_0[\phi]$ is made, 
the Dyson equation $\bm{G}^{-1}(1,2)=\bm{G}^{-1}_{0}[\phi](1,2)-\bm{\Sigma}(1,2)$ remains valid. This allows one to systematically construct the covariance equations for any given perturbative or self-consistent theory, as 
\begin{equation}
    \Gamma_\phi(1,2;3)=\gamma_\phi(1,2;3)-\frac{\delta \bm{\Sigma}(1,2)}{\delta \phi(3)}, \label{eq:covariance equation}
\end{equation}
where 
\begin{equation}
    \bm{\gamma}_{\phi}(1,2;3)\equiv\frac{\delta \bm{G}^{-1}_{0}[\phi](1,2)}{\delta \phi(3)}=\bm{K}_{Y}(1,2;3).\label{gamma}
\end{equation}
The covariance equations are governed solely by the kernel $K_Y$, and their computational cost is independent of the kernel choice.
Consequently, solving these equations is generally tractable, provided the approximate self-energy used in Eq.~(\ref{eq:covariance equation}) is not excessively complicated.\par

\subsection{\label{gwa}The GW and GW-covariance approximation}
The GW approximation, a non-perturbative method for calculating one-body Green's function, was respectively proposed by Migdal \cite{migdal_interaction_1958} and Hedin \cite{hedin_new_1965}.
To construct a spin-dependent GW approximation, one employs a generalized formalism \cite{aryasetiawan_generalized_2008}. 

For the action in Eq.~(\ref{action}), one can derive a set of equations called Hedin's equations. In these equations, the self-energy $\bm{\Sigma}$ comprises two terms, $\bm{\Sigma}(1,2)=\bm{\Sigma}_{H}(1,2)+\bm{\Sigma}'(1,2)$. 
The first is the Hartree self energy, 
\begin{equation}
    \bm{\Sigma}_{H}(1,2)=-\delta(1,2)\sum_{ab}\int\mathrm d(3)\bm{\sigma}^aV^{ab}(1,3)\mathrm{Tr}[\bm{\sigma}^b\bm{G}(3,3)].\label{hartree self energy}
\end{equation}
The second is 
\begin{equation}
    \bm{\Sigma}'(1,2)=\sum_{ab}\int\mathrm d(34)\bm{\sigma}^a\bm{G}(1,3)\bm{\Gamma}_{H}^b(3,2;4)W^{ba}(4,1),\label{gw}
\end{equation}
with 
\begin{eqnarray}
    &&[W^{-1}]^{ab}(1,2)=[V^{-1}]^{ab}(1,2)-P^{ab}(1,2),\label{winvs} \\
    &&P^{ab}(1,2)=
    -\int\mathrm d(34)\mathrm{Tr}\left[\bm{\sigma}^{a}\bm{G}(1,3)\bm{\Gamma}^{b}_{H}(3,4;2)\bm{G}(4,1)\right], 
    \label{polar}
\end{eqnarray}
where $\Gamma_H$ is the Hedin's vertex function.  
The Dyson equation $\bm{G}^{-1}(1,2)=\bm{G}^{-1}_{0}(1,2)-\bm{\Sigma}(1,2)$ and Eqs.~(\ref{hartree self energy}, \ref{gw}, \ref{winvs}, \ref{polar}) are precisely the Hedin's equations.
The formal definition of $\Gamma_H$ and details derivations are provided in Appendix~\ref{A}.
By introducing approximations order by order,
the Hedin's equations serve as a bridge connecting the exact many-body theory
and practical calculations.
Among these approximations,
the GW approximation retains the leading order term of the vertex function $\bm{\Gamma}_H^a(1,2;3)\simeq \bm{\sigma}^a\delta(1,2)\delta(1,3)$. 
Under GW approximation, Eqs.~(\ref{gw}, \ref{polar}) become 
\begin{eqnarray}
    &&\bm{\Sigma}_{GW}(1,2)=\sum_{ab}\bm{\sigma}^a\bm{G}(1,2)\bm{\sigma}^bW^{ba}(2,1),\label{app gw} \\
    &&P^{ab}(1,2)=-\mathrm{Tr}\left[\bm{\sigma}^{a}\bm{G}(1,2)\bm{\sigma}^{b}\bm{G}(2,1)\right],\label{app polar}
\end{eqnarray}
relatively. 
The combination of Dyson's equation and Eqs.~(\ref{hartree self energy}, \ref{app gw}, \ref{winvs}, \ref{app polar}) comprise the GW equations, and can be self-consistently solved to obtain the Green's function.

Applying the covariance theory on the GW approximation (substituting the GW approximate self-energy $\Sigma=\Sigma_{H}+\Sigma_{GW}$ into Eq.~(\ref{eq:covariance equation})), one obtains
\begin{equation}
    \Gamma_\phi=\gamma_\phi-\Gamma^{H}_{\phi}
    -\Gamma^{MT}_{\phi}-\Gamma^{AL}_{\phi}. \label{cgw}
\end{equation}
The $\Gamma^H_\phi$ is the functional derivative of the Hartree self energy,
\begin{equation}
    \begin{aligned}
         & \bm{\Gamma}^H_\phi(1,2;3)=\frac{\delta \bm{\Sigma}_H(1,2)}{\delta \phi(3)}                                          \\
         & =-\delta(1,2)\sum_{ab}\int\mathrm d(3)\bm{\sigma}^aV^{ab}(1,4)\mathrm{Tr}[\bm{\sigma}^b\bm{\Lambda}(4,4;3)]. \label{cgw h}
    \end{aligned}
\end{equation}
The $\Gamma^{MT}_\phi$ and $\Gamma^{AL}_\phi$ come from $\Sigma_{GW}$,
\begin{equation}
    \bm{\Gamma}^{MT}_\phi(1,2;3)+\bm{\Gamma}^{AL}_\phi(1,2;3)=\frac{\delta \bm{\Sigma}_{GW}(1,2)}{\delta \phi(3)}.
\end{equation}
They are respectively
\begin{equation}
    \bm{\Gamma}^{MT}_\phi(1,2;3)=\sum_{ab}\bm{\sigma}^a\bm{\Lambda}_\phi(1,2;3)\bm{\sigma}^bW^{ba}(2,1), \label{cgw mt}
\end{equation}
and
\begin{eqnarray}
    &&\bm{\Gamma}^{AL}_\phi(1,2;3)=-\sum_{abcd}\bm{\sigma}^a\bm{G}(1,2)\bm{\sigma}^b\int\mathrm d(45) W^{bc}(2,4)\Gamma^{Wcd}_\phi(4,5;3)W^{da}(5,1),\label{cgw al} \\
    &&\Gamma^{Wcd}_\phi(4,5;3)=\mathrm{Tr}\left[
        \bm{\sigma}^c\bm{\Lambda}_\phi(4,5;3)\bm{\sigma}^d\bm{G}(5,4) + \bm{\sigma}^c\bm{G}(4,5)\bm{\sigma}^d\bm{\Lambda}_\phi(5,4;3)
        \right].\label{cgw al GammaW}
\end{eqnarray}
Eqs.~(\ref{Lambda}, \ref{cgw}, \ref{gamma}, \ref{cgw h}, \ref{cgw mt} ,\ref{cgw al}, \ref{cgw al GammaW}) are the GW-covariance equations. Knowing the $G$ and $W$, they can be solved self-consistently like solving the GW equations to obtain $\Lambda_\phi$. The two-body correlation function is then given via Eq.~(\ref{correlation}).\par

\subsection{\label{model}Model}
We apply the theory on the 2D repulsive Hubbard model. 
The grand Hamiltonian of the Hubbard model is
\begin{equation}
    \hat{\mathcal{K}}=-\sum_{\langle i,j\rangle\alpha}\left(
        t_{ij}\hat{c}^\dagger_{i\alpha}\hat{c}_{j\alpha}+h.c.
        \right)-\mu\sum_{i\alpha}\hat{n}_{i\alpha}+U\sum_{i}\hat{n}_{i\uparrow}\hat{n}_{i\downarrow},
\end{equation}
where $i$ denotes the lattice site, $\alpha=\uparrow,\downarrow$ denotes the spin, $\hat{c}^\dagger_{i\alpha}$ ($\hat{c}_{i\alpha}$) creates (annihilates) a fermion with spin $\alpha$ on site $i$, $\hat{n}_{i\alpha}\equiv \hat{c}^\dagger_{i\alpha}\hat{c}_{i\alpha}$ denotes spin-resolved density operator, $t_{ij}$ denotes the hopping amplitude from site $j$ to site $i$ (with the property $t_{ij}=t_{ji}$), $\mu$ is the chemical potential, and $U>0$ denotes strength of the on-site repulsive interaction. The doping level of the system can be altered by adjusting the chemical potential, with $\mu$ being precisely $U/2$ in the case of a half-filled system.

To organize the Hamiltonian into a form suitable for implementation in Eq.~(\ref{action}), we employ the Fierz identity \cite{Nieves_Generalized_2004}, 
\begin{equation}
    U\hat{n}_{i\uparrow}\hat{n}_{i\downarrow}=\Delta\mu \sum_{\alpha}\hat{n}_{i \alpha}-\frac{1}{2}U_{\mathrm{ch}}\,\hat{S}^{0}_{i}\hat{S}^{0}_{i}-\frac{1}{2}U_{\mathrm{sp}}\sum_{a=x,y,z}\hat{S}^a_{i}\hat{S}^a_{i},
\end{equation}
where
\begin{equation}
    \frac{U_{\mathrm{ch}}}{U}=\frac{\alpha}{3},\qquad \frac{U_{\mathrm{sp}}}{U}=\frac{1-\alpha}{3},\qquad \frac{\Delta\mu}{U}=\frac{1}{2}+\frac{\alpha}{3}.
\end{equation}
The parameter $\alpha$ can take arbitrary values, leading to an ambiguity: although the exact solution should in principle be independent of $\alpha$, approximations yield results that depend on the specific choice thereof.
As shown in Ref.~\cite{Li_Quantum_2023}, the GW-covariance approach at $\alpha\simeq 0$ yields spin correlations that are in relatively good agreement with DQMC results.
Therefore, this parameter regime is appropriate for testing the symmetrization scheme.
For this case $\alpha=0$, the Hubbard Hamiltonian becomes
\begin{equation}
    \hat{\mathcal{K}}=-\sum_{\langle i,j\rangle\alpha}\left(
    t_{ij}\hat{c}^\dagger_{i\alpha}\hat{c}_{j\alpha}+h.c.
    \right)+\left(\frac{U}{2}-\mu\right)\sum_{i\alpha}\hat{n}_{i\alpha}-\frac{U}{6}\sum_{i}\sum_{a=x,y,z}\hat{S}^a_{i}\hat{S}^a_{i}. \label{Spin Hamiltonian}
\end{equation}
Preserving the spin $\mathrm{SU}(2)$ symmetry when doing approximation, we second-quantize the grand Hamiltonian Eq.~(\ref{Spin Hamiltonian}), which is equivalent to substituting
\begin{eqnarray}
    &&\bm{\mathcal{K}}_0(1,2)=\bm{\sigma}^0\delta(\tau_1,\tau_2)\left(-\,t_{\bm{x}_1,\bm{x}_2}-\mu\,\delta_{\bm{x}_1,\bm{x}_2}\right),\\
    &&V^{ab}(1,2)=\frac{U}{3}\delta(1,2)\sum_{c=x,y,z}\delta^{ac}\delta^{bc}
\end{eqnarray}
into the quadratic term Eq.~(\ref{kinetic matrix}) of the Matsubara action Eq.~(\ref{action}). \par

\subsection{\label{applia}Lattice representations}

\begin{figure}[htbp]
    \centering
    \includegraphics[width=.6\linewidth]{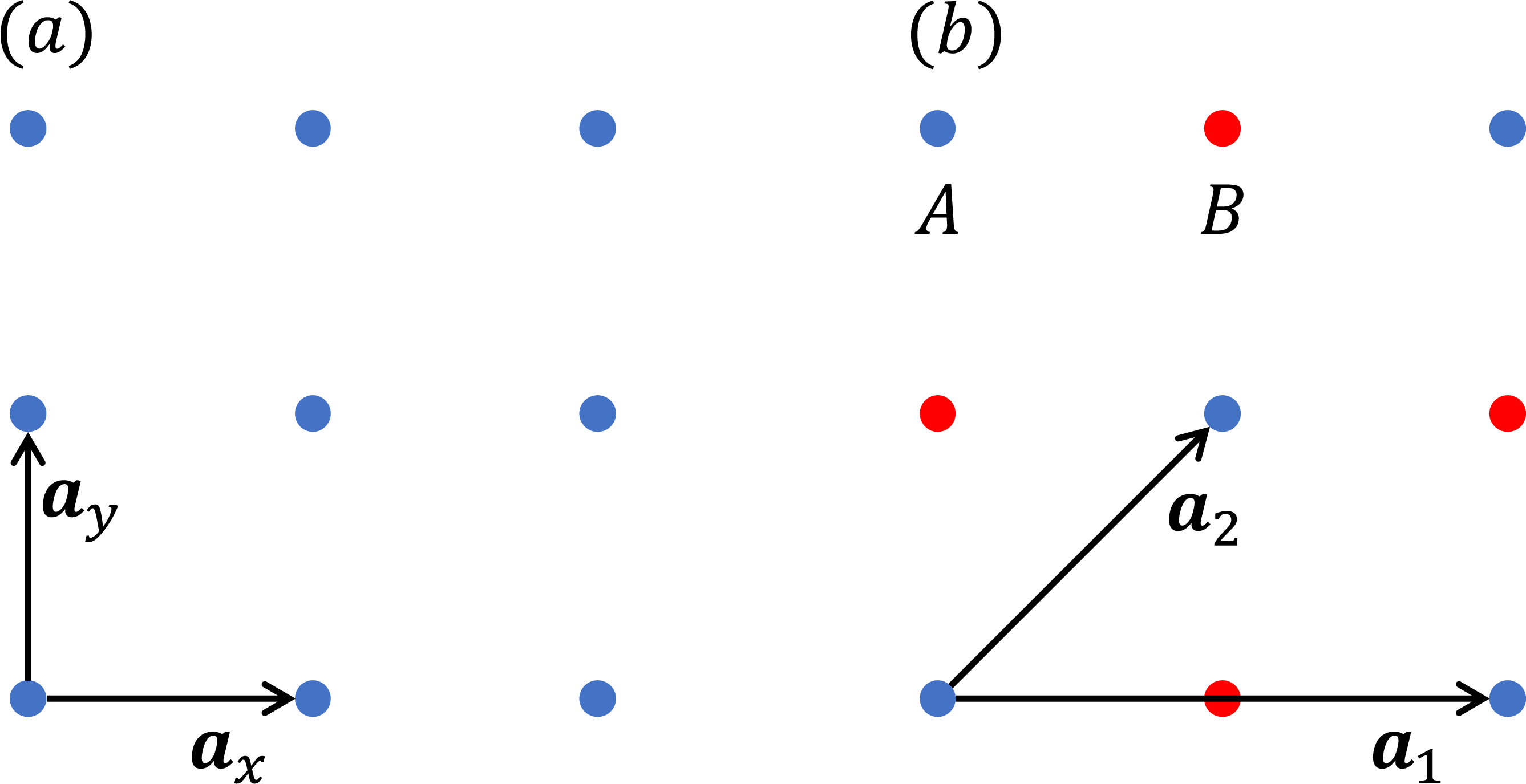}
    \caption{\label{fig:lattice} 
        Real-space lattice configurations for paramagnetic and antiferromagnetic solutions of the 2D Hubbard model. (a) Lattice configuration of the paramagnetic solution, with lattice vectors $a_x=(a,0)$ and $a_y=(0,a)$. All sites are equivalent. (b)  Lattice structure with AB sublattices for the antiferromagnetic solution with Néel order (A sublattice: blue; B sublattice: red), spanned by $a_1 = (2a,0)$ and $a_2 = (a,a)$.
    }
\end{figure}

In calculations, we restrict the system to the 2D square lattice. As shown in Fig.~\ref{fig:lattice} (a), the lattice sites located on $\bm{x}=n_{x}\bm{a}_x+n_{y}\bm{a}_y$ with periodic boundary condition, $a\equiv|\bm{a}_x| = |\bm{a}_y|$ is the distance between the two nearest lattice sites, and $n_{x},n_{y}\in\{0,\dots,N_x-1\}$ are integers. The reciprocal lattice is given by $\bm{k}=k_{x}\bm{b}_{x}+k_{y}\bm{b}_{y}$, where $\bm{a}_{i}\cdot\bm{b}_{j}=\delta_{ij}$ (here $i,j=x,y$), and $k_x,k_y\in\{2\pi i/N_x|i=0,\dots,N_x-1\}$. 

However, the AF order breaks the spatial translation invariance (precisely, the translation invariance on the 2D square lattice), thus one cannot apply the Fourier transformation directly on the space coordinate. 
To describe the structure of AF states, we introduce the A-B sublattice. 
As shown in Fig.~\ref{fig:lattice} (b), the lattice unit vectors of the A-B sublattice are $\bm{a}_{1}=2\bm{a}_{x}$ and $\bm{a}_{2}=\bm{a}_{x}+\bm{a}_{y}$. Each unit cell contains two lattice sites denoted by $l=A,B$, whose relative coordinates are $\bm{u}_A=\bm{0}$ and $\bm{u}_B=\bm{a}_{x}=\bm{a}_{1}/2$. To construct a lattice whose shape is consistent with the original 2D square lattice, we restrict the boundary condition of the A-B sublattice as $n_1\in\{0,\dots,N_1-1\}$ and $n_2\in\{0,\dots,N_2-1\}$, where $N_2=2N_1=N_x$. Coordinates of the lattice unit cells are $\bm{R}=n_1\bm{a}_{1}+n_2\bm{a}_{2}$, where $n_1,n_2$ are integers. The $\bm{R}$'s have the translation invariance on the A-B sublattice no matter whether the AF order exists. \par

Given any $\bm{x}$, there is one and only one set of $n_1,n_2\in \mathbb{Z}$ and $l=A,B$ such that $\bm{x}=\bm{R}+\bm{u}_{l}$ (with $\bm{R}=n_1\bm{a}_{1}+n_2\bm{a}_{2}$, $\bm{u}_A=\bm{0}$ and $\bm{u}_B=\bm{a}_{1}/2$). Thus, it's valid to replace the space coordinates $\bm{x}$ by these new quantum numbers $\bm{R},l$. 
Namely, using the new notations $l_1$ and $1=(\tau_1,\bm{R}_1)$ to replace the old one $1=(\tau_1,\bm{x}_1)$, and using $\sum_{l_1}\int\mathrm d(1)\,\dot{=}\sum_{l_1}\int_0^\beta\mathrm d\tau_1 \sum_{\bm{R}_1}$ to replace $\int\mathrm d(1)\,\dot{=}\int_0^\beta\mathrm d\tau_1 \sum_{\bm{x}_1}$. For example, the polarization function Eq.~(\ref{polar}) becomes
\begin{equation}
    P^{al_1,bl_2}(1,2)=\mathrm{Tr}\left[\bm{\sigma}^{a}\bm{G}^{l_1,l_2}(1,2)\bm{\sigma}^{b}\bm{G}^{l_2,l_1}(2,1)\right],
\end{equation}
where the Green's function $\bm{G}(1,2)$ with $1=(\tau_1,\bm{x}_1)=(\tau_1,\bm{R}_1+\bm{u}_{l_1})$ becomes $\bm{G}^{l_1l_2}(1,2)$ with $1=(\tau_1,\bm{R}_1)$ and $l_1,l_2=A,B$. Using such quantum numbers, theories can be formulated to be able to describe the AF broken phase.\par

The reciprocal A-B lattice is given by $\bm{k}=k_1\bm{b}_1+k_2\bm{b}_2$, where $\bm{b}_1=(\bm{b}_{x}-\bm{b}_{y})/2$, $\bm{b}_2=\bm{b}_{y}$ are reciprocal unit vectors, with $k_1\in\{2\pi i/N_1|i=0,\dots,N_1-1\}$ and $k_2\in\{2\pi i/N_2|i=0,\dots,N_2-1\}$. 
The Fourier transformation in A-B sublattice is defined as
\begin{eqnarray}
    &&\tilde{F}^{l_1l_2}(\omega_n,\bm{k})=
    \int_{0}^{\beta}\mathrm d\tau\sum_{\bm{R}}
    \mathrm{e}^{\mathrm{i}\omega_n\tau-\mathrm{i}\bm{k}\cdot\bm{R}}
    F^{l_1l_2}(\tau,\bm{R}),
    \label{eq:general fourier} \\
    &&F^{l_1l_2}(\tau,\bm{R})=
    \frac{1}{\beta N}\sum_{\omega_n,\bm{k}}
    \mathrm{e}^{-\mathrm{i}\omega_n\tau+\mathrm{i}\bm{k}\cdot\bm{R}}
    \tilde{F}^{l_1l_2}(\omega_n,\bm{k}),
    \label{eq:general inv fourier}
\end{eqnarray}
where $\beta$ is the inverse temperature, $N=N_1N_2$ is the number of unit cells of the A-B sublattice, and the translation invariance of the A-B sublattice, $F^{l_1l_2}(\tau_1-\tau_2,\bm{R}_1-\bm{R}_2)=F^{l_1l_2}(1,2)$, has been implied. The GW and GW-covariance equations in momentum space of the A-B sublattice are given in Appendix~\ref{B}. The Hubbard model can be implemented by the following kinetic term and interaction term,
\begin{eqnarray}
    &&\tilde{\bm{\mathcal{K}}}_0^{l_1l_2}(\omega_n,\bm{k})
    =\bm{\sigma}^0 \left(-\tilde{t}^{l_1l_2}_{\bm{k}}-\mu\,\delta_{l_1l_2}\right),\\
    &&\tilde{V}^{al_1,bl_2}(\omega_n,\bm{k})=\frac{U}{3}\delta_{l_1l_2}\sum_{c=x,y,z}\delta^{ac}\delta^{bc}.
\end{eqnarray}
In this work, we consider only nearest-neighbor hopping, represented on the reciprocal A-B lattice by
\begin{eqnarray}
    &&\tilde{t}^{AA}_{\bm{k}}=\tilde{t}^{BB}_{\bm{k}}=0, \\
    &&\tilde{t}^{AB}_{\bm{k}}=\tilde{t}^{BA\, *}_{\bm{k}}=2t\left[\cos\left( \frac{k_{1}}{2} \right)+\cos\left( k_{2}-\frac{k_{1}}{2} \right)\right]\mathrm{e}^{-\mathrm{i}k_1/2}.
\end{eqnarray}

\subsection{\label{calcus}Calculation of the Green's function and spin correlation}
The calculation consists of two steps: we first compute the physical quantities in the pseudo broken phase using a many-body method, and subsequently perform symmetrization of the results.
The Green's function $\tilde{\bm{G}}^{l_1l_2}(\mathrm{i}\omega_n,\bm{k})$ in the broken phase is obtained within the GW approximation, i.e., solving Eqs.~(\ref{eq:gw k 1}) to (\ref{eq:gw k 5}).
We then transform it back to real space and apply symmetrization, as detailed in Appendix~\ref{symmep}, using 
\begin{equation}
    \bar{\bm{G}}^{l_1l_2}(\bm{R}) = \frac{1}{2}\bm{\sigma}^{0}\left[
        G_{\uparrow\uparrow}^{l_1l_2}(\bm{R})+G_{\downarrow\downarrow}^{l_1l_2}(\bm{R})
    \right].
\end{equation}
% We can also express the symmetrized Green's function in the 2D square lattice representation, as follows,
We can also express the symmetrized Green's function as follows,
\begin{eqnarray}
    &&\bar{\bm{G}}(\bm{x}\in\text{$A$ lattice})=\frac{1}{2}\left[\bar{\bm{G}}^{AA}(\bm{x}) + \bar{\bm{G}}^{BB}(\bm{x})\right], \\
    &&\bar{\bm{G}}(\bm{x}\in\text{$B$ lattice})=\frac{1}{2}\left[\bar{\bm{G}}^{BA}(\bm{x}-\bm{a}_x) + \bar{\bm{G}}^{AB}(\bm{x}+\bm{a}_x)\right].
\end{eqnarray}\par

We are also concerned with the spin correlation function $\chi_\mathrm{sp}(1,2)\equiv \langle S^z(1)S^z(2) \rangle$. 
In the paramagnetic phase, $\chi_\mathrm{sp}=\chi_{S^{z}S^{z}}$, where $\chi_{S^{a}S^{b}}$ defined as Eq.~(\ref{corr_defin}) with $a,b=x,y,z$. 
In the pseudo AF phase, in contrast, the contribution of the AF order parameter should be considered separately, refer to Ref.~\cite{xiong_application_2025}.
Thus the spin correlation function becomes
\begin{equation}
    \tilde{\chi}_\mathrm{sp}(\mathrm{i}\omega_n,\bm{q})
    =\delta_{n,0}\delta_{\bm{q},\bm{Q}} \beta N_{\mathrm{sq}}|S^z_0|^2
    +(1-\delta_{n,0}\delta_{\bm{q},\bm{Q}})\tilde{\chi}_{S^{z}S^{z}}(\mathrm{i}\omega_n,\bm{q}),
\end{equation}
where $N_{\mathrm{sq}}$ is number of the points of the 2D square lattice, $\bm{Q}=\pi\bm{b}_x+\pi\bm{b}_y$ is the AF wave vector, and $S^z_0$ is the AF order (as shown in Eq.~(\ref{AF order})). 
Applying the symmetrization scheme, we derive the formula to calculate spin correlation function in the pseudo AF phase, as
\begin{equation}
    \tilde{\bar{\chi}}_\mathrm{sp}(\mathrm{i}\omega_n,\bm{q})
    =\delta_{n,0}\delta_{\bm{q},\bm{Q}}\frac{\beta N_{\mathrm{sq}}}{3}|S^z_0|^2
    +(1-\delta_{n,0}\delta_{\bm{q},\bm{Q}})\tilde{\bar{\chi}}_{S^{z}S^{z}}(\mathrm{i}\omega_n,\bm{q}).\label{eq:spin corr q}
\end{equation}
In the frequency-position space, it becomes
\begin{eqnarray}
    &&\bar{\chi}_\mathrm{sp}(\mathrm{i}\omega_n=0,\bm{x})
    =\frac{\beta}{3}|S^z_0|^2
    +\frac{1}{N_{\mathrm{sq}}}\sum_{\bm{q}\neq\bm{Q}}\mathrm{e}^{\mathrm{i}\bm{q}\cdot\bm{x}}\tilde{\bar{\chi}}_{S^{z}S^{z}}(\mathrm{i}\omega_n,\bm{q}), \label{eq:spin corr x 1} \\
    &&\bar{\chi}_\mathrm{sp}(\mathrm{i}\omega_n\neq 0,\bm{x})
    =\frac{1}{N_{\mathrm{sq}}}\sum_{\bm{q}}\mathrm{e}^{\mathrm{i}\bm{q}\cdot\bm{x}}\tilde{\bar{\chi}}_{S^{z}S^{z}}(\mathrm{i}\omega_n,\bm{q}). \label{eq:spin corr x 2}
\end{eqnarray}
Based on the methodology in Appendix~\ref{symmep}, we obtain the spin correlation function in the reciprocal 2D square lattice by that in the reciprocal A-B lattice using
\begin{eqnarray} 
    \tilde{\bar{\chi}}_{S^{z}S^{z}}(\mathrm{i}\omega_n, \bm{k})=\frac{1}{2}\left[
    \begin{gathered}
        \tilde{\bar{\chi}}_{S^{z}S^{z}}^{AA}(\mathrm{i}\omega_n, \bm{k})+\mathrm{e}^{\mathrm{i}\bm{k}\cdot \bm{a}_{x}}\tilde{\bar{\chi}}_{S^{z}S^{z}}^{AB}(\mathrm{i}\omega_n, \bm{k}) \\
        +\tilde{\bar{\chi}}_{S^{z}S^{z}}^{BB}(\mathrm{i}\omega_n, \bm{k})+\mathrm{e}^{-\mathrm{i}\bm{k}\cdot \bm{a}_{x}}\tilde{\bar{\chi}}_{S^{z}S^{z}}^{BA}(\mathrm{i}\omega_n, \bm{k})
    \end{gathered}
    \right],&& \label{eq:symmetrization chi k} \\
    \tilde{\bar{\chi}}_{S^{z}S^{z}}(\mathrm{i}\omega_n, \bm{k}+\bm{Q})=\frac{1}{2}\left[
    \begin{gathered}
        \tilde{\bar{\chi}}_{S^{z}S^{z}}^{AA}(\mathrm{i}\omega_n, \bm{k})-\mathrm{e}^{\mathrm{i}\bm{k}\cdot \bm{a}_{x}}\tilde{\bar{\chi}}_{S^{z}S^{z}}^{AB}(\mathrm{i}\omega_n, \bm{k}) \\
        +\tilde{\bar{\chi}}_{S^{z}S^{z}}^{BB}(\mathrm{i}\omega_n, \bm{k})-\mathrm{e}^{-\mathrm{i}\bm{k}\cdot \bm{a}_{x}}\tilde{\bar{\chi}}_{S^{z}S^{z}}^{BA}(\mathrm{i}\omega_n, \bm{k})
    \end{gathered}
    \right],&& \label{eq:symmetrization chi k+Q}
\end{eqnarray}
where $\tilde{\bar{\chi}}_{S^{z}S^{z}}^{l_1l_2}(\mathrm{i}\omega_n, \bm{k})$ is the symmetrized spin fluctuation as \begin{equation}
    \tilde{\bar{\chi}}_{S^{z}S^{z}}^{l_1l_2}(\mathrm{i}\omega_n, \bm{k})=\frac{1}{3}\left[
        \tilde{\chi}_{S^{x}S^{x}}^{l_1l_2}(\mathrm{i}\omega_n, \bm{k})
        +\tilde{\chi}_{S^{y}S^{y}}^{l_1l_2}(\mathrm{i}\omega_n, \bm{k})
        +\tilde{\chi}_{S^{z}S^{z}}^{l_1l_2}(\mathrm{i}\omega_n, \bm{k})
    \right].
\end{equation}
The fluctuations $\tilde{\chi}_{S^{a}S^{a}}^{l_1l_2}(\mathrm{i}\omega_n, \bm{k})$ for $a=0,x,y,z$ here are given by the GW-covariance theory (solving Eqs.~(\ref{eq:gw-c k 1}) to (\ref{eq:gw-c k 6}), the GW-covariance equations in momentum space). \par

From Eqs.~(\ref{eq:spin corr q}, \ref{eq:spin corr x 1}), we see that the spin correlation function $\chi_{\mathrm{sp}}$ contains contributions from both the pseudo AF order $|S^z_0|$ at $\mathrm{i}\omega_n=0,\bm{q}=\bm{Q}$ and spin fluctuations $\bar{\chi}_{S^zS^z}$ at other momenta.
As the contribution of order parameters is typically dominant, noticeable deviations in the correlation may stem from the unreliable results of many-body methods in estimating these order parameters.
In an attempt to determine the influence of the pseudo order, we therefore employ a separate estimate as 
\begin{equation}
    \tilde{\bar{\chi}}_{\mathrm{sp}}^{(\rm{sr})}(\mathrm{i}\omega_n=0,\bm{Q})
    =\beta N_{\mathrm{sq}} (2\rho-\rho^2)-\sum_{\mathrm{i}\omega_n\neq 0}\sum_{\bm{q}\neq\bm{Q}}\tilde{\bar{\chi}}_{S^{z}S^{z}}(\mathrm{i}\omega_n,\bm{q})-\sum_{\mathrm{i}\omega_n}\sum_{\bm{q}}\tilde{\chi}_{S^0S^0}(\mathrm{i}\omega_n, \bm{q}).\label{chi_sr_k0}
\end{equation}
This comes from the $\chi$-sum rule rooted in the Pauli exclusion principle (see Eq.~(\ref{chi sum rule appendix}) in Appendix~\ref{D}).
To avoid confusion, we will refer to the result obtained using Eqs.~(\ref{eq:spin corr q}, \ref{eq:spin corr x 1}) as the standard GW-covariance result, and designate this separate estimate (obtained by imposing the sum rule) as the $\chi$-constrained result.
We shall emphasize that this separate estimation serves as a case-specific approach that stands outside our systematic framework, i.e., the covariance theory and symmetrization scheme.
It is often neither necessary nor applicable when calculating other correlation functions that do not contain contributions from pseudo order parameters.
\par

\section{\label{chap:results}Results}
We investigated systems with intermediate-to-strong interactions at half filling, benchmarking our results against DQMC. 
To ensure the accuracy of DQMC results, we test the Trotter error of momentum distribution and charge correlation down to $\beta=20$ for $U=8$ system (see Appendix~\ref{C}). 
We focus on spin correlation with strong coupling $U=8$ at temperatures down to $\beta=16$, the corresponding Green's function results are also presented. 
Given the huge computational cost of DQMC, this benchmarking is conducted on $12\times 12$ lattices. 
Additionally, we present benchmark of spin correlation for $\beta = 8$ and $0 \leq U \leq 8$ on $16\times 16$ lattice, demonstrating the reliability of the symmetrization scheme across different interaction strengths and larger system sizes. 
We also investigated the spin correlation and the Green's function with intermediate coupling $U=4$ at temperatures down to $\beta=20$, but the results are presented in Appendix~\ref{results u4} rather than in this section. 

\subsection{Spin correlation function}

\begin{figure}[htb]
    \centering
    \includegraphics[width=0.95\linewidth]{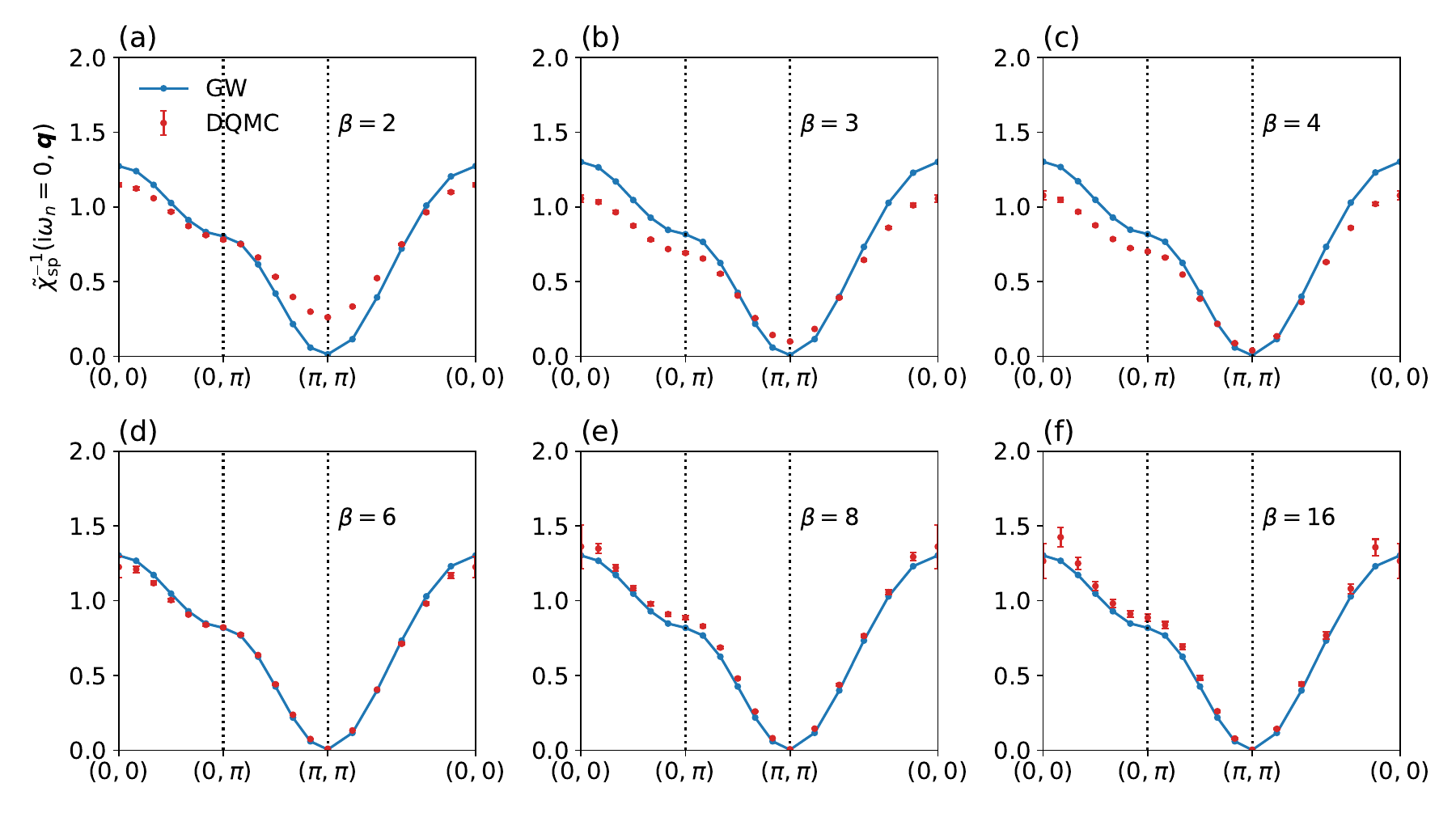}
    \caption{\label{fig:chi k w0}
        Momentum dependence of the inverse spin susceptibility for the half-filled Hubbard model on a $12\times12$ lattice with $U=8$. 
        Results are shown for decreasing temperatures, corresponding to inverse temperatures (a) $\beta=2$, (b) $\beta=3$, (c) $\beta=4$, (d) $\beta=6$, (e) $\beta=8$, and (f) $\beta=16$. 
        The momentum path follows $(0,0)\to(0,\pi)\to(\pi,\pi)\to(0,0)$. 
        Blue curves denote the symmetrized GW-covariance approximation, while red symbols with error bars show DQMC benchmarks. 
        A significant improvement in the agreement between GW-covariance and DQMC is observed upon cooling from $\beta=2$ to $\beta=6$. 
        At lower temperatures, the agreement remains largely unchanged, showing little evidence of any further enhancement.
    }
\end{figure}
The symmetrized GW-covariance and DQMC numerical results in $12\times 12$ lattice for the spin correlation function $\tilde\chi_\mathrm{sp}(\mathrm i\omega = 0,\bm{q})$ with $U=8$ at half filling are presented in Fig.~\ref{fig:chi k w0}. Qualitatively, the GW-covariance results are consistent with DQMC, exhibiting extrema at the AF wave vector $\bm{Q}=\pi\bm{b}_x+\pi\bm{b}_y$ and the zero momentum $\bm{k}=\bm{0}$, and relatively small slope at the anti-nodal point $\bm{k}^{AN}= \pi \bm{b}_{y}$. 
The AF solution of the GW equations emerges at about pseudo critical (crossover) temperature $T_{\rm c}\simeq 0.60$, or $\beta_{\rm c}\simeq 1.68$. 
Near the critical temperature $\beta_{\rm c}$, the GW-covariance results show relatively poor agreement with the DQMC results.
As the inverse temperature $\beta$ increases far away from $\beta_{\rm c}$, quantitatively, the our results gradually show more close agreement with DQMC. 
At low temperatures far from the critical region ($\beta\gtrsim 6$), the GW-covariance results are in good agreement with the DQMC benchmarks.\par

\begin{figure}[htb]
    \centering
    \includegraphics[width=0.95\linewidth]{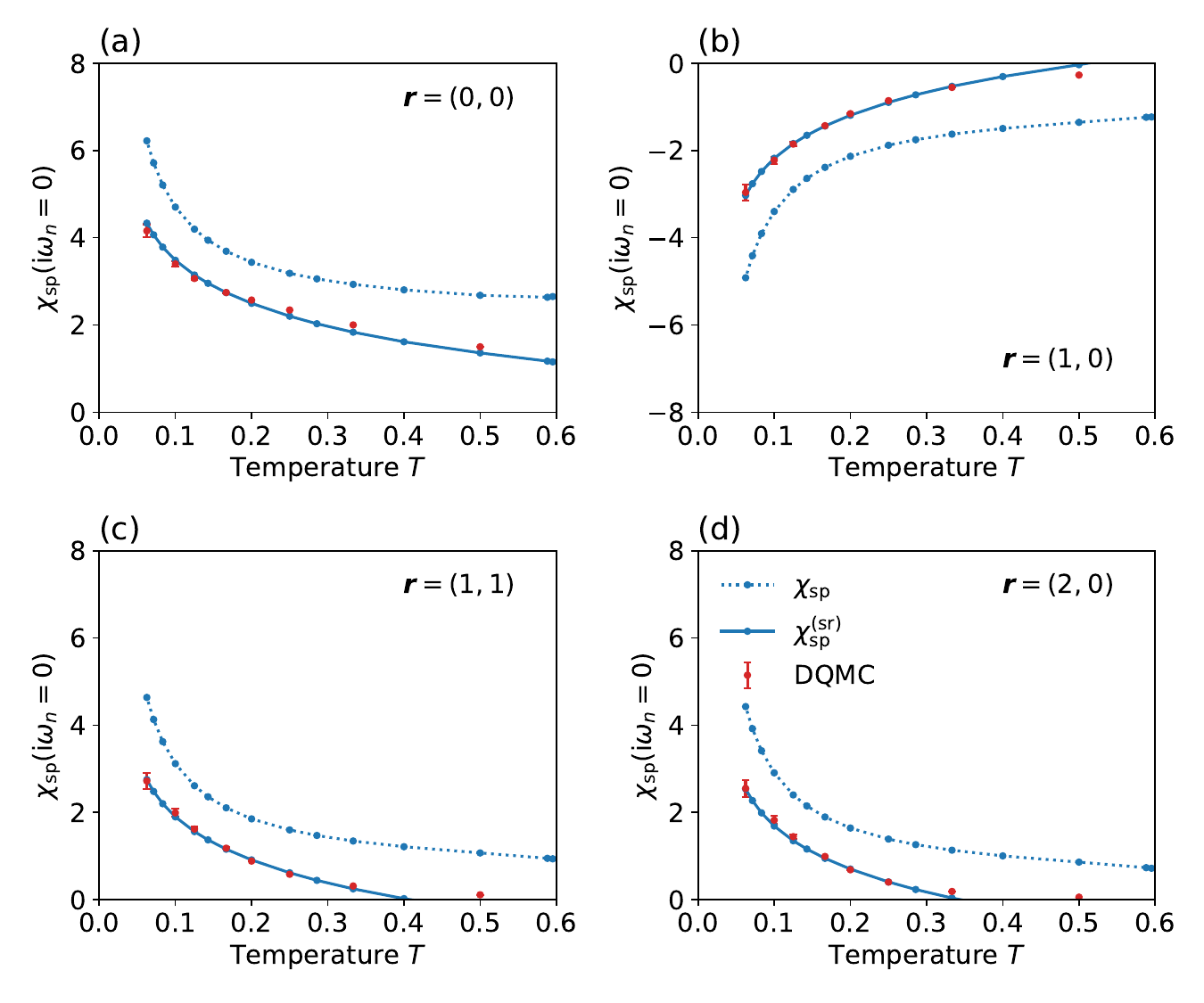}
    \caption{\label{fig:chi r w}
        Temperature dependence of the static spin correlation function in spatial space for the half-filled Hubbard model ($U=8$) on a $12\times12$ lattice. Panels (a)-(d) display results for lattice separations $\mathbf{r} = (0,0)$, $(1,0)$, $(1,1)$, and $(2,0)$, respectively. 
        The DQMC benchmarks (red points with error bars) are compared with the GW-covariance results.
        The $\chi_{\mathrm{sp}}$ (blue dashed curve) denotes the standard result and the $\chi_{\mathrm{sp}}^{\mathrm{(sr)}}$ (blue solid curve) denotes the $\chi$-constrained result.
    }
\end{figure}
As presented in Fig.~\ref{fig:chi r w}, we calculate the short-range behavior of the correlation function. 
Qualitatively the standard GW-covariance results are basically consistent with the DQMC results, but quantitatively the absolute values of GW-covariance are larger. 
This is consistent with our experience that GW tends to overestimate the influence of magnetic order on the system. 
The $\chi$-constrained result $\bar{\chi}_\mathrm{sp}^{(\mathrm{sr})}$ agrees very well with the DQMC at low temperature away from the pseudo critical $T_{\rm c}$, i.e., at $T\leq 0.25$.
Near $T_{\rm c}$, the short-range component of $\bar{\chi}_\mathrm{sp}^{(\mathrm{sr})}$ exhibits better behavior than its longer-range counterpart.
Specifically, the $T$-$\chi$ curves cross zero at $\bm{r}=\bm{a}_x$, $\bm{a}_x+\bm{a}_y$, and $2\bm{a}_x$ for $T > 0.5$, $0.4$, and $0.3$, respectively.
The correlation length, as calculated by our approach in the pseudo AF state, is overestimated near the critical region, which coincides with the poor behavior of the momentum-space results in this temperature region shown in Fig.~\ref{fig:chi k w0}.
Consequently, in this temperature region, the longer-range component is less reliable.
\par

\begin{figure}[htb]
    \centering
    \includegraphics[width=.8\linewidth]{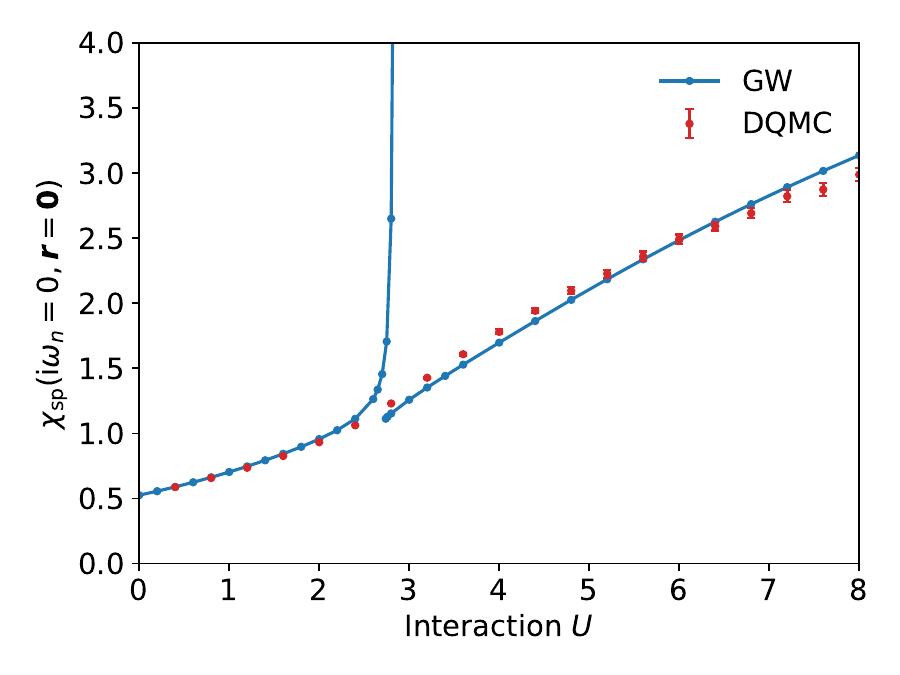}
    \caption{\label{fig:u_effect}
        Spin correlation function as a function of interaction strength $U$ for a $16 \times 16$ lattice at low temperature ($\beta = 8$). The DQMC results (red discrete points with error bars) exhibit a continuous evolution. In contrast, the GW-covariance approximation (blue curve) reveals a two-branch structure: a paramagnetic (Slater) branch for $U < U_c$ and a pseudo-antiferromagnetic (Mott-Heisenberg) branch for $U > U_c$, with a critical $U_c \simeq 2.8$ leading to a pronounced discontinuity. 
        The GW-covariance result on the antiferromagnetic branch corresponds to the $\chi$-constrained one.
    }
\end{figure}
We then turn to consider the influence of different interaction strengths $U$. 
As shown in Fig.~\ref{fig:u_effect}, in the $16\times 16$ system, the spin correlation function is calculated with $\chi_\mathrm{sp}(\mathrm i\omega_n=0,\bm r=\bm{0})$ as a representative. 
The temperature is fixed at $\beta = 8$, and the interaction strength $U$ gradually increases from zero to the typical strong correlation region $U = 8$. 
When the interaction $U < U_c$ ($U_c\sim 2.8$), the GW predicts paramagnetic phases. 
We identify it as the Slater branch.
When the interaction is relatively weak ($U \lesssim 2.5$), GW-covariance closely agrees with DQMC. 
However, when $2.5\lesssim U \lesssim U_c$, affected by the AF instability $\chi_\mathrm{sp}(\mathrm i\omega_n=0,\bm Q) \to \infty$, there is a serious overestimation for GW-covariance. 
When $U > U_c$, the GW predicts AF phases. 
We identify it as the Mott-Heisenberg branch.
Within the range we considered, the deviation from DQMC is not significant, indicating that the symmetrized GW-covariance theory performs well. 
This sheds light on the calculation of strongly correlated systems. \par

\FloatBarrier
\subsection{Green's function}
In this section, we calculate the Green's function at low temperature using the symmetrization scheme (see Appendix~\ref{symmep}).
We are concerned with the nodal point $\bm{k}^{N} = (\pi/2) (\bm{b}_{x}+\bm{b}_{y})$ and the anti-nodal point $\bm{k}^{AN} = \pi \bm{b}_{x}$ on the Fermi surface. 
\begin{figure}[htb]
    \centering
    \includegraphics[width=0.95\linewidth]{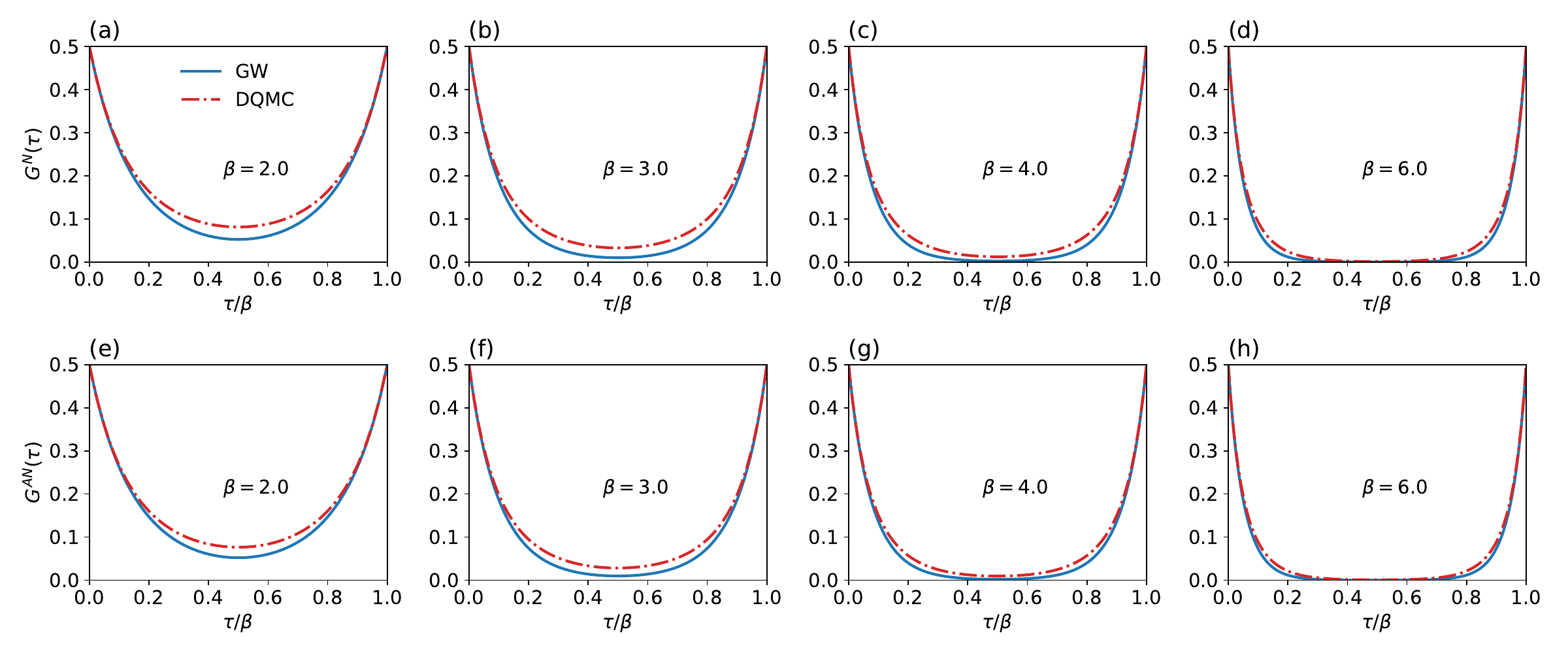}
    \caption{\label{fig:G tau}
        Imaginary-time Green's function $G(\mathbf{k},\tau)$ for the half-filled Hubbard model on a $12\times12$ lattice with $U=8$, comparing the symmetrized GW approximation (blue solid lines) and DQMC results (red dot-dashed lines). Panels (a)-(d) correspond to the nodal point $\mathbf{k} = (\pi/2, \pi/2)$, while panels (e)-(h) show the antinodal point $\mathbf{k} = (\pi, 0)$. The columns represent increasing inverse temperatures, i.e., (a,e) $\beta = 2$, (b,f) $\beta =3$, (c,g) $\beta =4$, and (d,h) $\beta =6$.
    }
\end{figure}
The benchmark is depicted in Fig.~\ref{fig:G tau},
within the spurious AF region. 
As the temperature decreases from $\beta=2$ near critical temperature
to deep AF region, 
the symmetrized Green's function calculated by GW
shows an increasingly closer agreement with
the results obtained from DQMC.
This tendency holds until extremely low temperature $\beta=16$,
which we do not show here.
Specifically, we are concerned with the
equal-time Green's function $G^{N/AN}(\tau=0)$
near and the minimal value of the Green's function $G^{N/AN}(\tau=\beta/2)$
near the Fermi surface.
$G^{N/AN}(\tau=0)$ corresponds to the momentum distribution of the density,
which is crucial for understanding the electronic structure of the system
\cite{Varney_Quantum_2009}.
On the other hand, $G^{N/AN}(\tau=\beta/2)$ can serve
as a proxy for spectral function
$A^{N/AN}(\omega=0)\simeq \frac{\beta}{2} G^{N/AN}(\tau=\beta/2)$,
helping us to probe the emergence of the pseudogap
\cite{wang_dc_2020}.
We find all of them become close to the DQMC results,
demonstrating the effectiveness of the symmetrized GW method
at extremely low temperatures.\par

\FloatBarrier
\section{\label{chap:criterion}The criterion for effectiveness of approximation method}
After benchmarking the symmetrization GW-covariance approximation with DQMC at half-filling, we expect to use the approximation to investigate doped regions of the Hubbard model, which exhibit various physical phenomena including CDW and superconducting states \cite{arovas_hubbard_2022,qin_hubbard_2022,rohringer_diagrammatic_2018,schafer_tracking_2021,Hao_Coexistence_Science2024}.
These systems are challenging for existing unbiased numerical methods.
For example, the DQMC suffers from the fermion sign problem and struggles to converge at low temperatures \cite{Hirsch_Two-dimensional_1985,Chang_recent_2015,SmoQyDQMC.jl,sun_delay_2024,iglovikov_geometry_2015}.
The DMRG is typically limited to cylindrical geometries, and its computational cost grows exponentially with sizes of 2D quantum systems, restricting it to very small system sizes \cite{White_Density_1992,White_Density-matrix_1993,Schollwock_density-matrix_2005}.
As a result, it is difficult to reliably benchmark many-body approximation methods in doped regions.
While the symmetrized GW-covariance method already performs well in benchmarks at half-filling, further evidence would strengthen the case for its reliability in doped regimes. 
Therefore, it would be beneficial to have a self-consistent criterion for assessing reliability, one that is independent of third-party data and based solely on the results of the approximation itself.  \par

A natural way to evaluate the effectiveness of an approximation is to examine the fundamental physical relations such as the conservation laws.
There are three fundamental relations, i.e., the FDR, the WTI and local momentum sum rules based on the Pauli exclusion principle.
The FDR provides an essential bridge between theory and experiment by relating fundamental two-body correlation functions to experimentally measurable response functions.
This is exemplified by the Kubo formula, which expresses electrical conductivity in terms of the current-current correlation function.
The WTI stems from the invariance of a system under a symmetry transformation, and reflects the conservation law corresponding to the symmetry, and is therefore important for the consistency checking of various theoretical methods.
For instance, the WTI corresponding to the $\mathrm{U}(1)$ gauge invariance reflects the current conservation law.
Furthermore, the $f$-sum rule, an implication of the WTI, is also used in experimental analysis.
Therefore, both the FDR and the WTI are crucial. 
Given that covariance theory \cite{Hui_Linear_PRB2023} now enables approaches that respect these two laws, we will now discuss the third relation.\par

It has been conjectured in Ref.~\cite{rohringer_diagrammatic_2018} that no theoretical approach for correlated electrons, except for the exact solution, should be able to fulfill both of the two requirements: (i) all conservation laws and the related sum rules, and (ii) sum rules for one- and two-particle Green's functions based on the Pauli principle.
Assuming that the conjecture holds, once the FDR and the WTI are satisfied, the Pauli principle becomes a straightforward criterion for assessing the reliability of an approximation method.
Specifically, a smaller deviation in the local momentum sum rules (based on the Pauli principle) for one- and two-particle Green's functions generally indicates a closer agreement with the exact theory.\par

\begin{figure}[htb]
    \centering
    \includegraphics[width=0.95\linewidth]{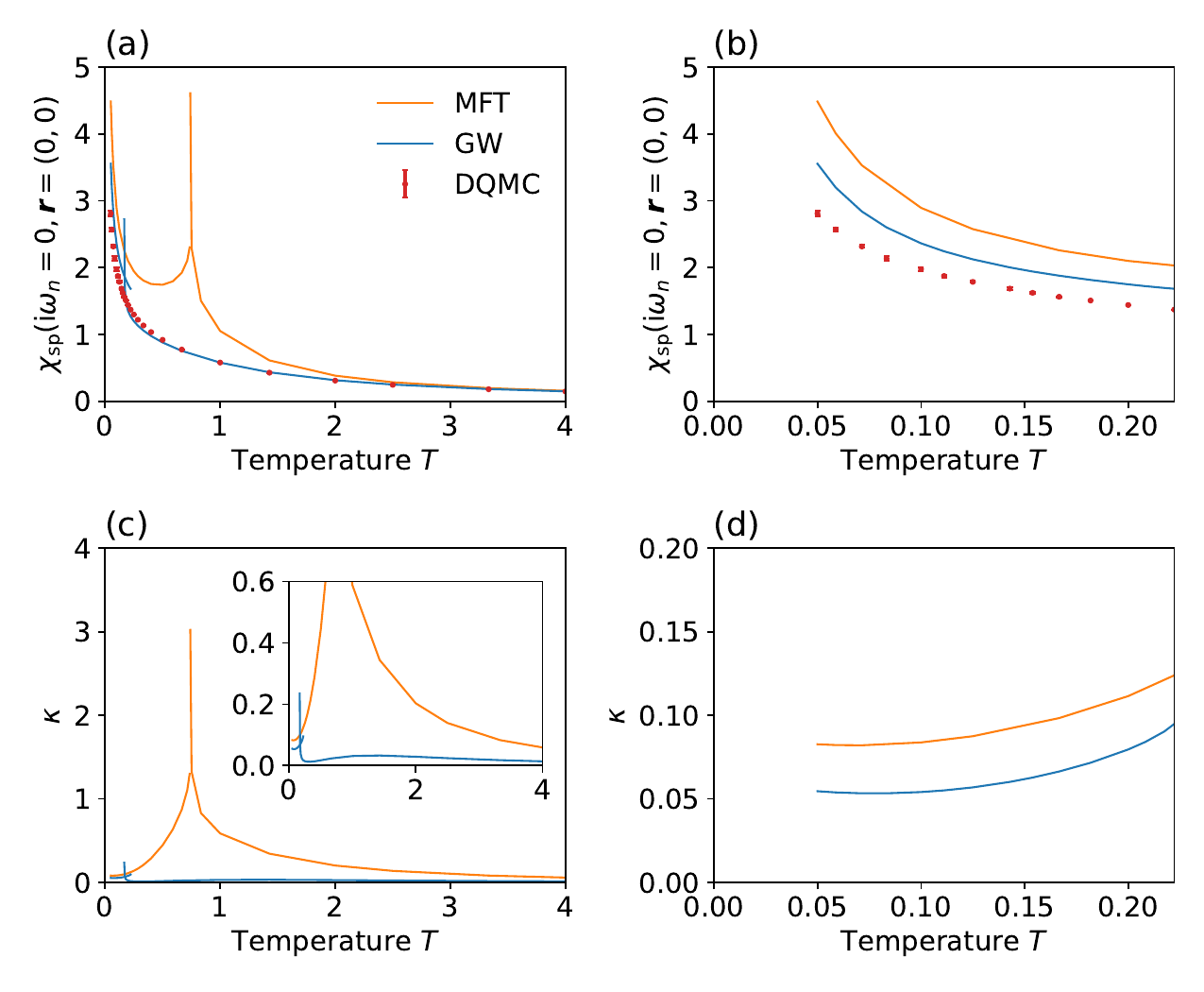}
    \caption{\label{fig:criterion}
        (a) Temperature dependence of the static spin susceptibility $\chi_{\mathrm{sp}}$, showing results from mean-field-covariance (orange, PM/AF branches), GW-covariance (blue, PM/AF branches), and DQMC (red points). 
        (b) Expanded view of the low-temperature range. 
        (c) Corresponding measure of the Pauli principle violation $\kappa$ over the full temperature range; the inset magnifies the small-$\kappa$ region. 
        (d) The $\kappa$ values in the low-temperature range shown in (b).
    }
\end{figure}
To verify the criterion, we compare the $\chi$-sum rule deviations of the symmetrized mean-field-covariance (MF-covariance) with those of the symmetrized GW approximation.
In previous sections, we use symmetrization scheme and the covariance theory based on the GW approximation to approach the correlation functions in the pseudo AF phase, respecting the FDR and the WTI.
The same framework can also be applied to mean-field theory to obtain correlation functions that obey the FDR and the WTI.
Indeed, due to the simplicity of the mean-field equation, this MF-covariance approach is equivalent to the RPA.
Figure~\ref{fig:criterion}a presents the two branches (AF and paramagnetic) of the spin correlation functions calculated using MF-covariance, GW-covariance, and DQMC.
The deviations from the $\chi$-sum rule, quantified by the parameter $\kappa$ (see Eq.~\ref{relative error of chi sum rule}), are shown in Fig.~\ref{fig:criterion}c.
For clarity, Fig.~\ref{fig:criterion}b and Fig.~\ref{fig:criterion}d show enlarged views of the respective subfigures.
Both the GW-covariance and MF-covariance results agree well with the DQMC benchmarks at extremely high and low temperatures that are far from the crossover region (i.e., near the pseudo AF critical point), and the corresponding $\kappa$ values are small ($<10\%$ for GW-covariance and $<20\%$ for MF-covariance) at these temperatures. 
However, in the crossover region, the MFT results deviate significantly from the DQMC benchmarks, and their $\kappa$ values go up to $>100\%$.
In contrast, the GW-covariance results in this region are much more reliable than those of MF-covariance.
The corresponding $\kappa$ values for the AF and paramagnetic branches remain below $10\%$ and $5\%$, respectively, except at points extremely close to the instability in the paramagnetic branch.
These results generally support our conjecture that a smaller deviation from the local momentum sum rules indicates greater reliability.

\FloatBarrier
\section{\label{chap:conclusion}Conclusion and discussion}
\subsection{\label{sec:summary}Summary}
In this work, we establish a general symmetrization scheme and test its effectiveness in the 2D Hubbard model within the GW and GW-covariance approximation.
In the pseudo AF phase predicted by the GW approximation, we calculate the single-particle Green's function and the spin correlation function.
By comparing with the DQMC method at strong coupling $U=8$ and intermediate coupling $U=4$, we confirm the scheme's effectiveness at sufficiently low temperatures.
Moreover, we calculate the correlation functions at $\beta=8$ for a range of interactions up to $U=8$, including the paramagnetic and pseudo AF phases.
These results demonstrate the reliability of our method.
Regarding the fundamental relations, the covariance theory ensures the the satisfaction of the FDR and WTI, while our numerical results indicate that at low temperatures far from the crossover region, deviations from the $\chi$-sum rule are below $10\%$ for $U=4$ and below $20\%$ for $U=8$.
We further conjecture a self-consistency criterion for many-body approximation methods: provided that the FDR and WTI are satisfied, the reliability of a method is determined by its degree of violation of the Pauli exclusion principle.
The degree of violation of the Pauli exclusion principle can be evaluated by deviations from the local momentum sum rules, such as the $\chi$-sum rule.
Our work offers a novel framework for exploring the rich physics of the 2D Hubbard model in the strong-coupling regime ($U\gtrsim 8$), which is relevant to real high-$T_\mathrm{c}$ cuprate superconductors.\par

\subsection{\label{sec:discussion}Outlook}

From the work and results presented in this paper, we can draw the following insights.
Even in the absence of continuous symmetry breaking in a 2D system, it is possible to select a group-symmetry-broken solution as the starting point for calculations.
This solution occurs at a suitably low temperature that is below the instability temperature of the group-invariant phase and hence is inaccessible from the group-invariant solution itself.
The example demonstrated here is that an AF instability temperature exists in a paramagnetic phase at half-filling for any positive $U$.
At temperatures below this, one should start from the AF solution and apply the symmetrization scheme, even though no spontaneous $\mathrm{SU}(2)$-symmetry-breaking solutions can physically persist for finite lattices or infinite lattices in dimensions lower or equal to two.
This symmetrized approach is not in contradiction with, but rather consistent with the Mermin-Wagner theorem.
After the symmetrization, any symmetry-broken physical quantities will become symmetry-invariant once again.
This coincides with the restoration of symmetry due to strong interactions of Goldstone modes in low dimensions.
Specifically, Goldstone modes in the Heisenberg universality class in 2D acquire a small mass due to the interaction, and long-range order is consequently destroyed.\par

The symmetrization scheme, proposed to address the unphysical phase-transition behavior in many-body approximation methods, has a notable limitation due to the inherent constraints of these approximations themselves.
In the crossover region near the pseudo-critical point, obtaining reliable results remains difficult on both the high- and low-temperature branches.
Specifically, the unphysical pseudo-critical point is accompanied by an anomalously large order-parameter correlation length in group-invariant solutions nearby, thereby rendering results on the high-temperature branch of the crossover region highly unreliable.
Meanwhile, in the pseudo-broken phase of the crossover region, the Higgs mass becomes very small, giving rise to large higher-order corrections.
This renders the region extremely non-perturbative, such that many-body approximation methods typically fail to provide reliable results.
Nevertheless, outside the crossover region, the symmetrized physical quantities obtained through this procedure remain reliable, as demonstrated by the physical quantities presented in this paper.
In particular, this approach can even provide a fairly reliable estimation of order parameter field correlations in the short-range regime.
At sufficiently low temperatures, where the correlation length becomes large, the method yields quite good approximations for spin-spin correlations even at a few lattice constants apart.
Although we primarily demonstrate parameter field correlations (spin-spin correlations) in this work, using this approach we also accurately calculate other physical quantities, such as the density-density correlations shown in the appendix.
Based on the evidence presented above, we speculate that, except for order parameter field (in this case, spin field) correlations, the method can also accurately compute other two-body correlations, such as density-density, superconducting pairing field, or current-current correlations, provided the system is not too close to critical or pseudo-critical phase transition lines.
Hence, our approach is promising for studying high-$T_{\mathrm{c}}$ superconductivity within the Hubbard model.\par

The mean-field results do not yield superconductivity.
However, detailed studies in Refs.~\cite{scholle_Comprehensive_2023,scholle_spiral_2024} have indicated that spin and density orders emerge at different doping levels, forming a rich variety of phases (such as spiral, stripes, and beat states).
According to Refs.~\cite{scholle_Comprehensive_2023,scholle_spiral_2024}, fluctuations of the Goldstone modes will restore the spin SU(2) symmetry for the 2D Hubbard model.
For approaches beyond mean-field theory, such as GW, GW-EDMFT, and TRILEX \cite{Vu_Trilex_PRB2017}, although the superconducting instability appears at $T_{\mathrm{c}}$, when cooling gradually from a sufficiently high temperature in the underdoped regime, the AF instability or other magnetic and density wave instabilities appear earlier at $T_{\mathrm{AF}} > T_{\mathrm{c}}$, thereby preventing the system from directly reaching $T_{\mathrm{c}}$ \cite{Vu_Trilex_PRB2017}.
However, within our symmetrization scheme, the $\mathrm{SU}(2)$-symmetrized superconducting instability can be calculated in the pseudo AF phase (or pseudogap phase), as well as in spiral or stripe phases with other spin and density orders, thereby enabling the determination of $T_{\mathrm{c}}$.
In the overdoped regime, in contrast, no such magnetic or density wave instabilities are expected, and the only instability that arises as the temperature is lowered is the superconducting instability.\par

The phase transitions in 2D systems with $\mathrm{O}(N)$ symmetry are discussed, for example, in David Tong's lecture \cite{tong_university_nodate}.
When $N=1$, it corresponds to the Ising type with discrete $Z_2$ symmetry, where phase transitions caused by discrete symmetry breaking are permitted.
When $N=2$, the Berezinskii-Kosterlitz-Thouless phase transition occurs, which is a topological phase transition without spontaneous continuous symmetry breaking.
When $N\geq 3$, there is no phase transition at finite temperature, but only a crossover occurs between the high-temperature paramagnetic phase and the fluctuating AF phase, neither of which exhibits long-range order.
Notably, the magnetism of universality class for 2D Hubbard model at half filling corresponds to the case of $N=3$.
It should be noted that a quantum phase transition occurs in 2D systems only at zero temperature for $N=3$.
Since low-temperature phenomena are closer to those at zero temperature, it is not surprising that a crossover exists between the low-temperature regime and the ``high-temperature phase''.
Although no true long-range order exists at finite temperatures, the correlation function is very large at sufficiently low temperatures \cite{Yu_Tensor_2014}.
Therefore, it is reasonably justified to use the long-range ordered (AF) phase as the starting point of approximation, while the symmetrization scheme should be adopted to ensure consistency with physical reality (i.e., the Mermin-Wagner theorem).\par

Despite the existence of continuous symmetry breaking in 3D systems at the thermodynamic limit, the symmetrization scheme proposed in this paper remains necessary for finite-sized systems at low temperatures, even in high spatial dimensions.
In studies of the 3D Hubbard model, many investigations start with the AF phase and yield physically reasonable results in the thermodynamic limit \cite{iskakov_phase_2022,Garioud_symmetry-Broken_2024}.
However, for finite lattice systems, the symmetrization scheme must be applied, and only the symmetrized results can be compared with Monte Carlo (MC) results from finite systems, since symmetry breaking is absent and the order parameter field is zero in any finite system.
The symmetry breaking occurs only in the thermodynamic limit. 
Nevertheless, finite-size scaling can be used to identify phase transitions in the thermodynamic limit and to determine whether long-range order exists, using two-body correlation functions of the order parameter field computed for finite lattices within many-body methods.
Since the order parameter field is always zero for finite lattices at any temperature, single-body averages of the order parameter field cannot be used to study the long-range order or spontaneous symmetry breaking in the thermodynamic limit from the finite $N$ analysis.\par

Numerous methods have been attempted to search for the superconducting phase in the low-temperature Hubbard model, and huge progress has been made in recent years.
Below, we provide a brief discussion of some recent developments. 
Note that, as this is not a review paper, we do not intend to present a comprehensive overview of recent advances; accordingly, the references quoted below constitute a selective and inevitably somewhat biased sample of recent works.\par

A substantial reduction in temperature has been achieved both with and without doping in cold-atom quantum simulations of the Hubbard model, as reported in Ref.~\cite{xu_neutral_2025}.
This progress allows access to interesting phases that had not previously been explored in cold-atom quantum simulators, such as spin or charge density wave states emerging at sufficiently low temperatures.
However, the temperature is still slightly high, with the maximum inverse temperature is only around $\beta = 10$ \cite{xu_neutral_2025}.
In contrast, the superconducting phase transition could occurs at $\beta \gtrsim 30$.\par

Huge progress has been made in the last four to five years in numerical studies of the Hubbard and $t-J$ model relevant to high-$T_{\mathrm{c}}$ superconductors.
The ground-state phase diagram of the $t-t'-J$ model has been studied using DMRG \cite{White_Density_1992,White_Density-matrix_1993,Schollwock_density-matrix_2005} calculations on cylinders of widths 6 and 8 in Refs.~\cite{jiang_ground_2021, gong_robust_2021, jiang_pairing_2022}.
These works report coexisting uniform AF and $d$-wave singlet pairing phases at low electron doping, while in the corresponding hole-doped region such behavior does not exist, which is inconsistent with experimental results.
However, upon increasing the cylinder width from 4 to 8, a significant strengthening of quasi-long-range superconducting correlations was observed in the hole-doped region \cite{jiang_superconducting_2023}.
Furthermore, using DMRG with substantially increased bond dimensions, the $d$-wave superconducting state also emerges on the hole-doped side at the optimal 1/8 doping in eight-leg systems of the $t-t'-J$ model \cite{lu_emergent_2024}.
The tangent space tensor renormalization group (tanTRG) method was applied to study the $t-t'-J$ model at finite temperature on cylinders up to width $W=6$ in Ref.~\cite{qu_phase_2024}, yielding the finite-temperature phase diagram of the model.
A clear finite-temperature phase transition to the $d$-wave superconducting state was identified in the electron-doped region, consistent with previous ground-state results for the $t-t'-J$ model.
Subsequently, tanTRG was applied to the Hubbard model without a $t'$ term to study at temperatures as low as $1/24$ (in unit of the hopping energy) \cite{li_tangent_2023}. 
However, no superconducting phase transition was found.
It would be of great interest for the authors of Ref.~\cite{li_tangent_2023} to investigate the Hubbard model with a $t'$ term to examine whether the finite-temperature superconducting phase transition occurs.
The ground state of the Hubbard model with a $t'$ term was studied on long six-leg square cylinders in Ref.~\cite{jiang_ground_2024}, where the $d$-wave superconducting phase was found only in the electron-doped region, consistent with results for the $t-t'-J$ model.
In Ref.~\cite{Hao_Coexistence_Science2024}, the ground state of the doped $t-t'-U$ Hubbard model on the square lattice was studied using a combination of DMRG and constrained path auxiliary field quantum Monte Carlo on width-4 and width-6 cylinders under the application of a finite global $d$-wave pairing field.
The work claims the presence of superconducting phases in both electron- and hole-doped regimes of the two-dimensional Hubbard model with next-nearest-neighbor hopping (i.e., with a $t'$ term), consistent with experimental results for high-$T_{\mathrm{c}}$ superconductivity.\par

DMRG and tanTRG can be used to study the Hubbard model at zero and finite temperature without a sign problem in contrast to DQMC.
However, the maximum cylinder width currently achievable with these methods is 8, and finite-size effects may still be significant for this width.
For finite-temperature studies of the Hubbard model, tanTRG can reach temperatures as low as 1/24 in unit of the hopping energy.
In the cuprates, however, the typical hopping energy is approximately 0.4eV, while the superconducting transition temperature at optimal hole doping ranges from 40K to 135K.
Expressed in units of the hopping energy, this corresponds to a transition temperature ranging from approximately 1/100 to 1/32.
Even the highest transition temperature remains relatively low compared to the hopping energy.
This temperature range (from 1/100 to 1/32) represents an intermediate regime that is notoriously challenging for DMRG-liked methods.
It remains challenging for DMRG or tanTRG to access even the highest transition temperature.\par

Despite the progress made with the numerical methods mentioned above, considerable challenges remain in solving the Hubbard model to understand the microscopic mechanism of cuprate high-$T_{\mathrm{c}}$ superconductivity.
We are exploring and pursuing a different path to investigate these problems, namely, a non-perturbative many-body method.
The method has passed the benchmark test against DQMC results for the Hubbard model without a $t'$ term at half-filling and very low temperatures.
Using orthogonal function expansion algorithms, many-body calculations converge exponentially as the number of imaginary time samples increases \cite{Shinaoka_Compressing_2017,Chikano_irbasis_2019,Gull_Chebyshev_2018,Dong_Legendre-spectral_2020}, allowing us to access larger lattices at extremely low temperatures while keeping computational costs acceptable.
For example, calculations of the GW approximation are feasible for $32\times 32$ and even $64\times 64$ lattices at $T = 1/100$.
We will use this method to study realistic models with the $t'$ term and away from half-filling (i.e., the doped Hubbard model).\par

We remind that the GW-covariance approximation is a non-perturbative and analytical method, which still falls within the framework of standard many-body theories such as HF and parquet approximations.
The $t-J$ model is a low-energy effective model derived from the Hubbard model in the strong repulsive interaction limit, where doubly occupied sites are projected out.
Since there is a double occupancy constraint, most many-body theories handle this via so-called slave-particle formalisms by ``fractionalizing'' the electron.
Many-body theories for the $t-J$ model are not standard in the usual sense, as the constraint requires the introduction of new particles.
In contrast to standard many-body theories, it remains unclear how to implement the WTI and FDR, or how to verify the local moment sum rule within such theories.
Nevertheless, results from these approaches have been successfully used to explain experimental observations in cuprate superconductors.
For example, Refs.~\cite{ma_low-temperature_2024, ma_correlation_2025} developed a theory based on the $t-J$ model that explains experimental resistivity data and the relation between the linear slope and the superconducting transition temperature.
In this paper, however, we concentrate on the original Hubbard model rather than its low-energy effective strong-coupling version, thereby allowing the use of standard many-body approaches to tackle those problems related to high-$T_{\mathrm{c}}$ cuprate superconductors.\par

We have applied many-body methods to study the doped Hubbard model.
In Ref.~\cite{su_effects_2025}, we used a PHF method to calculate the Hall coefficient at low temperatures and different doping levels using commonly adopted realistic parameters for the Hubbard model
The results are consistent with experimental data from both electron- and hole-doped cuprate materials.
In Ref.~\cite{xiong_application_2025}, we applied the GW-covariance method to study the negative-$U$ Hubbard model at quarter filling $n=0.5$.
The Bose-Einstein phase transition temperature was obtained for various $U$, and the Green's functions were calculated and compared with DQMC results, where no fermion sign problem occurs for negative $U$.
Thus, the GW-covariance method could be benchmarked against DQMC, and the results for the transition temperature and Green's functions were found to be consistent with DQMC up to intermediate values of $U$.
The pseudogap regime was also identified and investigated using a so-called post-GW approach.
In particular, the Goldstone mode, predicted by the WTI in the $\mathrm{U}(1)$ symmetry-breaking phase, was clearly identified, demonstrating that the theory is fully consistent with the constraints imposed by conservation laws.\par

We conclude this paper by quoting the question posed in Ref.~\cite{Hao_Coexistence_Science2024}: Is there any simple analytic theory of cuprate superconductivity in the style of Bardeen-Cooper-Schrieffer (BCS), or must we always resort to simulation?
In our view, the answer to this bold question is: an analytic many-body theory for cuprate superconductivity does exist, but numerical simulations are still essential to validate the correctness of the analytic theory.

\section*{Acknowledgements}

Authors are very grateful to B. Rosenstein, Mingpu Qing, Qiaoyi Li, Qingdong Jiang, Shiping Feng, Tao Wang, Wei Li, Xiaotian Zhang, Xinguo Ren, and Zhipeng Sun for valuable discussions and helps in numerical computations.
Authors are deeply grateful to open-source project SmoQyDQMC \cite{SmoQyDQMC.jl}. 
This code has been instrumental in applying DQMC results as benchmark within our research, significantly enhancing the efficiency and accuracy of our work.

% TODO: include author contributions
% \paragraph{Author contributions}
% This is optional. If desired, contributions should be succinctly described in a single short paragraph, using author initials.

% TODO: include funding information
\paragraph{Funding information}
This work is supported by the National Natural Science Foundation of China (Grant No.12174006 of Prof. Li's fund) and the High-performance Computing Platform of Peking University. 
H. H. acknowledges the support of the National Key R\&D Program of China (No. 2021YFA1401600), the National Natural Science Foundation of China (Grant No. 12474056).

\begin{appendix}

\section{\label{SuppleFormal}Supplementary materials on formalism}

\subsection{\label{A}Derivation of Hedin's equations}
Before the derivation, we denote the kinetic term and the interaction term of the action Eq.~(\ref{action}) by $\mathcal{S}_0$ and $\mathcal{S}_I$ for convenience. And we need to couple the external source 
\begin{equation}
    \mathcal{S}_{J}[\psi,\psi^*;J]\equiv \sum_{a}\int\mathrm d(3)\, J^{a}(3)S^{a}(3)
\end{equation}
to the system, $\mathcal{S} = \mathcal{S}_0+\mathcal{S}_I-\mathcal{S}_J$. 
Then, we start with the field translation invariance,
\begin{equation}
    \int \mathcal{D}[\psi,\psi ^{*}] \frac{\delta}{\delta \psi^{*}_{\alpha}(1)}\left[\psi^{*}_{\beta}(2)\mathrm{e}^{ -\mathcal{S}[\psi,\psi^{*}] }\right]=0.
\end{equation}
The equation above is equivalent to
\begin{equation}
    \delta_{\alpha\beta}\delta(1,2)+\left<\psi^{*}_{\beta}(2)\frac{\delta \mathcal{S}}{\delta \psi ^{*}_{\alpha}(1)}\right> =0.\label{field_tran_inv}
\end{equation}
According to the definition of the action Eq.~\ref{action}, 
\begin{eqnarray}
    &&\frac{\delta \mathcal{S}_{0}}{\delta \psi^{*}_{\alpha}(1)}=-\sum_{\gamma}\int\mathrm d(3)\,T_{\alpha\gamma}(1,3)\psi_{\gamma}(3)\label{dS_0}, \\
    &&\frac{\delta \mathcal{S}_{I}}{\delta \psi^{*}_{\alpha}(1)}=-\sum_{ab}\int d(34) \frac{\delta S^{a}(3)}{\delta \psi^{*}_{\alpha}(1)}V^{ab}(3,4)S^{b}(4)\label{dS_I}.
\end{eqnarray}
Using the trick described in Sec.~\ref{covart}, one has
\begin{equation}
    \frac{\delta (\mathcal{S}_{0}-\mathcal{S}_{J})}{\delta \psi^{*}_{\alpha}(1)}=-\sum_{\gamma}\int\mathrm d(3)\,T_{\alpha\gamma}[J](1,3)\psi_{\gamma}(3)\label{dS_0J},
\end{equation}
where 
\begin{eqnarray}
    &&\bm{T}[J](1,2)= \bm{T}(1,2)+\int d(3)\, J(3)\bm{K}_{S^a}(1,2;3), \\
    &&\bm{K}_{S^a}(1,2;3)=\bm{\sigma}^a\delta(1,2)\delta(1,3).
\end{eqnarray}
The combination of Eqs.~(\ref{field_tran_inv}, \ref{dS_0J}, \ref{dS_I}) gives the Dyson-Schwinger equation
\begin{equation}
    \begin{aligned}
        \delta_{\alpha\beta}\delta(1,2) & =\sum_{\gamma}\int\mathrm d(3)\,T_{\alpha\gamma}[J](1,3)G_{\gamma\beta}(3,2)                                            \\
                                        & +\sum_{ab}\int d(34) V^{ab}(3,4)\left< \psi^{*}_{\beta}(2)\frac{\delta S^{a}(3)}{\delta \psi^{*}_{\alpha}(1)}S^{b}(4) \right>.
    \end{aligned}\label{DysonSchwingerEquation}
\end{equation}
Making use of the equation
\begin{equation}
    \frac{\delta}{\delta J_{a}(1)}\langle F[\psi,\psi^*]\, \rangle=\left\langle F[\psi,\psi^*]\frac{\delta \mathcal{S}_{J}}{\delta J_{a}(1)} \right\rangle - \langle F[\psi,\psi^*]\, \rangle \left\langle \frac{\delta \mathcal{S}_{J}}{\delta J_{a}(1)} \right\rangle,
\end{equation}
where $F$ is any functional of $\psi,\psi^*$, and 
\begin{equation}
    \frac{\delta S^{a}(3)}{\delta \psi^{*}_{\alpha}(1)} = \delta(1,3)\sum_{\gamma}\sigma^{a}_{\alpha \gamma}\psi_{\gamma}(3), 
\end{equation}
one obtains that
\begin{equation}
    \begin{aligned}
        \left< \psi^{*}_{\beta}(2)\frac{\delta S^{a}(3)}{\delta \psi^{*}_{\alpha}(1)}S^{b}(4) \right> & = \delta(1,3)\sum_{\gamma}\sigma^{a}_{\alpha \gamma}\frac{\delta G_{\gamma\beta}(3,2)}{\delta J^{b}(4)} \\
        & +\delta(1,3)\sum_{\gamma}\sigma^{a}_{\alpha \gamma}G_{\gamma\beta}(3,2) \left< S^{b}(4) \right>.
    \end{aligned}\label{4_points_corr}
\end{equation}
Substituting the Eq.~(\ref{4_points_corr}) into the Dyson-Schwinger equation, Eq.~(\ref{DysonSchwingerEquation}), and then multiplying both sides by the inverse of the Green's function, one has
\begin{equation}
    \begin{aligned}
        G^{-1}_{\alpha\beta}(1,2) & = T_{\alpha \beta}[J](1,2)+\delta(1,2)\sum_{a}\sigma^{a}_{\alpha \beta}\sum_{b}\int d(4) V^{ab}(1,4) \left< S^{b}(4) \right>                   \\
                                  & +\sum_{ab}\sum_{\mu \nu}\sigma^{a}_{\alpha \mu} \int d(34) V^{ab}(1,3) \frac{\delta G_{\mu\nu}(1,4)}{\delta J^{b}(3)}G^{-1}_{\nu\beta}(4,2). 
    \end{aligned}
\end{equation}
Notice that $T[J=0]=G_0^{-1}$, thus the self energy is
\begin{equation}
    \begin{aligned}
        \Sigma_{\alpha\beta}(1,2) & =G^{-1}_{0\alpha\beta}(1,2)-G^{-1}_{\alpha \beta}(1,2)                                                                                         \\[0.5em]
                                  & =-\delta(1,2)\sum_{a}\sigma^{a}_{\alpha \beta}\left[J^{a}(1)+\sum_{b}\int d(4) V^{ab}(1,4) \left< S^{b}(4) \right>\right]                      \\
                                  & -\sum_{ab}\sum_{\mu \nu}\sigma^{a}_{\alpha \mu} \int d(34) V^{ab}(1,3) \frac{\delta G_{\mu\nu}(1,4)}{\delta J^{b}(3)}G^{-1}_{\nu\beta}(4,2), 
    \end{aligned}
\end{equation}
where the first term of the right hand side is the Hartree self energy
\begin{equation}
    \Sigma_{H\alpha\beta}(1,2)=-\delta(1,2)\sum_{a}\sigma^{a}_{\alpha \beta}\, v^{a}(1)\label{hartree_self_energy}
\end{equation}
with 
\begin{equation}
    v^{a}(1)\equiv J^{a}(1)+\sum_{b}\int d(4) V^{ab}(1,4) \left< S^{b}(4) \right>, \label{eff_pot}
\end{equation}
and the rest is denoted by $\Sigma'$ as 
\begin{equation}
    \Sigma'_{\alpha\beta}(1,2)=-\sum_{ab}\sum_{\mu \nu}\sigma^{a}_{\alpha \mu} \int d(34) V^{ab}(1,3) \frac{\delta G_{\mu\nu}(1,4)}{\delta J^{b}(3)}G^{-1}_{\nu\beta}(4,2). 
\end{equation}
Then, define the Hedin vertex $\Gamma_{H}$ and the functional $W$ as
\begin{eqnarray}
    &&\Gamma^a_{H\alpha\beta}(1,2;3)\equiv\frac{\delta G^{-1}_{\alpha\beta}(1,2)}{\delta v^{a}(3)},\label{Hedin_vertex_def} \\
    &&W^{ca}(5,1)\equiv\sum_{b}\int d(3)\frac{\delta v^{c}(5)}{\delta J^{b}(3)}V^{ab}(1,3),\label{W_def}
\end{eqnarray}
and using $-(\delta G/\delta J) G^{-1}=G(\delta G^{-1}/\delta J)=G(\delta G^{-1}/\delta v)(\delta v/\delta J)$, one finally has
\begin{equation}
    \Sigma'_{\alpha\beta}(1,2)=\sum_{ac}\sum_{\mu \nu}\sigma^{a}_{\alpha \mu} \int d(45)  G_{\mu\nu}(1,4)\Gamma^c_{H\nu\beta}(4,2;5)W^{ca}(5,1).\label{Sigma=GW}
\end{equation}
Applying $\delta /\delta J$ on the Eq.~(\ref{eff_pot}), one has
\begin{equation}
    \frac{\delta v^{c}(5)}{\delta J^{b}(3)}=\delta_{bc}\delta(3,5)+\frac{1}{2}\sum_{de}\sum_{\mu \nu}\sigma^{d}_{\mu \nu}\int d(46) V^{cd}(5,4)  \frac{\delta G_{\nu\mu}(4,4)}{\delta v^{e}(6)}\frac{\delta v^{e}(6)}{\delta J^{b}(3)}.
\end{equation}
And substituting the equation above into the definition of functional $W$, Eq.~(\ref{W_def}), one can prove that 
\begin{eqnarray}
    &&W^{ca}(5,1)=V^{ca}(5,1)+\sum_{de}\int d(46) V^{cd}(5,4)P^{de}(4,6)  W^{ea}(6,1),\label{W=V-P} \\
    &&P^{ab}(1,2)=-\frac{1}{2}\sum_{\mu\nu\alpha\beta}\int d(45) \, \sigma^{a}_{\mu \nu}G_{\nu\alpha}(1,4)\Gamma^b_{H\alpha\beta}(4,5;2)G_{\beta \mu}(5,1).\label{P=-GG}
\end{eqnarray}
The Eq.~(\ref{W=V-P}) is namely the equation $W^{-1}=V^{-1}-P$. The Eqs.~(\ref{hartree_self_energy}, \ref{Sigma=GW}, \ref{W=V-P}, \ref{P=-GG}) together with $G^{-1}=G_0^{-1}-\Sigma_H-\Sigma'$ form the Hedin Equations. Notice that 
\begin{equation}
\begin{aligned}
    \Gamma^a_{H\alpha\beta}(1,2;3) &\equiv \frac{\delta G^{-1}_{\alpha\beta}(1,2)}{\delta v^{a}(3)}=-\frac{\delta \Sigma_{H\alpha\beta}(1,2)}{\delta v^{a}(3)}-\frac{\delta \Sigma'_{\alpha\beta}(1,2)}{\delta v^{a}(3)} \\
    &=\sigma_{\alpha\beta}\delta(1,2)\delta(1,3)-\frac{\delta \Sigma'_{\alpha\beta}(1,2)}{\delta v^{a}(3)}, 
\end{aligned}
\end{equation}
and the GW approximation ignores the $\delta \Sigma'/\delta v$.

\subsection{\label{B}The GW and GW-covariance equations in momentum space}

On the reciprocal space of the A-B sublattice, the GW equations are 
\begin{eqnarray}
    &&[\tilde{\bm{G}}^{-1}]^{l_{1}l_{2}}(k)=\tilde{\bm{T}}^{l_{1}l_{2}}(k)-\tilde{\bm{\Sigma}}_{H}^{l_{1}l_{2}}(k)-\tilde{\bm{\Sigma}}_{GW}^{l_{1}l_{2}}(k),\label{eq:gw k 1} \\[0.5em]
    &&\tilde{\bm{\Sigma}}_{H}^{l_{1}l_{2}}(k)=-\delta_{l_{1}l_{2}}\sum_{a}\bm{\sigma}^{a}\sum_{b}\sum_{l_{3}}\tilde{V}^{al_{1},bl_{3}}(0)\,\mathrm{Tr}\left[\bm{\sigma}^{b}\bm{G}^{l_3l_3}(0)\right],\label{eq:gw k 2} \\
    &&\tilde{\bm{\Sigma}}_{GW}^{l_{1}l_{2}}(k)=\sum_{ab}\frac{1}{\beta N}\sum_{q}\bm{\sigma}^{a}\tilde{\bm{G}}^{l_{1}l_{2}}(q+k)\bm{\sigma}^{b}\tilde{W}^{bl_{2},al_{1}}(q),\label{eq:gw k 3} \\
    &&[\tilde{W}^{-1}]^{al_1, bl_2}(q)=[\tilde{V}^{-1}]^{al_1,bl_2}(q)-\tilde{P}^{al_1,bl_2}(q),\label{eq:gw k 4} \\[0.5em]
    &&\tilde{P}^{al_1,bl_2}(q)=-\frac{1}{\beta N}\sum_{k} \mathrm{Tr}\left[\bm{\sigma}^{a}\tilde{\bm{G}}^{l_{1}l_{2}}(k+q)\bm{\sigma}^{b}\tilde{\bm{G}}^{l_{2}l_{1}}(k)\right],\label{eq:gw k 5}
\end{eqnarray}
where $k=(\mathrm{i}\omega_n,\bm{k})$, $\omega_n$ is the Matsubara frequency, $\omega_n=2n\pi/\beta$ for boson and $\omega_n=(2n+1)\pi/\beta$ for fermion. \par

The Fourier transformation of the vertex-like functionals $\bm{\Lambda}(1-2,1-3)=\bm{\Lambda}(1,2;3)$ is defined by applying the Eqs.~(\ref{eq:general fourier}, \ref{eq:general inv fourier}) on $1-2$ and $2-3$ respectively. Thus, the correlation function is
\begin{equation}
    \tilde{\chi}_{XY}(q)=\frac{1}{\beta N}\sum_{k} \sum_{l_{1}l_{2}}\mathrm{Tr}\left[\tilde{\bm{K}}_{X}^{l_{1}l_{2}}(k,-q)\tilde{\bm{\Lambda}}_{\phi}^{l_{2}l_{1}}(k-q,q)\right],
\end{equation}
and the GW-covariance equations are 
\begin{eqnarray}
    &&\tilde{\bm{\Gamma}}_{\phi}^{l_{1}l_{2}}(k,q)=\left(\tilde{\bm{\gamma}}_{\phi}-\tilde{\bm{\Gamma}}_{\phi}^{H}-\tilde{\bm{\Gamma}}_{\phi}^{MT}-\tilde{\bm{\Gamma}}_{\phi}^{AL}\right)^{l_{1}l_{2}}(k,q),\label{eq:gw-c k 1} \\[0.3em]
    &&\tilde{\bm{\Gamma}}_{\phi}^{H\, l_{1}l_{2}}(k,q)=-\delta_{l_{1}l_{2}}\sum_{c}\bm{\sigma}^{c}\sum_{b}\sum_{l_{4}}\tilde{V}^{cl_{1}, bl_{4}}(q) \frac{1}{\beta N}\sum_{k^{\prime}}\mathrm{Tr}\left[\bm{\sigma}^{b}\tilde{\bm{\Lambda}}^{l_{4}l_{4}}_{\phi}(k^{\prime},q)\right],\label{eq:gw-c k 2} \\
    &&\tilde{\bm{\Gamma}}_{\phi}^{MT\, l_{1}l_{2}}(k,q)= \sum_{cb} \frac{1}{\beta N}\sum_{q^{\prime}} \bm{\sigma}^{c}\tilde{\bm{\Lambda}}_{\phi}^{l_{1}l_{2}}(q^{\prime}+k,q)\bm{\sigma}^{b}\tilde{W}^{bl_{2},cl_{1}}(q^{\prime}),\label{eq:gw-c k 3} \\
    &&\begin{aligned}
        \tilde{\bm{\Gamma}}_{\phi}^{AL\, l_{1}l_{2}}(k,q)=-\sum_{abcd}\sum_{l_{4}l_{5}} \frac{1}{\beta N}\sum_{p} &\bm{\sigma}^{a}\tilde{\bm{G}}^{l_{1}l_{2}}(k+p+q)\bm{\sigma}^{b} \\
        &\times\tilde{W}^{bl_{2}, cl_{4}}(p+q)\tilde{\Gamma}_{\phi}^{W cl_{4},dl_{5}}(p,q)\tilde{W}^{dl_{5}, al_{1}}(p), 
    \end{aligned}\label{eq:gw-c k 4} \\
    &&\begin{aligned}
        \tilde{\Gamma}_{\phi}^{W dl_{4}, el_{5}}(p,q)= \frac{1}{\beta N}\sum_{k} \mathrm{Tr}&\left[\bm{\sigma}^{d}\tilde{\bm{\Lambda}}_{\phi}^{l_{4}l_{5}}(k+p,q)\bm{\sigma}^{e}\tilde{\bm{G}}^{l_{5}l_{4}}(k)\right. \\
        &\left. +\, \bm{\sigma}^{d}\tilde{\bm{G}}^{l_{4}l_{5}}(k+p+q)\bm{\sigma}^{e}\tilde{\bm{\Lambda}}_{\phi}^{l_{5}l_{4}}(k,q)\right],
    \end{aligned}\label{eq:gw-c k 5} \\
    &&\tilde{\bm{\Lambda}}_{\phi}^{l_{1}l_{2}}(k,q)=-\sum_{l_{4}l_{5}}\tilde{\bm{G}}^{l_{1}l_{4}}(k+q)\tilde{\bm{\Gamma}}_{\phi}^{l_{4}l_{5}}(k,q)\tilde{\bm{G}}^{l_{5}l_{2}}(k). \label{eq:gw-c k 6}
\end{eqnarray}

\subsection{\label{symmep}Symmetrization for $\mathrm{SU}(2)$ broken AF phase}
The spatial distribution of the AF order parameter can be expressed as follows. 
Without loss of generality, 
it is assumed that the direction of the magnetic moment $\langle\bm S(\bm x)\rangle$ is along the $z$-axis. 
\begin{equation}
    \langle S^x(\bm x)\rangle=\langle S^y(\bm x)\rangle = 0,\quad
    \left< S^z(\bm{x}) \right> =S^z_0\,\mathrm{e}^{\mathrm{i}\bm{Q}
    \cdot\bm{x}}= \sum_{\bm{k}\in\mathcal{B}}S^z_0\delta_{\bm{k},\bm{Q}}\mathrm{e}^{\mathrm{i}\bm{k}
    \cdot\bm{x}}, \label{AF order}
\end{equation}
where $S^z_0$ is amplitude of the AF wave, $\bm{Q}=\pi\bm{b}_{x}+\pi\bm{b}_{y}$ is the AF wave vector, and we denote the reciprocal lattice of 2D square lattice as $\mathcal{B}$ for clarity. The amplitude of AF order breaks $\mathrm{SU}(2)$ invariance and the wave vector $\bm{Q}$ is related to the breaking of phase translation invariance. We only need to restore the continuous symmetry, $\mathrm{SU}(2)$ symmetry. However, specially for the AF order, the effect of the wave vector ceases to exist when the amplitude disappears, the translation invariance caused by the AF order happens to be automatically restored. For the $\mathrm{SU}(N)$ group, the integrations (see Eq.~(\ref{haar measure})) of the fundamental representation are
\begin{equation}
    \int\mathrm dU\, 
    [U^*]^{\ \alpha'_1}_{\alpha_1}U^{\ \beta'_1}_{\beta_1}
    =\frac{1}{N}\delta^{\alpha'_1 \beta'_1}\delta_{\alpha_1 \beta_1}
    \label{eq:integral 1}
\end{equation}
and
\begin{equation}
    \begin{aligned}
        \int\mathrm dU\
        [U^*]^{\ \alpha'_1}_{\alpha_1}[U^*]^{\ \alpha'_2}_{\alpha_2}
        U^{\ \beta'_1}_{\beta_1} & U^{\ \beta'_2}_{\beta_2}
        = \frac{1}{N^2-1}\left(
        \delta^{\alpha'_1\beta'_1}\delta^{\alpha'_2\beta'_2}
        \delta_{\alpha_1\beta_1}\delta_{\alpha_2\beta_2}
        +\delta^{\alpha'_1\beta'_2}\delta^{\alpha'_2\beta'_1}
        \delta_{\alpha_1\beta_2}\delta_{\alpha_2\beta_1}
        \right)                      \\
        &- \frac{1}{(N^2-1)N}\left(
        \delta^{\alpha'_1\beta'_2}\delta^{\alpha'_2\beta'_1}
        \delta_{\alpha_1\beta_1}\delta_{\alpha_2\beta_2}
        +\delta^{\alpha'_1\beta'_1}\delta^{\alpha'_2\beta'_2}
        \delta_{\alpha_1\beta_2}\delta_{\alpha_2\beta_1}
        \right).
    \end{aligned}
    \label{eq:integral 2}
\end{equation}
As a consequence, the symmetrized Green's function defined in Eq.~(\ref{eq:define G}) is given by
\begin{equation}
    \begin{aligned}
        \bar{G}_{\alpha_1\alpha_2}(1,2)
         & = \overline{\langle\psi^*_{\alpha_2}(2)\psi_{\alpha_1}(1)\rangle}
        = \int\mathrm dU\, 
        [U^*]^{\ \alpha'_1}_{\alpha_1}U^{\ \alpha'_2}_{\alpha_2}
        \langle\psi^*_{\alpha_2'}(2)\psi_{\alpha_1'}(1)\rangle               \\
         & = \frac{1}{2}\delta_{\alpha_1\alpha_2}\left[
            \langle\psi^*_\uparrow(2)\psi_\uparrow(1)\rangle
            +\langle\psi^*_\downarrow(2)\psi_\downarrow(1)\rangle
        \right]                                                              \\
         & = \frac{1}{2}\delta_{\alpha_1\alpha_2}\left[
        G_{\uparrow\uparrow}(1,2)+G_{\downarrow\downarrow}(1,2)
        \right].
    \end{aligned}
    \label{eq:symmetric G}
\end{equation}
And for the spin-$z$ correlation defined in Eq.~(\ref{corr_defin}), 
the symmetrization gives
\begin{equation}
    \chi_\mathrm{sp}(1,2)\equiv\bar{\chi}_{S^{z}S^{z}}(1,2)=\frac{1}{3}\sum_{b=x,y,z}\chi_{S^{b}S^{b}}(1,2).
    \label{eq:symmetric chi}
\end{equation}\par

Actually, in this case, the translation invariance has been restored after the symmetrization.
To clarify this, we use a symmetry of a $z$-polarized pure AF system:
\begin{equation}
\begin{aligned}
    \langle \psi^*_{\alpha_1}(\bm{x}_1)&\dots\psi^*_{\alpha_n}(\bm{x}_n)\psi_{\beta_1}(\bm{x}'_1)\dots\psi_{\beta_m}(\bm{x}'_m) \rangle \\
    &=\langle \mathrm{i}\psi^*_{\bar{\alpha}_1}(\bm{x}_1+\bm{a}_{x})\dots\mathrm{i}\psi^*_{\bar{\alpha}_n}(\bm{x}_n+\bm{a}_{x})\mathrm{i}\psi_{\bar{\beta}_1}(\bm{x}'_1+\bm{a}_{x})\dots\mathrm{i}\psi_{\bar{\beta}_m}(\bm{x}'_m+\bm{a}_{x}) \rangle, 
\end{aligned}
\end{equation}
where the $\bar{\alpha}$ means flipping the spin, i.e., $\bar{\alpha}=\downarrow$ when $\alpha=\uparrow$.
This symmetry transformation $\psi_{\alpha}(\bm{x})\to \mathrm{i}\psi_{\bar{\alpha}}(\bm{x}+\bm{a}_{x})$ is combination of a rotation of $\pi$ around the $x$-axis and a translation of the smallest unit along the $x$-axis.
For the Green's function, it is quite simple that
\begin{equation}
    G_{\alpha_1\alpha_2}(1,2)=G_{\bar{\alpha}_1\bar{\alpha}_2}(1+\bm{a}_{x},2+\bm{a}_{x}).\label{eq:af symmetry G}
\end{equation}
Then one can prove that
\begin{equation}
    \bar{G}_{\alpha_1\alpha_2}(1,2)=\bar{G}_{\alpha_1\alpha_2}(1+\bm{a}_{x},2+\bm{a}_{x}).
    \label{eq:G translational invariance}
\end{equation}
For the spin correlation, one can first consider how the spin field operator changes. For spin-$z$ operator, 
\begin{equation}
    \begin{aligned}
        S^{z}(1) & =\sum_{\alpha\beta}\psi^{*}_{\alpha}(1)\sigma^{z}_{\alpha\beta}\psi_{\beta}(1) \\
        & \to\sum_{\alpha\beta}\psi^{*}_{\bar{\alpha}}(1+\bm{a}_{x})\sigma^{z}_{\alpha\beta}\psi_{\bar{\beta}}(1+\bm{a}_{x})=-S^{z}(1+\bm{a}_{x}), 
    \end{aligned}
\end{equation}
similarly, $S^{x}(1)\to S^{x}(1+\bm{a}_{x}), S^{y}(1)\to -S^{y}(1+\bm{a}_{x})$. As a result, (for $b=x,y,z$), 
\begin{equation}
    \chi_{S^{b}S^{b}}(1,2)=\chi_{S^{b}S^{b}}(1+\bm{a}_{x},2+\bm{a}_{x}). 
\end{equation}
Thus, it is clear that
\begin{equation}
    \bar{\chi}_{S^{z}S^{z}}(1,2)=\bar{\chi}_{S^{z}S^{z}}(1+\bm{a}_{x},2+\bm{a}_{x}).
    \label{eq:chi translational invariance}
\end{equation}\par

Given that translational invariance has been shown to be restored (refer to Eq.~(\ref{eq:G translational invariance}) and Eq.~(\ref{eq:chi translational invariance})), it is now valid to apply the Fourier transform (over the whole 2D square lattice rather than the A-B sublattice) to these symmetrized quantities.
This allows us to relate the 2D square reciprocal lattice to the A-B reciprocal lattice and to perform the symmetrization directly in momentum space, which may be convenient in some cases.
The trick is to do the following.
Through Eq.~(\ref{eq:G translational invariance}), we can express the Green's function as
\begin{equation}
    \bar{\bm{G}}(\bm{x}_1,\bm{x}_2)=\frac{1}{2}\left[\bar{\bm{G}}(\bm{x}_1,\bm{x}_2)+\bar{\bm{G}}(\bm{x}_1+\bm{a}_{x},\bm{x}_2+\bm{a}_{x})\right], \label{eq:symmetric G 1}
\end{equation}
where, for convenience, the imaginary time $\tau$ is omitted.
Without losing generality, one sets $\bm{x}_1=\bm{R}+\bm{u}_{l}$ and $\bm{x}_2=\bm{0}$, and obtains
\begin{equation}
    \bar{\bm{G}}(\bm{R}+\bm{u}_{l},\bm{0})=\frac{1}{2}\left[\bar{\bm{G}}(\bm{R}+\bm{u}_{l},\bm{0})+\bar{\bm{G}}(\bm{R}+\bm{u}_{l}+\bm{a}_{x},\bm{a}_{x})\right].\label{eq:symmetric G 2}
\end{equation}
Substituting $\bm{u}_{A}=\bm{0}$ and $\bm{u}_{B}=\bm{a}_x$, Eq.~(\ref{eq:symmetric G 2}) yields
\begin{eqnarray}
    && \bar{\bm{G}}(\bm{R},\bm{0})=\frac{1}{2}\left[\bar{\bm{G}}(\bm{R},\bm{0})+\bar{\bm{G}}(\bm{R}+\bm{a}_{x},\bm{a}_{x})\right], \\
    && \bar{\bm{G}}(\bm{R}+\bm{a}_x,\bm{0})=\frac{1}{2}\left[\bar{\bm{G}}(\bm{R}+\bm{a}_x,\bm{0})+\bar{\bm{G}}(\bm{R}+\bm{a}_1,\bm{a}_{x})\right].
\end{eqnarray}
The Green's functions in the right hand side can be expressed by A-B lattice representation (see Sec.\ref{applia}), as
\begin{eqnarray}
    && \bar{\bm{G}}(\bm{R},\bm{0})=\frac{1}{2}\left[\bar{\bm{G}}^{AA}(\bm{R},\bm{0})+\bar{\bm{G}}^{BB}(\bm{R},\bm{0})\right], \\
    && \bar{\bm{G}}(\bm{R}+\bm{a}_x,\bm{0})=\frac{1}{2}\left[\bar{\bm{G}}^{BA}(\bm{R},\bm{0})+\bar{\bm{G}}^{AB}(\bm{R}+\bm{a}_1,\bm{0})\right].
\end{eqnarray}
Then, the Fourier transformation (on the 2D square lattice) is
\begin{equation}
    \begin{aligned}
        \tilde{\bar{\bm{G}}}(\bm{k}\in\mathcal{B}) & =\sum_{\bm{x}}\mathrm{e}^{-\mathrm{i}\bm{k}\cdot \bm{x}}\bar{\bm{G}}(\bm{x})=\sum_{\bm{R}}\mathrm{e}^{-\mathrm{i}\bm{k}\cdot \bm{R}}\bar{\bm{G}}(\bm{R})+\mathrm{e}^{-\mathrm{i}\bm{k}\cdot \bm{a}_{x}}\sum_{\bm{R}}\mathrm{e}^{-\mathrm{i}\bm{k}\cdot \bm{R}}\bar{\bm{G}}(\bm{R}+\bm{a}_{x}) \\
        & =\frac{1}{2}\sum_{\bm{R}}\mathrm{e}^{-\mathrm{i}\bm{k}\cdot\bm{R}}
        \left[
        \bar{\bm{G}}^{AA}(\bm{R})+\bar{\bm{G}}^{BB}(\bm{R})
        +\mathrm{e}^{\mathrm{i}\bm{k}\cdot \bm{a}_{x}}\bar{\bm{G}}^{AB}(\bm{R})
        +\mathrm{e}^{-\mathrm{i}\bm{k}\cdot \bm{a}_{x}}\bar{\bm{G}}^{BA}(\bm{R})
        \right],\label{eq:symmetric G 1 AB FT}
    \end{aligned}
\end{equation}
where $\bar{\bm{G}}(\bm{x})$ means $\bar{\bm{G}}(\bm{x}, \bm{0})$ and $\bar{\bm{G}}^{l_1l_2}(\bm{R})$ means $\bar{\bm{G}}^{l_1l_2}(\bm{R}, \bm{0})$. 
Denote the momentum space of the A-B sublattice by $\mathcal{B}_{\mathrm{AB}}$. If $\bm{k}\in\mathcal{B}_{\mathrm{AB}}$, the right hand side of the Eq.~(\ref{eq:symmetric G 1 AB FT}) is just the Fourier transform in the A-B sublattice defined as Eq.~(\ref{eq:general fourier}). One can divide the $\mathcal{B}$ into $\mathcal{B}_{\mathrm{AB}}$ and $\{\bm{k}+\bm{Q}|\bm{k}\in\mathcal{B}_{\mathrm{AB}}\}$. Then, since $\bm{Q}\cdot (n_{1}\bm{a}_{1}+n_{2}\bm{a}_{2})=2\pi(n_{1}+n_{2})$ and $\bm{Q}\cdot \bm{a}_{x}=\pi$, for $\bm{k}\in\mathcal{B}_{\mathrm{AB}}$ one has 
\begin{eqnarray}
    &&\tilde{\bar{\bm{G}}}(\bm{k})=\frac{1}{2}\left[
    \tilde{\bar{\bm{G}}}^{AA}(\bm{k})+\tilde{\bar{\bm{G}}}^{BB}(\bm{k})+\mathrm{e}^{\mathrm{i}\bm{k}\cdot \bm{a}_{x}}\tilde{\bar{\bm{G}}}^{AB}(\bm{k})+\mathrm{e}^{-\mathrm{i}\bm{k}\cdot \bm{a}_{x}}\tilde{\bar{\bm{G}}}^{BA}(\bm{k})
    \right],\label{eq:symmetrization G k} \\
    &&\tilde{\bar{\bm{G}}}(\bm{k}+\bm{Q})=\frac{1}{2}\left[
    \tilde{\bar{\bm{G}}}^{AA}(\bm{k})+\tilde{\bar{\bm{G}}}^{BB}(\bm{k})-\mathrm{e}^{\mathrm{i}\bm{k}\cdot \bm{a}_{x}}\tilde{\bar{\bm{G}}}^{AB}(\bm{k})-\mathrm{e}^{-\mathrm{i}\bm{k}\cdot \bm{a}_{x}}\tilde{\bar{\bm{G}}}^{BA}(\bm{k})
    \right].\label{eq:symmetrization G k+Q}
\end{eqnarray}
Similar formula also holds for two-body correlation functions.
Substituting
\begin{equation}
    \tilde{\bar{\bm{G}}}^{l_1l_2}(\bm{k}) = \frac{1}{2}\bm{\sigma}^{0}\left[
        \tilde{G}_{\uparrow\uparrow}^{l_1l_2}(\bm{k})+\tilde{G}_{\downarrow\downarrow}^{l_1l_2}(\bm{k})
        \right],
\end{equation}
which is the Eq.~(\ref{eq:symmetric G}) in the A-B sublattice reciprocal space, into the Eqs.~(\ref{eq:symmetrization G k}, \ref{eq:symmetrization G k+Q}), we accomplish the symmetrization directly in momentum space. And one can also go back to the position space directly from the Eqs.~(\ref{eq:symmetrization G k}, \ref{eq:symmetrization G k+Q}) by
\begin{eqnarray}
    &&\bar{\bm{G}}(\bm{R})=\frac{1}{2N}\sum_{\bm{k}\in\mathcal{B}_{\mathrm{AB}}}\mathrm{e}^{\mathrm{i}\bm{k}\cdot \bm{R}}\left[\tilde{\bar{\bm{G}}}(\bm{k})+\tilde{\bar{\bm{G}}}(\bm{k}+\bm{Q})\right], \\
    &&\bar{\bm{G}}(\bm{R}+\bm{a}_{x})=\frac{1}{2N}\sum_{\bm{k}\in\mathcal{B}_{\mathrm{AB}}}\mathrm{e}^{\mathrm{i}\bm{k}\cdot \bm{R}}\mathrm{e}^{\mathrm{i}\bm{k}\cdot \bm{a}_{x}}\left[ \tilde{\bar{\bm{G}}}(\bm{k}) - \tilde{\bar{\bm{G}}}(\bm{k}+\bm{Q}) \right],
\end{eqnarray}
where $2N$ is number of lattice points of the 2D square lattice. Regarding the correlation function, an analogous methodology is employed, the details of which are omitted herein for conciseness. 

\section{\label{C}Effectiveness of DQMC under extreme conditions}
\begin{figure}[htb]
    \centering
    \includegraphics[width=0.95\linewidth]{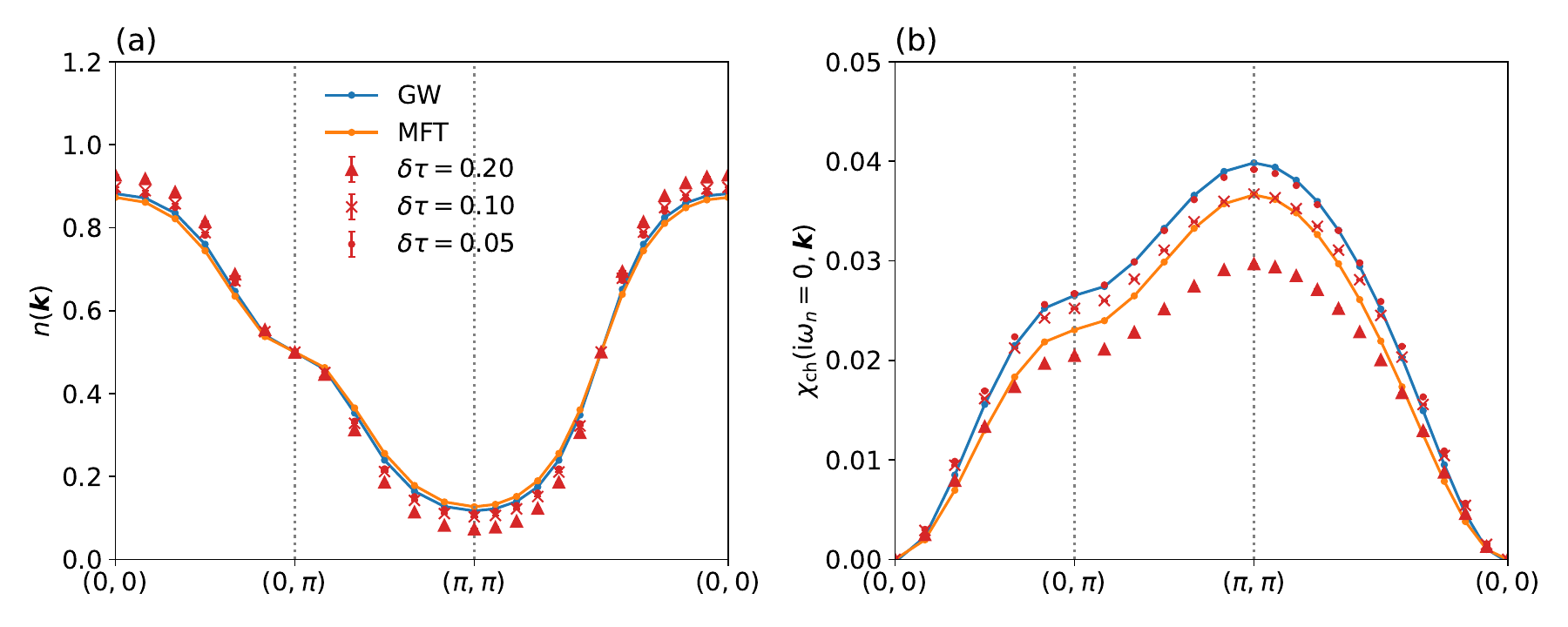}
    \caption{\label{fig:b20u8}
    Panels (a) and (b) display the momentum distribution $n(\bm{k})$ and charge correlation function $\chi_\mathrm{ch}$ relatively, which are presented at an extremely low temperatures $\beta=20$ and a typically strong interaction strength $U=8$ on a $12\times 12$ lattice. 
    The results from the GW-covariance and the mean-field-covariance approaches are compared with those from DQMC using different Trotter parameter $\Delta\tau=0.20,0.10,0.05$.
    The variations over $\Delta\tau$ is neglectable compared with the errorbar.
    }
\end{figure}
We investigate an extreme condition, 
specifically an extremely low temperature $\beta=20$ 
and $U=8$ in the regime of strong correlations. 
Within this domain, DQMC results exhibit poor convergence as the system size increases, 
and the Trotter error arising from the finite $\Delta\tau$
cannot be neglected \cite{schafer_tracking_2021}. 
To address the challenge of the first difficulty, 
we focus on a finite-sized system, 
namely the lattice size is $12\times 12$. 
Additionally, we calculate results for various $\Delta\tau$ values to estimate the magnitude of the Trotter error. 
Notably, in this extreme condition, 
the computational cost of many-body methods, 
exemplified by the GW-covariance and MF-covariance approximation, 
remains within an acceptable range. \par

We calculated the momentum distribution $n(\bm{k})$, 
which is consistent with that in Ref.~\cite{Varney_Quantum_2009}
\begin{equation}
    n(\bm{k})=\frac{1}{N}\sum_{\bm{r}_1,\bm{r}_2}\sum_\sigma
    e^{\mathrm i\bm{k}\cdot(\bm{r}_i-\bm{r}_j)}
    \langle\hat c^\dagger_\sigma(\bm{r}_i)\hat c_\sigma(\bm{r}_j)\rangle, 
\end{equation}
and the charge correlations $\chi_\mathrm{ch}\equiv \chi_{S^0 S^0}$ (see Eq.~(\ref{corr_defin})) at $\mathrm i\omega_n=0$ in the momentum space. 
These two represent single-particle properties and two-particle properties, respectively. 
% These properties all respect $\mathrm{SU}(2)$ symmetry. 
As shown in Fig.~\ref{fig:b20u8}, $\Delta \tau=0.1$ for the momentum distribution $n(\bm{k})$ and $\Delta \tau=0.05$ for the charge correlation function $\chi_\mathrm{ch}(\mathrm i\omega_n=0, \bm{k})$ are small enough to make the DQMC simulations converge. 
When comparing the $\chi_\mathrm{ch}$ calculated by GW-covariance, MF-covariance and DQMC with $\Delta\tau =0.1$, one will mistakenly believe that the MF-covariance exhibits better behavior than the GW-covariance, although the fact shown by DQMC with $\Delta\tau=0.05$ is that the GW-covariance perfactly matches the DQMC. 
Thus, it is necessary to ensure that the DQMC data used for benchmarking converges. 

\section{\label{results u4}Results at half filling with intermediate coupling $U=4$}
\begin{center}
\begin{figure}[p]
    \centering
    \includegraphics[width=0.95\linewidth]{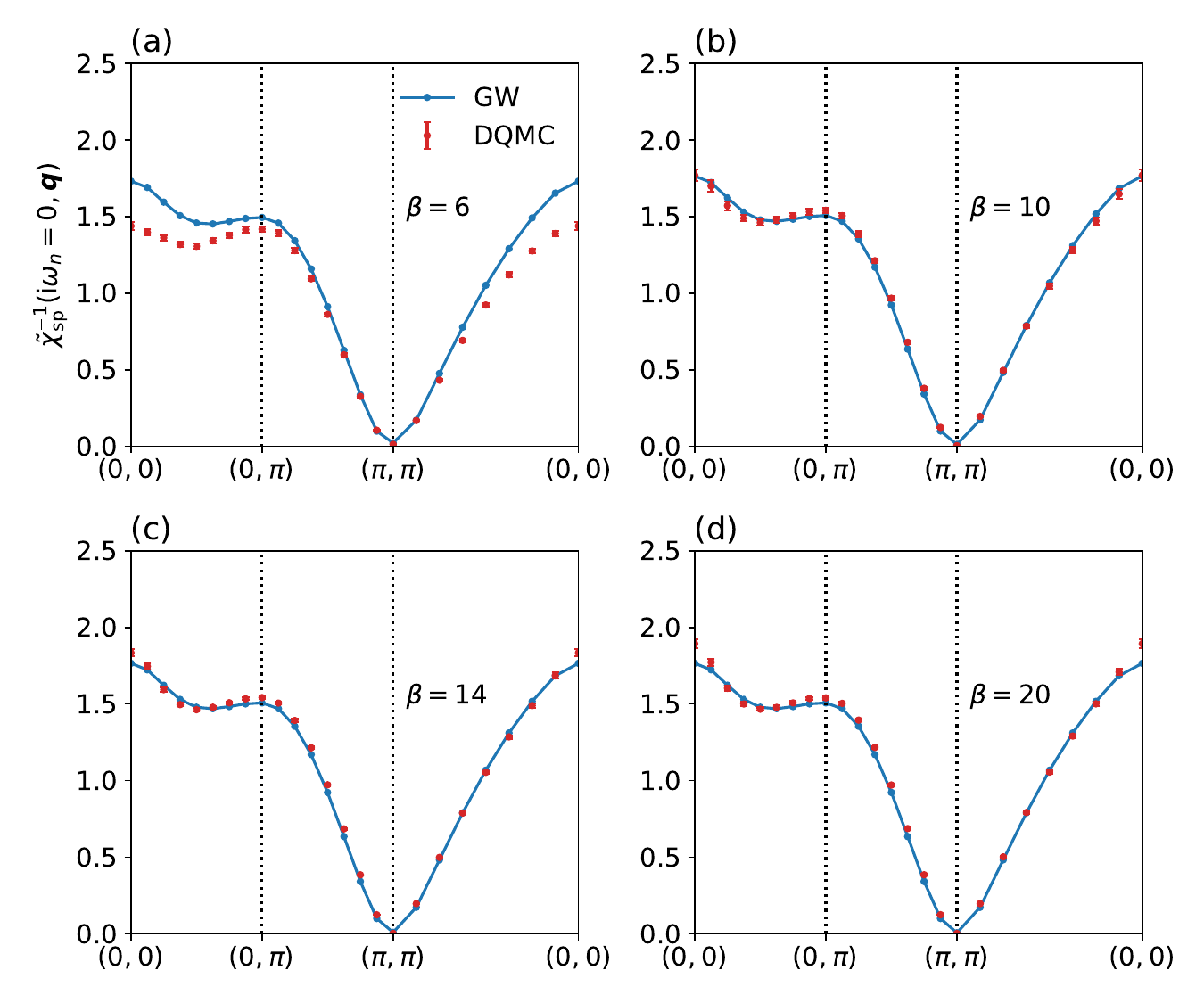}
    \caption{\label{fig:chi k w0 u4}
        Momentum dependence of the inverse spin susceptibility for the half-filled Hubbard model on a $16\times16$ lattice with $U=4$. 
        Results are shown for decreasing temperatures, corresponding to inverse temperatures (a) $\beta=6$, (b) $\beta=10$, (c) $\beta=14$, and (d) $\beta=20$. The momentum path follows $(0,0)\to(0,\pi)\to(\pi,\pi)\to(0,0)$. 
        Blue curves denote the symmetrized GW-covariance approximation, while red symbols with error bars show DQMC benchmarks. 
        A significant improvement in the agreement between GW-covariance and DQMC is observed upon cooling from $\beta=6$ to $\beta=10$. 
        At lower temperatures, the agreement remains largely unchanged, showing little evidence of any further enhancement.
    }
\end{figure}
\begin{figure}[p]
    \centering
    \includegraphics[width=0.95\linewidth]{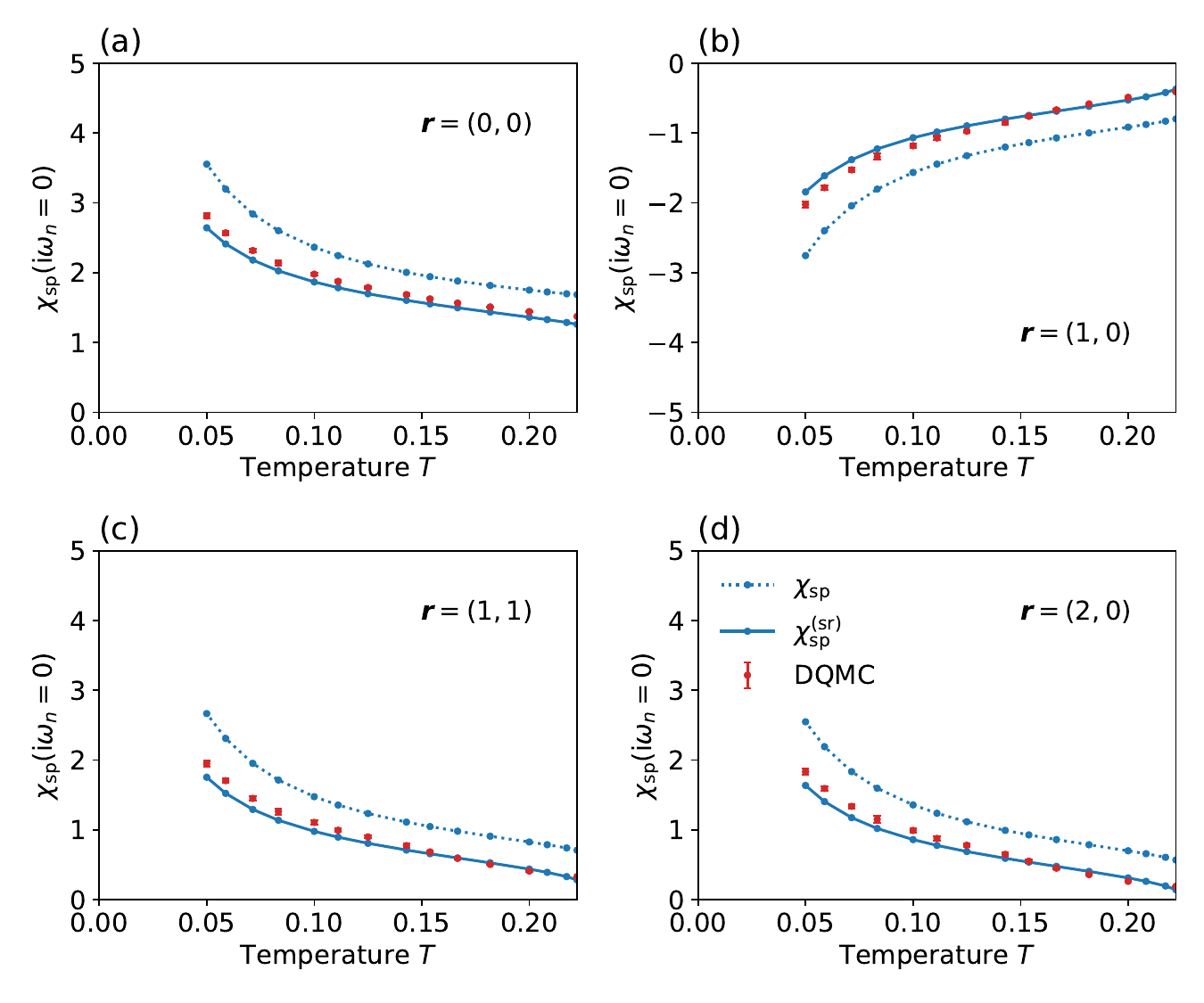}
    \caption{\label{fig:chi r w u4}
        Temperature dependence of the static spin correlation function in spatial space for the half-filled Hubbard model ($U=4$) on a $16\times16$ lattice. Panels (a)-(d) display results for lattice separations $\mathbf{r} = (0,0)$, $(1,0)$, $(1,1)$, and $(2,0)$, respectively. 
        The DQMC benchmarks (red points with error bars) are compared with the GW-covariance results.
        The $\chi_{\mathrm{sp}}$ (blue dashed curve) denotes the standard result and the $\chi_{\mathrm{sp}}^{\mathrm{(sr)}}$ (blue solid curve) denotes the $\chi$-constrained result.
    }
\end{figure}
\begin{figure}[p]
    \centering
    \includegraphics[width=0.95\linewidth]{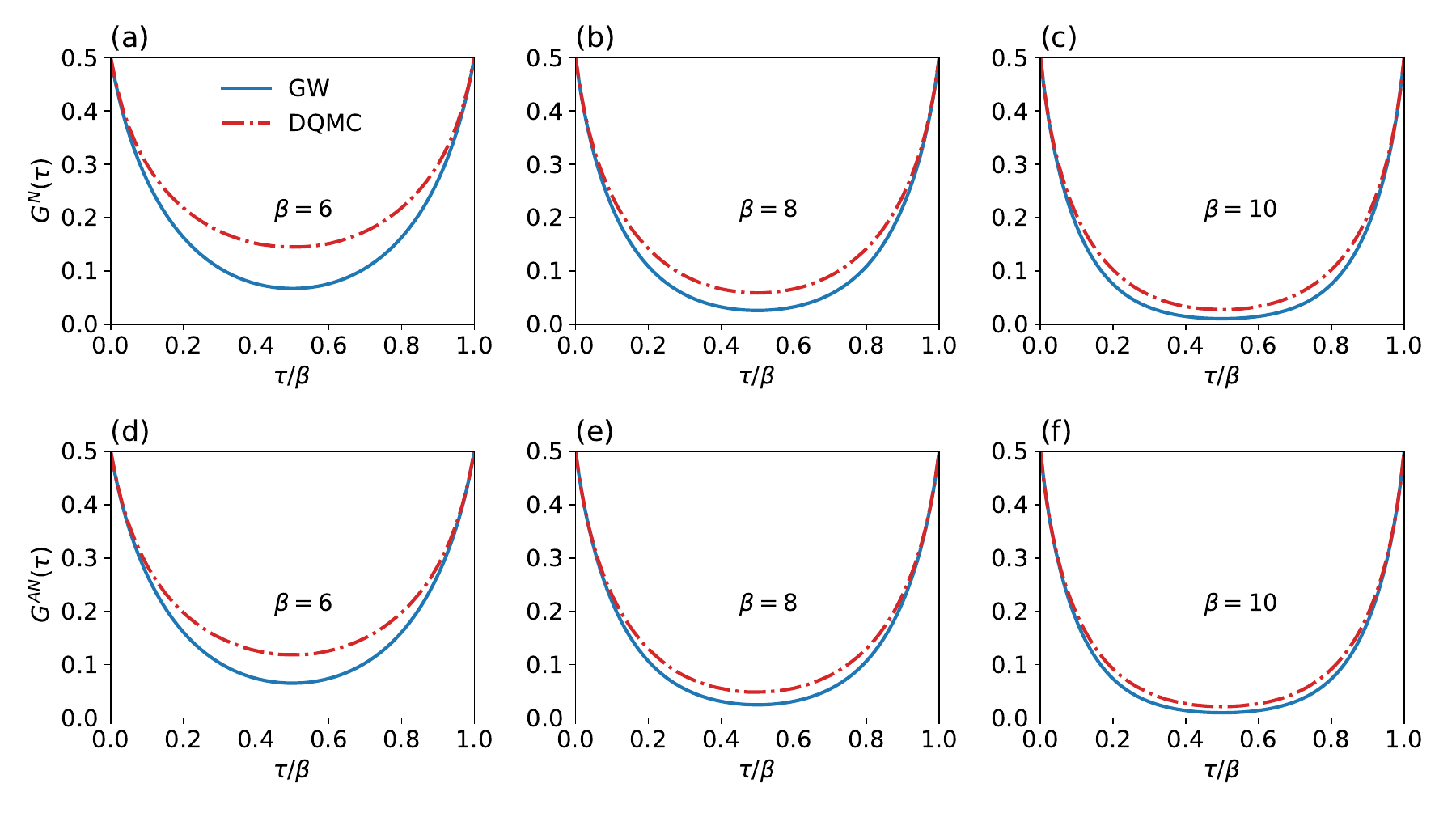}
    \caption{\label{fig:G tau u4}
        Imaginary-time Green's function $G(\mathbf{k},\tau)$ for the half-filled Hubbard model on a $16\times16$ lattice with $U=4$, comparing the symmetrized GW approximation (blue solid lines) and DQMC results (red dot-dashed lines). Panels (a)-(c) correspond to the nodal point $\mathbf{k} = (\pi/2, \pi/2)$, while panels (d)-(f) show the antinodal point $\mathbf{k} = (\pi, 0)$. The columns represent increasing inverse temperatures, i.e., (a,d) $\beta = 6$, (b,e) $\beta =8$, and (c,f) $\beta =10$.
    }
\end{figure}
\end{center}
We also investage the intermediate coupling $U=4$ system at half filling on $16\times 16$ lattice, results are shown in Fig.~\ref{fig:chi k w0 u4}, Fig.~\ref{fig:chi r w u4} and Fig.~\ref{fig:G tau u4} relatively. 
The $U=4$ results show very similar properties to Fig.~\ref{fig:chi k w0}, Fig.~\ref{fig:chi r w} and Fig.~\ref{fig:G tau}. 
For example, at low temperature away from the pseudo critical temperature $\beta_{\rm c}\simeq 0.22$, the GW agree with the corresponding DQMC benchmark.
This further indicates that the effectiveness of the symmetrization scheme in the strong coupling $U=8$ system is not accidental. \par

\FloatBarrier
\section{\label{D}On the $\chi$-sum rule and Pauli exclusion principle}

In the section, we present the derivation of the $\chi$-sum rule provided in Ref.~\cite{Vilk_Non-Perspective_1997} and discuss it.
Here we denote the spin-resolved density operator as $\hat{n}_{\alpha}(1)$ with $\alpha=\uparrow,\downarrow$ for clarity. 
The charge density operator is $\hat{n}(1)=\hat{n}_{\uparrow}(1)+\hat{n}_{\downarrow}(1)$, and the spin-z density operator is $\hat{S}^z(1)=\hat{n}_{\uparrow}(1)-\hat{n}_{\downarrow}(1)$. 
Using this notation, the onsite charge function at $\tau=0$ becomes
\begin{equation}
\begin{aligned}
    \chi_{\mathrm{ch}}(1,1) &\equiv 
    \langle \hat{n}(1)\hat{n}(1) \rangle_{\mathrm{C}}
    =\langle \hat{n}(1)\hat{n}(1) \rangle -\rho^2 \\
    &=\langle \hat{n}_{\uparrow}(1)\hat{n}_{\uparrow}(1) \rangle
    +\langle \hat{n}_{\downarrow}(1)\hat{n}_{\downarrow}(1) \rangle
    +2\langle \hat{n}_{\uparrow}(1)\hat{n}_{\downarrow}(1) \rangle -\rho^2, 
\end{aligned}
\end{equation}
where $\rho=\langle\hat{n}\rangle$ is the charge density. The spin correction becomes
\begin{equation}
\begin{aligned}
    \chi_{\mathrm{sp}}(1,1) &\equiv \langle \hat{S}^z(1)\hat{S}^z(1) \rangle \\
    &=\langle \hat{n}_{\uparrow}(1)\hat{n}_{\uparrow}(1) \rangle
    +\langle \hat{n}_{\downarrow}(1)\hat{n}_{\downarrow}(1) \rangle
    -2\langle \hat{n}_{\uparrow}(1)\hat{n}_{\downarrow}(1) \rangle.
\end{aligned}
\end{equation}
Combining these two equations, one has
\begin{equation}
    \chi_{\mathrm{ch}}(1,1)+\chi_{\mathrm{sp}}(1,1)
    =2\langle \hat{n}_{\uparrow}(1)\hat{n}_{\uparrow}(1) \rangle
    +2\langle \hat{n}_{\downarrow}(1)\hat{n}_{\downarrow}(1) \rangle-\rho^2.
\end{equation}
The Pauli exclusion principle indicates that the eigenvalues of a fermion density operator $\hat{n}_{\alpha}(1)$ are either $0$ or $1$. As a results, $\langle\hat{n}_{\alpha}(1)\hat{n}_{\alpha}(1)\rangle=\langle\hat{n}_{\alpha}(1)\rangle$, thus
\begin{equation}
    \chi_{\mathrm{ch}}(\tau=0,\bm{r}=\bm{0})+\chi_{\mathrm{sp}}(\tau=0,\bm{r}=\bm{0})=2\rho - \rho^2.\label{chi sum rule appendix}
\end{equation}
Eq.~(\ref{chi sum rule appendix}) is commonly known as the $\chi$-sum rule. The left hand side of Eq.~(\ref{chi sum rule appendix}) are two particle properties $\chi_{\mathrm{ch}}$ and $\chi_{\mathrm{sp}}$, while the right hand side is the charge density $\rho$, which a single particle property. \par

\begin{figure}[htb]
    \centering
    \includegraphics[width=.8\linewidth]{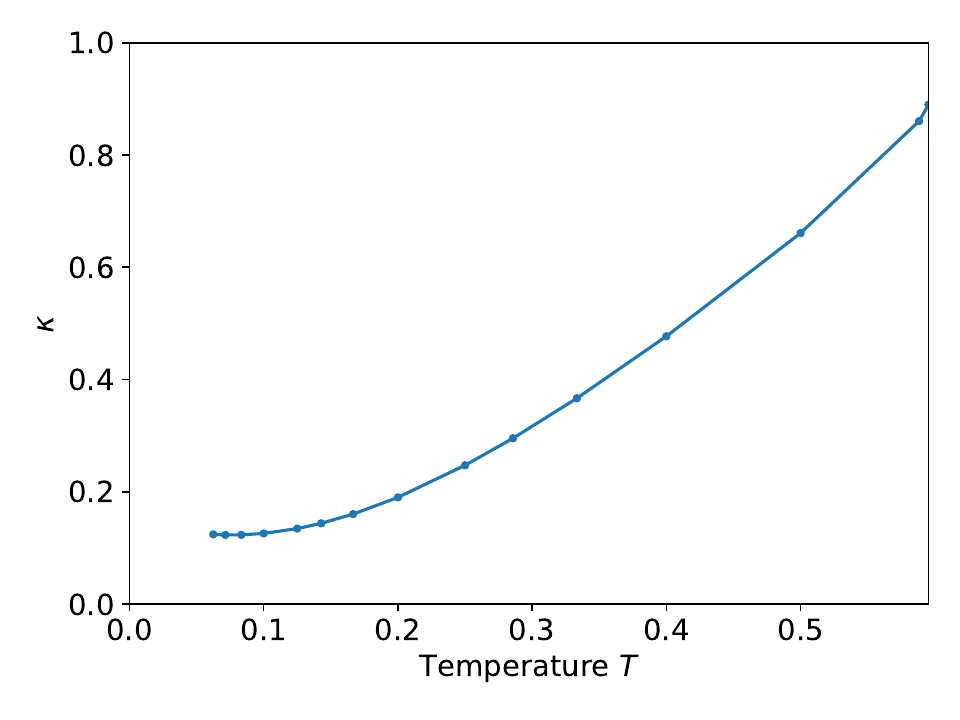}
    \caption{\label{fig:break chi sum rule}
        Deviation of $\chi$-sum rule, quantitatively judged by $\kappa$ as defined in Eq.~(\ref{relative error of chi sum rule}), of the pseudo AF solutions calculated by GW-covariance approach in $U=8$ half-filling Hubbard model on the $12\times 12$ lattice. 
    }
\end{figure}
It is meaningful to examine the degree to which our approach violates the Pauli exclusion principle.
However, we cannot directly quantify this property.
As a proxy, we investigate the deviation from the $\chi$-sum rule
by introducing
\begin{equation}
    \kappa\equiv \left|\, \frac{\chi_{\mathrm{ch}}(\tau=0,\bm{r}=\bm{0})+\chi_{\mathrm{sp}}(\tau=0,\bm{r}=\bm{0})}{2\rho - \rho^2}-1 \,\right|, \label{relative error of chi sum rule}
\end{equation}
which is the relative error between two- and single-particle properties in Eq.~(\ref{chi sum rule appendix}). 
The $\chi$-sum rule is one representation of the local momentum sum rules based on the Pauli exclusion principle, and thus is a necessary condition for a system to obey this principle.
We calculate $\kappa$, which reflects the deviations from the $\chi$-sum rule, for a strongly coupled system ($U=8$) at half-filling on a $12\times12$ lattice.
The results are presented in Fig.~\ref{fig:break chi sum rule}.
As shown in Fig.~\ref{fig:break chi sum rule}, $\kappa$ decreases from $\sim 90\%$ to $\sim 10\%$ as the temperature drops below the pseudo critical temperature $T_{\mathrm{c}}$.
This trend in $\kappa$ appears to correlate with the reliability of the results, supporting our conjecture in Sec.~\ref{chap:criterion}. 
In the region near $T_{\rm c}$ ($0.3<T<T_{\rm c}$), where the $\chi$-sum rule violation is large ($\kappa$ is high), the results of the symmetrized GW-covariance approximation (shown in Fig.~\ref{fig:chi r w}) are less reliable.
Conversely, at the deep low temperatures far from $T_{\mathrm{c}}$, where the $\chi$-sum rule is better satisfied ($\kappa$ is small), the symmetrized GW-covariance results exhibit better agreement with the DQMC benchmarks.

\FloatBarrier
\end{appendix}

% TODO:
% Provide your bibliography here. You have two options:

% FIRST OPTION - write your entries here directly, following the example below, including Author(s), Title, Journal Ref. with year in parentheses at the end, followed by the DOI number.
%\begin{thebibliography}{99}
%\bibitem{1931_Bethe_ZP_71} H. A. Bethe, {\it Zur Theorie der Metalle. i. Eigenwerte und Eigenfunktionen der linearen Atomkette}, Zeit. f{\"u}r Phys. {\bf 71}, 205 (1931), \doi{10.1007\%2FBF01341708}.
%\bibitem{arXiv:1108.2700} P. Ginsparg, {\it It was twenty years ago today... }, \url{http://arxiv.org/abs/1108.2700}.
%\end{thebibliography}

% \bibliographystyle{SciPost_bibstyle} % Include this style file here only if you are not using our template

\bibliography{4submit.bib}

\begin{thebibliography}{10}
\providecommand{\url}[1]{\texttt{#1}}
\providecommand{\urlprefix}{URL }
\expandafter\ifx\csname urlstyle\endcsname\relax
  \providecommand{\doi}[1]{doi:\discretionary{}{}{}#1}\else
  \providecommand{\doi}{doi:\discretionary{}{}{}\begingroup
  \urlstyle{rm}\Url}\fi
\providecommand{\eprint}[2][]{\url{#2}}

\bibitem{arovas_hubbard_2022}
D.~P. Arovas, E.~Berg, S.~A. Kivelson and S.~Raghu,
\newblock \emph{The {Hubbard} {Model}},
\newblock Annual Review of Condensed Matter Physics \textbf{13}(Volume 13,
  2022), 239 (2022),
\newblock \doi{https://doi.org/10.1146/annurev-conmatphys-031620-102024}.

\bibitem{qin_hubbard_2022}
M.~Qin, T.~Schäfer, S.~Andergassen, P.~Corboz and E.~Gull,
\newblock \emph{The {Hubbard} {Model}: {A} {Computational} {Perspective}},
\newblock Annual Review of Condensed Matter Physics \textbf{13}(1), 275 (2022),
\newblock \doi{10.1146/annurev-conmatphys-090921-033948}.

\bibitem{rohringer_diagrammatic_2018}
G.~Rohringer, H.~Hafermann, A.~Toschi, A.~A. Katanin, A.~E. Antipov, M.~I.
  Katsnelson, A.~I. Lichtenstein, A.~N. Rubtsov and K.~Held,
\newblock \emph{Diagrammatic routes to nonlocal correlations beyond dynamical
  mean field theory},
\newblock Reviews of Modern Physics \textbf{90}(2), 025003 (2018),
\newblock \doi{10.1103/RevModPhys.90.025003}.

\bibitem{schafer_tracking_2021}
T.~Sch\"afer, N.~Wentzell, F.~\ifmmode~\check{S}\else \v{S}\fi{}imkovic, Y.-Y.
  He, C.~Hille, M.~Klett, C.~J. Eckhardt, B.~Arzhang, V.~Harkov, F.~m. c.-M.
  Le~R\'egent, A.~Kirsch, Y.~Wang \emph{et~al.},
\newblock \emph{Tracking the footprints of spin fluctuations: A multimethod,
  multimessenger study of the two-dimensional hubbard model},
\newblock Phys. Rev. X \textbf{11}, 011058 (2021),
\newblock \doi{10.1103/PhysRevX.11.011058}.

\bibitem{Hao_Coexistence_Science2024}
H.~Xu, C.-M. Chung, M.~Qin, U.~Schollwöck, S.~R. White and S.~Zhang,
\newblock \emph{Coexistence of superconductivity with partially filled stripes
  in the hubbard model},
\newblock Science \textbf{384}(6696), eadh7691 (2024),
\newblock \doi{10.1126/science.adh7691},
\newblock \eprint{https://www.science.org/doi/pdf/10.1126/science.adh7691}.

\bibitem{Sakai_Nonperturbative_2023}
S.~Sakai,
\newblock \emph{Nonperturbative calculations for spectroscopic properties of
  cuprate high-temperature superconductors},
\newblock Journal of the Physical Society of Japan \textbf{92}(9), 092001
  (2023),
\newblock \doi{10.7566/JPSJ.92.092001},
\newblock \eprint{https://doi.org/10.7566/JPSJ.92.092001}.

\bibitem{Scalapino_a_common_2012}
D.~J. Scalapino,
\newblock \emph{A common thread: The pairing interaction for unconventional
  superconductors},
\newblock Rev. Mod. Phys. \textbf{84}, 1383 (2012),
\newblock \doi{10.1103/RevModPhys.84.1383}.

\bibitem{Gull_Continuous-time_2011}
E.~Gull, A.~J. Millis, A.~I. Lichtenstein, A.~N. Rubtsov, M.~Troyer and
  P.~Werner,
\newblock \emph{Continuous-time monte carlo methods for quantum impurity
  models},
\newblock Rev. Mod. Phys. \textbf{83}, 349 (2011),
\newblock \doi{10.1103/RevModPhys.83.349}.

\bibitem{Maier_Dynamical_2018}
T.~A. Maier,
\newblock \emph{Dynamical mean-field and dynamical cluster approximation based
  theory of superconductivity},
\newblock In E.~Pavarini, E.~Koch, A.~I. Lichtenstein, D.~Vollhardt, {Institute
  for Advanced Simulation} and {Forschungszentrum Jülich}, eds., \emph{{DMFT}:
  From infinite dimensions to real materials}, no. Band/volume 8 in Schriften
  des Forschungszentrums Jülich Reihe Modeling and Simulation, chap.~13.
  Forschungszentrum, Zentralbibliothek,
\newblock ISBN 978-3-95806-313-6,
\newblock Meeting Name: Autumn School on Correlated Electrons (2018).

\bibitem{kubo_Statistical_I_1957}
R.~Kubo,
\newblock \emph{Statistical-mechanical theory of irreversible processes. i.
  general theory and simple applications to magnetic and conduction problems},
\newblock Journal of the Physical Society of Japan \textbf{12}(6), 570 (1957),
\newblock \doi{10.1143/JPSJ.12.570},
\newblock \eprint{https://doi.org/10.1143/JPSJ.12.570}.

\bibitem{kubo_Statistical_II_1957}
R.~Kubo, M.~Yokota and S.~Nakajima,
\newblock \emph{Statistical-mechanical theory of irreversible processes. ii.
  response to thermal disturbance},
\newblock Journal of the Physical Society of Japan \textbf{12}(11), 1203
  (1957),
\newblock \doi{10.1143/JPSJ.12.1203},
\newblock \eprint{https://doi.org/10.1143/JPSJ.12.1203}.

\bibitem{Callen_Irreversibility_1951}
H.~B. Callen and T.~A. Welton,
\newblock \emph{Irreversibility and generalized noise},
\newblock Phys. Rev. \textbf{83}, 34 (1951),
\newblock \doi{10.1103/PhysRev.83.34}.

\bibitem{Mahan_Many-particle_2013}
G.~D. Mahan,
\newblock \emph{Many-particle physics},
\newblock Springer Science \& Business Media (2013).

\bibitem{stefanucci_nonequilibrium_2013}
G.~Stefanucci and R.~v. Leeuwen,
\newblock \emph{Nonequilibrium many-body theory of quantum systems: a modern
  introduction},
\newblock Cambridge University Press, Cambridge,
\newblock ISBN 978-0-521-76617-3 (2013).

\bibitem{Benfatto_Ward_2005}
L.~Benfatto, S.~G. Sharapov, N.~Andrenacci and H.~Beck,
\newblock \emph{Ward identity and optical conductivity sum rule in the
  $d$-density wave state},
\newblock Phys. Rev. B \textbf{71}, 104511 (2005),
\newblock \doi{10.1103/PhysRevB.71.104511}.

\bibitem{Vilk_Non-Perspective_1997}
{Y.M. Vilk} and {A.-M.S. Tremblay},
\newblock \emph{Non-perturbative many-body approach to the hubbard model and
  single-particle pseudogap},
\newblock J. Phys. I France \textbf{7}(11), 1309 (1997),
\newblock \doi{10.1051/jp1:1997135}.

\bibitem{Bergeron_Optical_2011}
D.~Bergeron, V.~Hankevych, B.~Kyung and A.-M.~S. Tremblay,
\newblock \emph{Optical and dc conductivity of the two-dimensional hubbard
  model in the pseudogap regime and across the antiferromagnetic quantum
  critical point including vertex corrections},
\newblock Phys. Rev. B \textbf{84}, 085128 (2011),
\newblock \doi{10.1103/PhysRevB.84.085128}.

\bibitem{Tremblay_Improved_2023}
C.~Gauvin-Ndiaye, C.~Lahaie, Y.~M. Vilk and A.-M.~S. Tremblay,
\newblock \emph{Improved two-particle self-consistent approach for the
  single-band hubbard model in two dimensions},
\newblock Phys. Rev. B \textbf{108}, 075144 (2023),
\newblock \doi{10.1103/PhysRevB.108.075144}.

\bibitem{Vilk_Antiferromagnetic_2024}
Y.~M. Vilk, C.~Lahaie and A.-M.~S. Tremblay,
\newblock \emph{Antiferromagnetic pseudogap in the two-dimensional hubbard
  model deep in the renormalized classical regime},
\newblock Phys. Rev. B \textbf{110}, 125154 (2024),
\newblock \doi{10.1103/PhysRevB.110.125154}.

\bibitem{Hirsch_Two-dimensional_1985}
J.~E. Hirsch,
\newblock \emph{Two-dimensional hubbard model: Numerical simulation study},
\newblock Phys. Rev. B \textbf{31}, 4403 (1985),
\newblock \doi{10.1103/PhysRevB.31.4403}.

\bibitem{Chang_recent_2015}
C.-C. Chang, S.~Gogolenko, J.~Perez, Z.~Bai and R.~T. Scalettar,
\newblock \emph{Recent advances in determinant quantum monte carlo},
\newblock Philosophical Magazine \textbf{95}(12), 1260 (2015),
\newblock \doi{10.1080/14786435.2013.845314}.

\bibitem{SmoQyDQMC.jl}
B.~Cohen-Stead, S.~M. Costa, J.~Neuhaus, A.~T. Ly, Y.~Zhang, R.~Scalettar,
  K.~Barros and S.~Johnston,
\newblock \emph{{SmoQyDQMC.jl: A flexible implementation of determinant quantum
  Monte Carlo for Hubbard and electron-phonon interactions}},
\newblock SciPost Phys. Codebases p.~29 (2024),
\newblock \doi{10.21468/SciPostPhysCodeb.29}.

\bibitem{sun_delay_2024}
F.~Sun and X.~Y. Xu,
\newblock \emph{Delay update in determinant quantum monte carlo},
\newblock Phys. Rev. B \textbf{109}, 235140 (2024),
\newblock \doi{10.1103/PhysRevB.109.235140}.

\bibitem{iglovikov_geometry_2015}
V.~I. Iglovikov, E.~Khatami and R.~T. Scalettar,
\newblock \emph{Geometry dependence of the sign problem in quantum monte carlo
  simulations},
\newblock Phys. Rev. B \textbf{92}, 045110 (2015),
\newblock \doi{10.1103/PhysRevB.92.045110}.

\bibitem{Varney_Quantum_2009}
C.~N. Varney, C.-R. Lee, Z.~J. Bai, S.~Chiesa, M.~Jarrell and R.~T. Scalettar,
\newblock \emph{Quantum monte carlo study of the two-dimensional fermion
  hubbard model},
\newblock Phys. Rev. B \textbf{80}, 075116 (2009),
\newblock \doi{10.1103/PhysRevB.80.075116}.

\bibitem{Prokofev_Polaron_1998}
N.~V. Prokof'ev and B.~V. Svistunov,
\newblock \emph{Polaron problem by diagrammatic quantum monte carlo},
\newblock Phys. Rev. Lett. \textbf{81}, 2514 (1998),
\newblock \doi{10.1103/PhysRevLett.81.2514}.

\bibitem{Kris_Diagrammatic_2010}
K.~{Van Houcke}, E.~Kozik, N.~Prokof’ev and B.~Svistunov,
\newblock \emph{Diagrammatic monte carlo},
\newblock Physics Procedia \textbf{6}, 95 (2010),
\newblock \doi{https://doi.org/10.1016/j.phpro.2010.09.034},
\newblock Computer Simulations Studies in Condensed Matter Physics XXI.

\bibitem{Rossi_Determinant_2017}
R.~Rossi,
\newblock \emph{Determinant diagrammatic monte carlo algorithm in the
  thermodynamic limit},
\newblock Phys. Rev. Lett. \textbf{119}, 045701 (2017),
\newblock \doi{10.1103/PhysRevLett.119.045701}.

\bibitem{Paiva_Critical_2004}
T.~Paiva, R.~R. dos Santos, R.~T. Scalettar and P.~J.~H. Denteneer,
\newblock \emph{Critical temperature for the two-dimensional attractive hubbard
  model},
\newblock Phys. Rev. B \textbf{69}, 184501 (2004),
\newblock \doi{10.1103/PhysRevB.69.184501}.

\bibitem{Lenihan_Evaluating_2022}
C.~Lenihan, A.~J. Kim, F.~\ifmmode~\check{S}\else \v{S}\fi{}imkovic and
  E.~Kozik,
\newblock \emph{Evaluating second-order phase transitions with diagrammatic
  monte carlo: N\'eel transition in the doped three-dimensional hubbard model},
\newblock Phys. Rev. Lett. \textbf{129}, 107202 (2022),
\newblock \doi{10.1103/PhysRevLett.129.107202}.

\bibitem{Garioud_symmetry-Broken_2024}
R.~Garioud, F.~\ifmmode~\check{S}\else \v{S}\fi{}imkovic, R.~Rossi, G.~Spada,
  T.~Sch\"afer, F.~Werner and M.~Ferrero,
\newblock \emph{Symmetry-broken perturbation theory to large orders in
  antiferromagnetic phases},
\newblock Phys. Rev. Lett. \textbf{132}, 246505 (2024),
\newblock \doi{10.1103/PhysRevLett.132.246505}.

\bibitem{migdal_interaction_1958}
A.~Migdal,
\newblock \emph{Interaction between electrons and lattice vibrations in a
  normal metal},
\newblock Sov. Phys. JETP \textbf{7}(6), 996 (1958).

\bibitem{hedin_new_1965}
L.~Hedin,
\newblock \emph{New method for calculating the one-particle green's function
  with application to the electron-gas problem},
\newblock Phys. Rev. \textbf{139}, A796 (1965),
\newblock \doi{10.1103/PhysRev.139.A796}.

\bibitem{mermin_absence_1966}
N.~D. Mermin and H.~Wagner,
\newblock \emph{Absence of {Ferromagnetism} or {Antiferromagnetism} in {One}-
  or {Two}-{Dimensional} {Isotropic} {Heisenberg} {Models}},
\newblock Physical Review Letters \textbf{17}(22), 1133 (1966),
\newblock \doi{10.1103/PhysRevLett.17.1133}.

\bibitem{Yu_Tensor_2014}
J.~F. Yu, Z.~Y. Xie, Y.~Meurice, Y.~Liu, A.~Denbleyker, H.~Zou, M.~P. Qin,
  J.~Chen and T.~Xiang,
\newblock \emph{Tensor renormalization group study of classical $xy$ model on
  the square lattice},
\newblock Phys. Rev. E \textbf{89}, 013308 (2014),
\newblock \doi{10.1103/PhysRevE.89.013308}.

\bibitem{schmoll_classical_2021}
P.~Schmoll, A.~Kshetrimayum, J.~Eisert, R.~Orús and M.~Rizzi,
\newblock \emph{{The classical two-dimensional Heisenberg model revisited: An
  $SU(2)$-symmetric tensor network study}},
\newblock SciPost Phys. \textbf{11}, 098 (2021),
\newblock \doi{10.21468/SciPostPhys.11.5.098}.

\bibitem{Ueda_Tensor_2022}
A.~Ueda and M.~Oshikawa,
\newblock \emph{Tensor network renormalization study on the crossover in
  classical heisenberg and {${\mathrm{RP}}^{2}$} models in two dimensions},
\newblock Phys. Rev. E \textbf{106}, 014104 (2022),
\newblock \doi{10.1103/PhysRevE.106.014104}.

\bibitem{Burgelman_Contrasting_2023}
L.~Burgelman, L.~Devos, B.~Vanhecke, F.~Verstraete and L.~Vanderstraeten,
\newblock \emph{Contrasting pseudocriticality in the classical two-dimensional
  heisenberg and {${\mathrm{RP}}^{2}$} models: Zero-temperature phase
  transition versus finite-temperature crossover},
\newblock Phys. Rev. E \textbf{107}, 014117 (2023),
\newblock \doi{10.1103/PhysRevE.107.014117}.

\bibitem{Shenker_Monte-Carlo_1980}
S.~H. Shenker and J.~Tobochnik,
\newblock \emph{Monte carlo renormalization-group analysis of the classical
  heisenberg model in two dimensions},
\newblock Phys. Rev. B \textbf{22}, 4462 (1980),
\newblock \doi{10.1103/PhysRevB.22.4462}.

\bibitem{Tomita_Finite-size_2014}
Y.~Tomita,
\newblock \emph{Finite-size scaling analysis of pseudocritical region in
  two-dimensional continuous-spin systems},
\newblock Phys. Rev. E \textbf{90}, 032109 (2014),
\newblock \doi{10.1103/PhysRevE.90.032109}.

\bibitem{Stanley_Possibility_1966}
H.~E. Stanley and T.~A. Kaplan,
\newblock \emph{Possibility of a phase transition for the two-dimensional
  {Heisenberg} model},
\newblock Phys. Rev. Lett. \textbf{17}, 913 (1966),
\newblock \doi{10.1103/PhysRevLett.17.913}.

\bibitem{Adler_High-temperature_1993}
J.~Adler, C.~Holm and W.~Janke,
\newblock \emph{High-temperature series analyses of the classical {Heisenberg}
  and {XY} models},
\newblock Physica A: Statistical Mechanics and its Applications
  \textbf{201}(4), 581 (1993).

\bibitem{Jevicki_Ground_PLB1977}
A.~Jevicki,
\newblock \emph{On the ground state and infrared divergences of goldstone
  bosons in two dimensions},
\newblock Physics Letters B \textbf{71}(2), 327 (1977),
\newblock \doi{https://doi.org/10.1016/0370-2693(77)90229-5}.

\bibitem{Elitzur_applicability_1983}
S.~Elitzur,
\newblock \emph{The applicability of perturbation expansion to two-dimensional
  goldstone systems},
\newblock Nuclear Physics B \textbf{212}(3), 501 (1983),
\newblock \doi{https://doi.org/10.1016/0550-3213(83)90682-X}.

\bibitem{David_Cancellations_CMP1981}
F.~David,
\newblock \emph{Cancellations of infrared divergences in the two-dimensional
  non-linear $\sigma$ models},
\newblock Communications in Mathematical Physics \textbf{81}, 149 (1981),
\newblock \doi{10.1007/BF01208892}.

\bibitem{Eilenberger_Thermodynamic_1967}
G.~Eilenberger,
\newblock \emph{Thermodynamic fluctuations of the order parameter in {Type-II}
  superconductors near the upper critical field ${H}_{c2}$},
\newblock Phys. Rev. \textbf{164}, 628 (1967),
\newblock \doi{10.1103/PhysRev.164.628}.

\bibitem{Maki_Thermodynamic_1971}
K.~Maki and H.~Takayama,
\newblock \emph{Thermodynamic fluctuation of the order parameter in the vortex
  state of type ii superconductors},
\newblock Progress of Theoretical Physics \textbf{46}(6), 1651 (1971),
\newblock \doi{10.1143/PTP.46.1651},
\newblock
  \eprint{https://academic.oup.com/ptp/article-pdf/46/6/1651/5453312/46-6-1651.pdf}.

\bibitem{Rosenstein_First-principles_1999}
B.~Rosenstein,
\newblock \emph{First-principles theory of fluctuations in vortex liquids and
  solids},
\newblock Phys. Rev. B \textbf{60}, 4268 (1999),
\newblock \doi{10.1103/PhysRevB.60.4268}.

\bibitem{Li_Thermal_2001}
D.~Li and B.~Rosenstein,
\newblock \emph{Thermal fluctuation correction to magnetization and specific
  heat of vortex solids in {type-II} superconductors},
\newblock Phys. Rev. B \textbf{65}, 024514 (2001),
\newblock \doi{10.1103/PhysRevB.65.024514}.

\bibitem{Li_Thermal_1999}
D.~Li and B.~Rosenstein,
\newblock \emph{Thermal fluctuations and disorder effects in vortex lattices},
\newblock Phys. Rev. B \textbf{60}, 10460 (1999),
\newblock \doi{10.1103/PhysRevB.60.10460}.

\bibitem{Li_Melting_2002}
D.~Li and B.~Rosenstein,
\newblock \emph{Melting of the vortex lattice in high$\ensuremath{-}{T}_{c}$
  superconductors},
\newblock Phys. Rev. B \textbf{65}, 220504 (2002),
\newblock \doi{10.1103/PhysRevB.65.220504}.

\bibitem{Hikami_Magnetic-flux-lattice_1991}
S.~Hikami, A.~Fujita and A.~I. Larkin,
\newblock \emph{Magnetic-flux-lattice melting in a strong magnetic field},
\newblock Phys. Rev. B \textbf{44}, 10400 (1991),
\newblock \doi{10.1103/PhysRevB.44.10400}.

\bibitem{Li_Supercooled_2004}
D.~Li and B.~Rosenstein,
\newblock \emph{Supercooled vortex liquid and quantitative theory of melting of
  the flux-line lattice in {type-II} superconductors},
\newblock Phys. Rev. B \textbf{70}, 144521 (2004),
\newblock \doi{10.1103/PhysRevB.70.144521}.

\bibitem{Rosenstein_Ginzburg-Landau_2010}
B.~Rosenstein and D.~Li,
\newblock \emph{Ginzburg-landau theory of {type II} superconductors in magnetic
  field},
\newblock Rev. Mod. Phys. \textbf{82}, 109 (2010),
\newblock \doi{10.1103/RevModPhys.82.109}.

\bibitem{Kokubo_Dynamic_2007}
N.~Kokubo, T.~Asada, K.~Kadowaki, K.~Takita, T.~G. Sorop and P.~H. Kes,
\newblock \emph{Dynamic ordering of driven vortex matter in the peak effect
  regime of amorphous moge films and
  {$2\mathrm{H}\text{\ensuremath{-}}\mathrm{Nb}{\mathrm{Se}}_{2}$} crystals},
\newblock Phys. Rev. B \textbf{75}, 184512 (2007),
\newblock \doi{10.1103/PhysRevB.75.184512}.

\bibitem{Koshelev_Melting_2019}
A.~E. Koshelev, K.~Willa, R.~Willa, M.~P. Smylie, J.-K. Bao, D.~Y. Chung, M.~G.
  Kanatzidis, W.-K. Kwok and U.~Welp,
\newblock \emph{Melting of vortex lattice in the magnetic superconductor
  ${\mathrm{rbeufe}}_{4}{\mathrm{as}}_{4}$},
\newblock Phys. Rev. B \textbf{100}, 094518 (2019),
\newblock \doi{10.1103/PhysRevB.100.094518}.

\bibitem{rosenstein_mean_2019}
B.~Rosenstein, D.~Li, T.~Ma and H.~C. Kao,
\newblock \emph{Mean field theory of short-range order in strongly correlated
  low dimensional electronic systems},
\newblock Physical Review B \textbf{100}(12), 125140 (2019),
\newblock \doi{10.1103/PhysRevB.100.125140}.

\bibitem{Hui_Linear_PRB2023}
H.~Li, Z.~Sun, Y.~Su, H.~Lin, H.~Huang and D.~Li,
\newblock \emph{Linear response functions respecting {Ward-Takahashi} identity
  and fluctuation-dissipation theorem within the {$GW$} approximation},
\newblock Phys. Rev. B \textbf{107}, 085106 (2023),
\newblock \doi{10.1103/PhysRevB.107.085106}.

\bibitem{xiao_observation_2004}
Z.~L. Xiao, O.~Dogru, E.~Y. Andrei, P.~Shuk and M.~Greenblatt,
\newblock \emph{Observation of the vortex lattice spinodal in
  ${\mathrm{nbse}}_{2}$},
\newblock Phys. Rev. Lett. \textbf{92}, 227004 (2004),
\newblock \doi{10.1103/PhysRevLett.92.227004}.

\bibitem{thakur_exploring_2005}
A.~D. Thakur, S.~S. Banerjee, M.~J. Higgins, S.~Ramakrishnan and A.~K. Grover,
\newblock \emph{Exploring metastability via third harmonic measurements in
  single crystals of $2h\text{\ensuremath{-}}{\mathrm{nbse}}_{2}$ showing an
  anomalous peak effect},
\newblock Phys. Rev. B \textbf{72}, 134524 (2005),
\newblock \doi{10.1103/PhysRevB.72.134524}.

\bibitem{Adesso_Transition_2006}
M.~G. Adesso, D.~Uglietti, R.~Fl\"ukiger, M.~Polichetti and S.~Pace,
\newblock \emph{Transition between the bragg glass and the disordered phase in
  ${\mathrm{nb}}_{3}\mathrm{Sn}$ detected by third harmonics of the ac magnetic
  susceptibility},
\newblock Phys. Rev. B \textbf{73}, 092513 (2006),
\newblock \doi{10.1103/PhysRevB.73.092513}.

\bibitem{Li_Fluctuation_2012}
Q.~Li,
\newblock \emph{Fluctuations in Bose-Einstein Condensates and Type II
  superconductors},
\newblock Phd thesis, Peking University (2012).

\bibitem{Aubert_Invariant_2003}
S.~Aubert and C.~S. Lam,
\newblock \emph{Invariant integration over the unitary group},
\newblock Journal of Mathematical Physics \textbf{44}(12), 6112 (2003),
\newblock \doi{10.1063/1.1622448}.

\bibitem{hashimoto_energy_2014}
M.~Hashimoto, I.~M. Vishik, R.-H. He, T.~P. Devereaux and Z.-X. Shen,
\newblock \emph{Energy gaps in high-transition-temperature cuprate
  superconductors},
\newblock Nature Physics \textbf{10}(7), 483 (2014),
\newblock \doi{10.1038/nphys3009}.

\bibitem{altland_condensed_2010}
A.~Altland and B.~Simons,
\newblock \emph{Condensed matter field theory},
\newblock Cambridge University Press, 2nd ed edn.,
\newblock ISBN 978-0-521-76975-4.

\bibitem{negele_quantum_2018}
J.~W. Negele and H.~Orland,
\newblock \emph{Quantum Many-Particle Systems},
\newblock {CRC} Press, 1 edn.,
\newblock ISBN 978-0-429-49792-6,
\newblock \doi{10.1201/9780429497926}.

\bibitem{stoof_ultracold_2009}
H.~T.~C. Stoof, K.~B. Gubbels and D.~B.~M. Dickerscheid,
\newblock \emph{Ultracold quantum fields},
\newblock Theoretical and mathematical physics. Springer,
\newblock ISBN 978-1-4020-8762-2 978-1-4020-8763-9,
\newblock {OCLC}: ocn310153808.

\bibitem{aryasetiawan_generalized_2008}
F.~Aryasetiawan and S.~Biermann,
\newblock \emph{Generalized {Hedin}'s {Equations} for {Quantum} {Many}-{Body}
  {Systems} with {Spin}-{Dependent} {Interactions}},
\newblock Physical Review Letters \textbf{100}(11), 116402 (2008),
\newblock \doi{10.1103/PhysRevLett.100.116402}.

\bibitem{Nieves_Generalized_2004}
J.~F. Nieves and P.~B. Pal,
\newblock \emph{Generalized fierz identities},
\newblock American Journal of Physics \textbf{72}(8), 1100 (2004),
\newblock \doi{10.1119/1.1757445},
\newblock
  \eprint{https://pubs.aip.org/aapt/ajp/article-pdf/72/8/1100/10126610/1100_1_online.pdf}.

\bibitem{Li_Quantum_2023}
H.~Li,
\newblock \emph{Quantum Many-Body Self-Consistent Theory and Its Applications},
\newblock Phd thesis, Peking University (2023).

\bibitem{xiong_application_2025}
J.~Xiong, H.~Li, Y.~Su and D.~Li,
\newblock \emph{Application of many-body nonperturbative theories to the
  three-dimensional attractive hubbard model},
\newblock Phys. Rev. B \textbf{112}, 064509 (2025),
\newblock \doi{10.1103/973p-gphs}.

\bibitem{wang_dc_2020}
W.~O. Wang, J.~K. Ding, B.~Moritz, E.~W. Huang and T.~P. Devereaux,
\newblock \emph{{DC} {Hall} coefficient of the strongly correlated {Hubbard}
  model},
\newblock npj Quantum Materials \textbf{5}(1), 51 (2020),
\newblock \doi{10.1038/s41535-020-00254-w}.

\bibitem{White_Density_1992}
S.~R. White,
\newblock \emph{Density matrix formulation for quantum renormalization groups},
\newblock Phys. Rev. Lett. \textbf{69}, 2863 (1992),
\newblock \doi{10.1103/PhysRevLett.69.2863}.

\bibitem{White_Density-matrix_1993}
S.~R. White,
\newblock \emph{Density-matrix algorithms for quantum renormalization groups},
\newblock Phys. Rev. B \textbf{48}, 10345 (1993),
\newblock \doi{10.1103/PhysRevB.48.10345}.

\bibitem{Schollwock_density-matrix_2005}
U.~Schollw\"ock,
\newblock \emph{The density-matrix renormalization group},
\newblock Rev. Mod. Phys. \textbf{77}, 259 (2005),
\newblock \doi{10.1103/RevModPhys.77.259}.

\bibitem{scholle_Comprehensive_2023}
R.~Scholle, P.~M. Bonetti, D.~Vilardi and W.~Metzner,
\newblock \emph{Comprehensive mean-field analysis of magnetic and charge orders
  in the two-dimensional hubbard model},
\newblock Phys. Rev. B \textbf{108}, 035139 (2023),
\newblock \doi{10.1103/PhysRevB.108.035139}.

\bibitem{scholle_spiral_2024}
R.~Scholle, W.~Metzner, D.~Vilardi and P.~M. Bonetti,
\newblock \emph{Spiral to stripe transition in the two-dimensional hubbard
  model},
\newblock Phys. Rev. B \textbf{109}, 235149 (2024),
\newblock \doi{10.1103/PhysRevB.109.235149}.

\bibitem{Vu_Trilex_PRB2017}
J.~Vu\ifmmode \check{c}\else \v{c}\fi{}i\ifmmode \check{c}\else
  \v{c}\fi{}evi\ifmmode~\acute{c}\else \'{c}\fi{}, T.~Ayral and O.~Parcollet,
\newblock \emph{{TRILEX} and {$GW$+EDMFT} approach to $d$-wave
  superconductivity in the {Hubbard} model},
\newblock Phys. Rev. B \textbf{96}, 104504 (2017),
\newblock \doi{10.1103/PhysRevB.96.104504}.

\bibitem{tong_university_nodate}
D.~Tong,
\newblock \emph{University of {Cambridge} {Part} {III} {Mathematical} {Tripos}}
  (2017).

\bibitem{iskakov_phase_2022}
S.~Iskakov and E.~Gull,
\newblock \emph{Phase transitions in partial summation methods: {Results} from
  the three-dimensional {Hubbard} model},
\newblock Physical Review B \textbf{105}(4), 045109 (2022),
\newblock \doi{10.1103/PhysRevB.105.045109},
\newblock Publisher: American Physical Society.

\bibitem{xu_neutral_2025}
M.~Xu, L.~H. Kendrick, A.~Kale, Y.~Gang, C.~Feng, S.~Zhang, A.~W. Young,
  M.~Lebrat and M.~Greiner,
\newblock \emph{A neutral-atom hubbard quantum simulator in the cryogenic
  regime},
\newblock Nature pp. 1--7 (2025),
\newblock \doi{10.1038/s41586-025-09112-w}.

\bibitem{jiang_ground_2021}
S.~Jiang, D.~J. Scalapino and S.~R. White,
\newblock \emph{Ground-state phase diagram of the t-t'-j model},
\newblock Proceedings of the National Academy of Sciences \textbf{118}(44),
  e2109978118 (2021).

\bibitem{gong_robust_2021}
S.~Gong, W.~Zhu and D.~Sheng,
\newblock \emph{Robust d-wave superconductivity in the square-lattice t-j
  model},
\newblock Physical Review Letters \textbf{127}(9), 097003 (2021).

\bibitem{jiang_pairing_2022}
S.~Jiang, D.~J. Scalapino and S.~R. White,
\newblock \emph{Pairing properties of the t-t'-t''-j model},
\newblock Physical Review B \textbf{106}(17), 174507 (2022).

\bibitem{jiang_superconducting_2023}
H.-C. Jiang, S.~A. Kivelson and D.-H. Lee,
\newblock \emph{Superconducting valence bond fluid in lightly doped eight-leg
  t-j cylinders},
\newblock Physical Review B \textbf{108}(5), 054505 (2023).

\bibitem{lu_emergent_2024}
X.~Lu, F.~Chen, W.~Zhu, D.~N. Sheng and S.-S. Gong,
\newblock \emph{Emergent superconductivity and competing charge orders in
  hole-doped square-lattice t-j model},
\newblock Physical review letters \textbf{132}(6), 066002 (2024).

\bibitem{qu_phase_2024}
D.-W. Qu, Q.~Li, S.-S. Gong, Y.~Qi, W.~Li and G.~Su,
\newblock \emph{Phase diagram, d-wave superconductivity, and pseudogap of the
  t-t'-j model at finite temperature},
\newblock Physical Review Letters \textbf{133}(25), 256003 (2024).

\bibitem{li_tangent_2023}
Q.~Li, Y.~Gao, Y.-Y. He, Y.~Qi, B.-B. Chen and W.~Li,
\newblock \emph{Tangent space approach for thermal tensor network simulations
  of the 2d hubbard model},
\newblock Physical Review Letters \textbf{130}(22), 226502 (2023).

\bibitem{jiang_ground_2024}
Y.-F. Jiang, T.~P. Devereaux and H.-C. Jiang,
\newblock \emph{Ground-state phase diagram and superconductivity of the doped
  hubbard model on six-leg square cylinders},
\newblock Physical Review B \textbf{109}(8), 085121 (2024).

\bibitem{Shinaoka_Compressing_2017}
H.~Shinaoka, J.~Otsuki, M.~Ohzeki and K.~Yoshimi,
\newblock \emph{Compressing green's function using intermediate representation
  between imaginary-time and real-frequency domains},
\newblock Phys. Rev. B \textbf{96}, 035147 (2017),
\newblock \doi{10.1103/PhysRevB.96.035147}.

\bibitem{Chikano_irbasis_2019}
N.~Chikano, K.~Yoshimi, J.~Otsuki and H.~Shinaoka,
\newblock \emph{irbasis: Open-source database and software for
  intermediate-representation basis functions of imaginary-time green's
  function},
\newblock Computer Physics Communications \textbf{240}, 181 (2019),
\newblock \doi{https://doi.org/10.1016/j.cpc.2019.02.006}.

\bibitem{Gull_Chebyshev_2018}
E.~Gull, S.~Iskakov, I.~Krivenko, A.~A. Rusakov and D.~Zgid,
\newblock \emph{Chebyshev polynomial representation of imaginary-time response
  functions},
\newblock Phys. Rev. B \textbf{98}, 075127 (2018),
\newblock \doi{10.1103/PhysRevB.98.075127}.

\bibitem{Dong_Legendre-spectral_2020}
X.~Dong, D.~Zgid, E.~Gull and H.~U.~R. Strand,
\newblock \emph{Legendre-spectral dyson equation solver with super-exponential
  convergence},
\newblock The Journal of Chemical Physics \textbf{152}(13), 134107 (2020),
\newblock \doi{10.1063/5.0003145}.

\bibitem{ma_low-temperature_2024}
X.~Ma, M.~Zeng, H.~Guo and S.~Feng,
\newblock \emph{Low-temperature ${T}^{2}$ resistivity in the underdoped
  pseudogap phase versus $t$-linear resistivity in the overdoped strange-metal
  phase of cuprate superconductors},
\newblock Phys. Rev. B \textbf{110}, 094520 (2024),
\newblock \doi{10.1103/PhysRevB.110.094520}.

\bibitem{ma_correlation_2025}
X.~Ma, M.~Zeng, H.~Guo and S.~Feng,
\newblock \emph{Correlation between the strength of low-temperature t-linear
  resistivity and tc in overdoped electron-doped cuprate superconductors},
\newblock Philosophical Magazine \textbf{0}(0), 1 (2025),
\newblock \doi{10.1080/14786435.2025.2468808},
\newblock \eprint{https://doi.org/10.1080/14786435.2025.2468808}.

\bibitem{su_effects_2025}
Y.~Su, H.~Li, H.~Huang and D.~Li,
\newblock \emph{Effects of the pseudogap and the fermi surface on the rapid
  hall-coefficient changes in cuprates},
\newblock Phys. Rev. B \textbf{111}, 064518 (2025),
\newblock \doi{10.1103/PhysRevB.111.064518}.

\end{thebibliography}

\nolinenumbers

\end{document}